\documentclass[11pt]{article}
\usepackage{comment,soul}
\usepackage[hidelinks]{hyperref}
\usepackage{enumitem}
\usepackage[normalem]{ulem}
\usepackage{amsmath,amssymb,epsf,cite,graphicx,subfigure}
\usepackage{braket,tikzsymbols}
\usepackage[bbgreekl]{mathbbol}
\usepackage{comment}
\usepackage[export]{adjustbox}
\usepackage{dsfont}
\usepackage{faktor}
\usepackage{mathrsfs}
\usepackage{float}
\usepackage{empheq}
\usepackage{tikz}

\usepackage{comment}

\numberwithin{equation}{section}
\newcommand{\nn}{\nonumber}

\definecolor{airforceblue}{rgb}{0.36, 0.54, 0.66}
\newcommand{\beq}{\begin{equation}}
\newcommand{\eeq}{\end{equation}}

\DeclareMathOperator{\Vol}{vol}

\begin{document}
\baselineskip=15.5pt
\pagestyle{plain}
\setcounter{page}{1}
%--------+---------+---------+---------+---------+---------+---------+

\begin{center}
{\LARGE \bf Runaways in the no-boundary wavefunction}
\vskip 1cm

\textbf{Jordan Cotler$^{1,a}$, Prateksh Dhivakar$^{2,b}$, William Harvey$^{2,c}$, \\ Victor Ivo$^{3,d}$, Kristan Jensen$^{2,e}$}

\vspace{0.5cm}

{\it ${}^1$Department of Physics, Harvard University, Cambridge, MA 02138, USA \\}
{\it ${}^2$Department of Physics and Astronomy, University of Victoria, Victoria, BC V8W 3P6, Canada\\}
{\it ${}^3$Jadwin Hall, Princeton University, Princeton, NJ 08540, USA \\}

\vspace{0.3cm}

{\tt  ${}^a$jcotler@fas.harvard.edu, ${}^b$pratekshd@uvic.ca, ${}^c$wharvey@uvic.ca, ${}^d$vi6083@princeton.edu, ${}^e$kristanj@uvic.ca\\}

\medskip

\end{center}

\vskip1cm

\begin{center}
{\bf Abstract}
\end{center}
\hspace{.3cm} 

We study the no-boundary wavefunction of Einstein gravity with positive cosmological constant. We identify families of weakly curved bulk saddles, including fillings of anisotropic late-time universes $\mathbb{S}^p\times\mathbb{S}^q$, whose weights grow exponentially with geometric parameters of the late-time boundary. These runaways threaten the normalizability of the no-boundary state within gravitational effective field theory. Whether these saddles contribute remains open. Many of the runaway families we study violate the Kontsevich-Segal-Witten criterion, although several are minus asymptotically hyperbolic geometries whose one-loop determinants appear well-defined for pure gravity. For $\mathbb{S}^1\times\mathbb{S}^q$ universes with $q\equiv 0\pmod4$, we construct a contour integral model that excludes the runaway at large $\mathbb{S}^q$.

\newpage

\tableofcontents

%--------------------------------------
\section{Introduction}
%--------------------------------------

The no-boundary proposal of Hartle and Hawking~\cite{Hartle:1983ai} uses the gravitational path integral (GPI) to prepare a natural state of the universe. Given data on a spacelike slice, the path integral over regular, generally complex metrics that smoothly fill it in defines the no-boundary wavefunction as a functional of that data. In this work we consider the no-boundary state in $d$-dimensional Einstein gravity with positive cosmological constant, focusing on the wavefunction at late times, where saddles asymptote to de Sitter space.

For quantum fields on a fixed de Sitter background, the analogous path integral produces the Bunch--Davies state, a well-defined normalizable state that underlies the standard prediction of primordial fluctuations. It is not clear whether these properties survive when the metric itself becomes dynamical, i.e.~whether the no-boundary prescription defines a healthy, normalizable quantum state of gravity. Our main goal is to address this question within semiclassical gravity.

To discuss normalizability we require a measure on late-time data, for which the ultralocal measure on boundary geometries provides the natural semiclassical choice. This measure has been discussed in several recent works in Einstein~\cite{Chakraborty:2023yed,Godet:2024ich,Turiaci:2025xwi,Cotler:2025gui} and dilaton gravity~\cite{Maldacena:2019cbz,Cotler:2019dcj,Cotler:2024xzz,Cotler:2026jdz}. We highlight two recent results concerning the norm of the no-boundary state that rely on it. First, the na\"{i}vely dominant contribution to the wavefunction, where the late-time slice is approximately a round sphere and the saddle point geometry is empty de Sitter space (suitably capped off smoothly in imaginary time), gives a vanishing contribution to the norm at one loop~\cite{Cotler:2025gui} thanks to a noncompact $SO(d,1)$ residual gauge symmetry of the norm.\footnote{The analogous one-loop norm is nonzero in deformed dS JT gravity~\cite{Cotler:2026jdz} and likely nonzero for gravity coupled to a slowly-rolling inflaton.} Second, in quantum gravity, unlike in quantum field theory, the data specifying a late-time state includes topology as well as local metric data. For a toroidal late-time slice with fixed moduli, infinitely many saddle-point geometries fill in the same boundary data. There is evidence that in $d=3$ the resulting saddle sum makes the torus contribution divergent~\cite{Godet:2024ich}, as presaged in~\cite{Castro:2012gc}.

The wavefunction is therefore labeled by the topology of the late-time slice. Semiclassically, different topologies behave like different superselection sectors, and the norm of the no-boundary state is a sum over topologies, each term an integral over the metrics with that topology.

In this work we study the no-boundary wavefunction on late-time slices of several topologies. The geometries that fill in these slices are generally complex, and a given slice can admit several bulk fillings. Our most basic goal is to find these geometries, sort them into branches, understand how their weights $|e^{iS}|^2$ behave, and ask whether they contribute to the wavefunction.

Some of this is known in $d=4$. When the late-time slice is $\mathbb{S}^1\times\mathbb{S}^2$, the wavefunction is largest when the $\mathbb{S}^1$ is much larger than the $\mathbb{S}^2$. There the filling approaches the Nariai solution~\cite{Conti:2014uda, Maldacena:2019cbz}, and the distribution is normalizable once loop effects are included~\cite{Turiaci:2025xwi}. When instead the $\mathbb{S}^2$ is large, the tree-level distribution stays bounded~\cite{Maldacena:2019cbz}. Earlier studies found saddles that would make it grow without bound there~\cite{Anninos:2012ft,Banerjee:2013mca}, but these have the sign of the action that would also give the round $\mathbb{S}^3$ the tunneling initial condition rather than the no-boundary one. The wavefunction has also been studied for squashed three-spheres~\cite{Bobev:2016sap,Janssen:2019sex,Lehners:2024kus,Hertog:2024nbh} and toroidal universes~\cite{Godet:2025bju}. Squashed spheres show a new feature, first found in~\cite{Janssen:2019sex}. A family of Taub-Bolt saddles has weights that grow without bound at large squashing, when the Hopf fiber is large compared to the base $\mathbb{S}^2$. If these saddles contributed, $|\Psi|^2\sim |e^{iS}|^2$ would run away. The authors of~\cite{Janssen:2019sex} set them aside because they would make the state non-normalizable. Two later results give reasons to discard them. In a minisuperspace model, Stokes phenomena in the lapse integral remove them~\cite{Lehners:2024kus}, and they violate the Kontsevich-Segal-Witten (KSW) criterion on allowable complex metrics~\cite{Kontsevich:2021dmb,Witten:2021nzp,Hertog:2024nbh}, at least for contours that respect their symmetry. We find the same behavior when the late-time slice is a lens space $L(p,1)=\mathbb{S}^3/\mathbb{Z}_p$. A Taub-Bolt family again has weights that grow without bound at large squashing, and, for contours that respect the symmetry, KSW excludes it beyond the same squashing as for $\mathbb{S}^3$.

There are many other runaway directions. We find several families of them, in dimensions from $d=4$ up to $d=11$, the largest we consider. The most prominent occurs when the late time slice is $\mathbb{S}^p\times\mathbb{S}^q$, in the limit of large anisotropy $\varphi=R_q/R_p$, provided $d\geq 6$.

Most of the runaway saddles have the same simple form. Along a suitable complex time contour, the metric is minus a real Euclidean AdS metric, that is, minus a real asymptotically hyperbolic Einstein metric. This lets us anticipate the runaways. In even $d$ the weight of such a saddle is fixed by the renormalized volume $V_{\rm ren}$ of the hyperbolic filling,
\beq
\label{E:introWeight}
 \ln|e^{iS}|=(-1)^{d/2}\,\frac{(d-1)V_{\rm ren}}{8\pi G}=(-1)^{d/2}\,I_{\rm AdS}\,,
\eeq
where $I_{\rm AdS}$ is the renormalized action of the same filling in Euclidean AdS gravity with the same radius. So in $d=6$ and $d=10$ the no-boundary weight of the saddle equals its weight $e^{-I_{\rm AdS}}$ in the AdS partition function, while in $d=4$ and $d=8$ it equals the inverse. Now take the late-time slice to be $\mathbb{S}^1\times\mathbb{S}^{d-2}$ or a torus, and shrink the circle. In AdS language this is the limit of high temperature $T$, where large black holes and black branes dominate with $I_{\rm AdS}=F/T\propto-V_{d-2}\,T^{d-2}/G$, $V_{d-2}$ being the volume of the large directions. So in $d\equiv2\pmod4$ these saddles have weights that grow as $\exp(c\,V_{d-2}T^{d-2}/G)$ with $c>0$. In $d\equiv0\pmod4$ they are suppressed, and the runaways come instead from fillings with large positive $I_{\rm AdS}$. In odd $d$ the weight is fixed by the Weyl anomaly of the late-time metric rather than by $V_{\rm ren}$~\cite{Henningson:1998gx,Cotler:2026flq}.

We find the following families of runaways.
\begin{itemize}
\item $\mathbb{S}^p\times\mathbb{S}^q$ with $p,q\geq2$, at large anisotropy. The filling in which the $\mathbb{S}^p$ caps off has $\ln|e^{iS}|^2\simeq 2c\,\varphi^q/G$. Through $d=11$, where we stopped, $c>0$ exactly when $q\equiv0,3\pmod4$, so every dimension $6\leq d\leq11$ has at least one runaway. The simplest example is $\mathbb{S}^2\times\mathbb{S}^3$ in $d=6$, at large $\mathbb{S}^3$. In $d=5$, $\mathbb{S}^2\times\mathbb{S}^2$ has no runaway.
\item $\mathbb{S}^1\times\mathbb{S}^q$ and tori with a small circle, for $q\equiv0\pmod4$, i.e.~$d=6,10,\hdots$. These are the large black holes and black branes described above.
\item Hyperbolic fillings in $d=4$. When the late-time slice is a particular closed hyperbolic three-manifold, there are infinitely many smooth hyperbolic fillings of it, with $\ln|e^{iS}|^2=S_0\,\chi$ where $\chi$, the Euler characteristic of the filling, is unbounded. Here the boundary is fixed, and the topology of the filling grows.
\item Hyperbolic boundaries in $d=5$, such as products of hyperbolic surfaces. The Weyl anomaly predicts arbitrarily large weights, provided that suitable fillings exist; we have not constructed them.
\item Lens spaces in $d=4$, where we extend the Taub-Bolt analysis of~\cite{Hertog:2024nbh} to the case when the boundary is a quotient of the sphere. The part of the family whose weights grow without bound violates KSW. These saddles also have a KSW allowable range with an explicit contour, so they consist of new saddles that respect KSW.
\end{itemize}
The saddles in the first three families and the lens-space saddles are weakly curved, with curvatures of order the de Sitter scale, so their growth occurs within the regime of validity of gravitational effective field theory (EFT).

Suppose a runaway saddle contributes to the wavefunction. Using the semiclassical ultralocal measure, the contribution of, say, the $\mathbb{S}^p\times\mathbb{S}^q$ sector to the norm is schematically
\beq
 \int d\mu(\varphi)\,\mathcal N(\varphi)\,|\Psi(\varphi)|^2\,,
\eeq
where $d\mu(\varphi)$ is the induced measure on the anisotropy and $\mathcal N(\varphi)$ comes from integrating over the fluctuations orthogonal to it. Both enter the logarithm of the integrand at $O(G^0)$, while the runaway enters at order $1/G$. They cannot cancel the growth while the loop expansion is under control, and so the norm diverges. In the hyperbolic example, the problem is instead the sum over fillings at fixed boundary, which diverges unless their contributions cancel. Either way, if these saddles contribute, the no-boundary state is not normalizable within gravitational EFT, and this threatens the no-boundary proposal itself.

However, a saddle need not contribute. The wavefunction is an integral over complex metrics along some integration cycle, and a saddle contributes only if its thimble appears in that cycle with a nonzero coefficient. Earlier work on squashed spheres makes this point. There, the choice of lapse contour decides which Taub-Bolt saddles survive~\cite{Lehners:2024kus}. 

One possible rule is the KSW criterion\footnote{See \cite{Caminiti:2026efx} for an alternative criterion}, which restricts the complex metrics allowed in the GPI so that quantum fields on them have convergent path integrals. It is not clear to us that it should be trusted as a rule. The all-minus saddles above violate it, yet in pure gravity a constant rotation of the metric fluctuations makes their one-loop determinants as well defined as those of the corresponding AdS problem. In AdS gravity, Mahajan and Singhi~\cite{Mahajan:2025bzo} found in a minisuperspace model of the GPI that the integration contour excludes all of the complex AdS-Schwarzschild black holes, including those just below the minimum black-hole temperature, which KSW allows. So KSW is not sufficient for a saddle to contribute. Nevertheless, KSW excludes the product-sphere and toroidal runaway families, and the lens-space runaways within the class of symmetry-preserving contours. It would also allow some families with bounded weights, such as the lens-space Taub-Bolt saddles in a range of squashing, and exclude others, such as the small all-minus black holes that dominate the $\mathbb{S}^1\times\mathbb{S}^2$ wavefunction at large $\mathbb{S}^2$.

The no-boundary state is ordinarily understood to include a sum over the possible spatial topologies of the universe. If KSW is imposed, very few no-boundary saddles survive. In $d=4$ the allowable saddles we know of fill in the round and squashed $\mathbb{S}^3$, with both NUT and bolt fillings, and $\mathbb{S}^1\times\mathbb{S}^2$ near the Nariai endpoint, each in a range of the boundary moduli~\cite{Hertog:2024nbh} (see also Section~\ref{S:S1contour}). To these we add the lens spaces $L(p,1)$ with $p\geq2$, in a range of squashing that includes the round metric as discussed in Section~\ref{S:lens}. Thus, although the no-boundary proposal formally involves a sum over spatial topologies, only a few families of saddles are currently known to satisfy KSW. Although we do not give a complete classification of KSW-allowable saddles, our analysis raises the question of how much of the usual sum over spatial topologies survives the KSW restriction.

There is a related tension with the common expectation that the norm of the no-boundary state is given by a Euclidean gravitational path integral summed over closed topologies. Every real positive-definite Euclidean metric satisfies KSW. Yet it is not clear that the closed Euclidean saddles can be obtained by gluing together bra and ket no-boundary saddles that individually satisfy KSW and admit the required late-time Lorentzian continuation. In particular, there is no obvious correspondence between the topologies in the Euclidean sum and those of the bra-ket geometries constructed from the known allowable fillings. Even when a Euclidean saddle and a geometry obtained by gluing a no-boundary bra and ket have the same topology, their contributions to the Euclidean gravitational path integral and the no-boundary norm need not agree. Recent work~\cite{Cotler:2026jdz} shows that these contributions can differ already at one loop, because the two calculations treat gravitational gauge symmetries differently.

We can make more progress on the question of which saddles actually contribute to the no-boundary wavefunction. For the $\mathbb{S}^1\times\mathbb{S}^q$ and toroidal slices, we can say more, because the cap radius of the circle gives a simple candidate family of off-shell metrics. These are geometries that solve Einstein's equations everywhere except for a conical singularity where the circle caps off. Following~\cite{Maldacena:2019cbz, Mahajan:2025bzo}, we write the wavefunction as an integral over the cap radius, i.e., as an integral over this off-shell family, and ask which saddles appear on a given convergent contour. A natural contour, which includes the Nariai point, drops the runaway large black holes and black branes. It also drops the lower Nariai saddle, apart from a Stokes-dependent range in $d=5$, and KSW excludes that saddle too. Where the two can be compared, the contour model is the more permissive. It keeps every saddle that KSW allows, and it also keeps the upper Nariai saddle beyond the KSW bound and, in $d=4,5$, the small all-minus black holes, all of which fail KSW. This contour is not the one natural in AdS. On the continuation of the AdS contour, which converges exactly when $q\equiv0\pmod4$, the large black hole does contribute. So in this model a normalizable state needs an integration cycle that differs from the continued AdS one, and the natural de Sitter contour is such a cycle. It is only a toy model. The off-shell family and the contour are assumptions; requiring the Nariai saddle to contribute does not fix the contour, and for $p,q>1$ we have no comparable off-shell family.

Where does this leave us? In our view, the problem is real. Runaway saddles are common. In even $d$ they follow from the simple relation~\eqref{E:introWeight} between no-boundary weights and AdS actions, and we find them in many families. We do not know a fundamental principle that removes them. KSW removes several of these families, at the cost of also removing saddles that a natural contour keeps. Where we can compare the two, they agree on the runaway and on the lower Nariai saddle. Our contour model removes the runaway it can address, but we cannot yet derive its contour from the GPI or extend it beyond circle caps. Settling this question requires a better understanding of the integration cycle of the gravitational path integral.

The remainder of this manuscript is organized as follows. In Section~\ref{S:setup} we set up the problem, review the KSW criterion, and relate saddle weights to renormalized volumes and $\mathsf{Q}$-curvature. In Section~\ref{S:S1} we study the $\mathbb{S}^1\times\mathbb{S}^q$ wavefunction, where the saddles are known in closed form. There we present the contour model, study loop effects near the Nariai endpoint for general $q$, and study loop effects at large $\mathbb{S}^2$ in $d=4$, where the thermal graviton gas makes the distribution grow again until the gas backreacts strongly on the geometry. In Section~\ref{S:S2S2} we construct the $\mathbb{S}^2\times\mathbb{S}^2$ saddles in $d=5$ numerically. They include an infinite nested sequence of complex branches, and we find that the Nariai endpoints are local maxima of the distribution even though the corresponding Euclidean saddles have a breathing instability~\cite{Ivo:2025yek,Shi:2025xxx}. In Section~\ref{S:anisotropy} we compute the large-anisotropy weights for $\mathbb{S}^p\times\mathbb{S}^q$ through $d=11$. In Section~\ref{S:otherGeometries} we present the other families of runaways, namely hyperbolic boundaries, toroidal black branes, hyperbolic fillings at fixed boundary, and lens spaces. In Section~\ref{S:inflaton} we comment on coupling to a slowly rolling inflaton, and we conclude in Section~\ref{S:discuss}. The appendices construct the nested branches analytically and give explicit KSW-allowable contours for the $\mathbb{S}^1\times\mathbb{S}^q$ and lens-space saddles.

%--------------------------------------
\section{Setup}
\label{S:setup}
%--------------------------------------

The Einstein-Hilbert action with positive cosmological constant is 
\begin{align}
\begin{split}
	S &= S_{\rm EH} + S_{\rm GH} + S_{\rm CT}\,,
	\\
	S_{\rm EH} & = \frac{1}{16\pi G}\int d^dx \sqrt{-g}(R-2\Lambda) \,, \qquad \Lambda = \frac{(d-1)(d-2)}{2L^2}
	\\
	S_{\rm GH} &= - \frac{1}{8\pi G} \int d^{d-1}x \sqrt{h} \,K \,, 
\end{split}
\end{align}
where $h$ is the boundary metric on the late-time slice, $K$ is the trace of its extrinsic curvature, $S_{\rm CT}$ are the counterterms we add in order to obtain a finite tree-level wavefunction on the late-time slice\footnote{These counterterms redefine the basis states $\langle\!\langle \gamma|$ corresponding to fixed boundary data by a phase.} and henceforth we fix the de Sitter radius to unity, $L=1$.

The late-time boundary conditions are
\begin{align}
\label{gammaeq}
	ds^2 = - dt^2 + \left( e^{2t} \gamma_{ij}(x) + O(1)\right)dx^i dx^j\,,
\end{align}
with $\gamma_{ij}$ the fixed late-time data. The GPI with the early-time boundary condition that spacetime caps off smoothly, inevitably in complex time, and this late-time boundary condition, computes the late-time no-boundary wavefunction
\beq
	\Psi_{\rm HH}[\gamma] = \langle\!\langle \gamma|\text{HH}\rangle\!\rangle = \int \frac{[dg]}{\text{diff}}\, e^{iS}\,,
\eeq
where the boundary conditions are implicit in the measure. These equalities are understood to be formal insofar as the far right-hand-side only exists in the EFT approximation.

%--------------------------------------
\subsection{Review: round $\mathbb{S}^{d-1}$ wavefunction}
%--------------------------------------

Let us review how to use these definitions to obtain the tree-level wavefunction when the late-time slice is a round sphere. In this case the $d$-dimensional geometry is
\beq
	ds^2 = -dt^2 + \cosh^2(t) d\Omega_{d-1}^2\,,
\eeq
for $t$ traversing a complex time contour connecting $t=\frac{\pi i}{2}$ to $t\to\infty$. This is necessarily a complex metric. The Hartle-Hawking time contour is the union of two segments: (i.) the interval $t = i \tau$ from $\tau = \frac{\pi}{2}$ to $\tau=0$, along which the geometry is a Euclidean hemisphere whose equator is reached as $\tau \to 0$, and (ii.) a Lorentzian segment $t\geq 0$. The Lorentzian portion of the geometry is the future half of empty global de Sitter space.

The Lorentzian contour produces a real contribution to the on-shell action $S$. The imaginary part, which controls the no-boundary distribution $|\Psi|^2$, comes entirely from the Euclidean interval. Using $R = d(d-1)$ and $d^dx \sqrt{-g} = i d\tau d^{d-1}\Omega \cos^{d-1}(\tau)$ we have
\beq
	S = -i \frac{S_0}{2} + (\text{real})\,, \qquad S_0 =\frac{(d-1)\text{vol}(\mathbb{S}^d)}{8\pi G}= \frac{\text{vol}(\mathbb{S}^{d-2})}{4G} \,,
\eeq
where $S_0$ is at once the tree-level static patch entropy of the cosmological horizon, and minus the Euclidean action $S_{\rm E,sphere}$ of the round $d$-sphere saddle of Euclidean gravity. So
\beq
	|\Psi_{\rm HH}|^2 \sim e^{S_0} = e^{-S_{\rm E,sphere}}\,.
\eeq

\begin{figure}[t]
\begin{center}
    \includegraphics[width=5.5in]{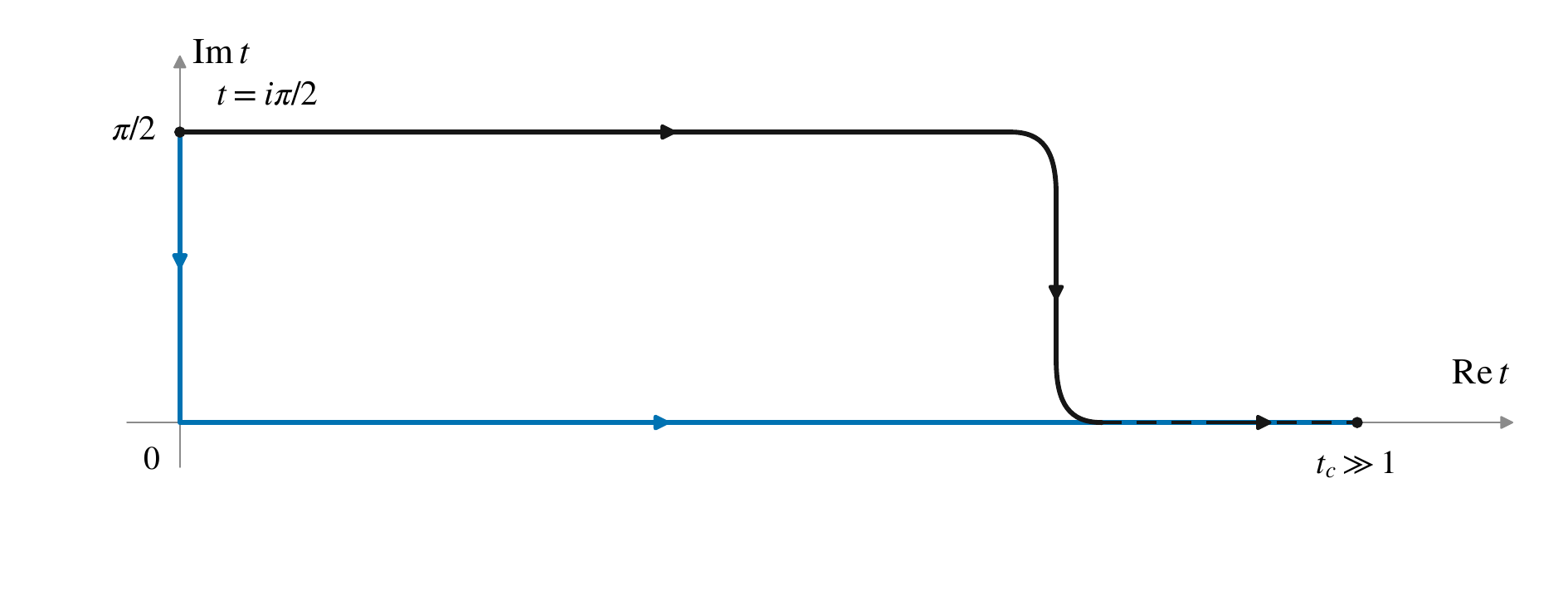}
    \caption{\label{F:HHcontour} The Hartle-Hawking (blue) and Maldacena (black) contours for the bulk geometry that prepares the late-time no-boundary wavefunction on a round sphere.}
\end{center}
\end{figure}

A more convenient contour for our purposes, which will have an analogue in our $\mathbb{S}^p\times\mathbb{S}^q$ components, is the Maldacena contour. For this one complex time is the union of a contour $t = \rho + \frac{\pi i}{2}$ where $\rho$ goes from $0$ to a large value, where the geometry has entered an asymptotic region, at which the contour dips down to real time so that the geometry becomes Lorentzian. See Fig.~\ref{F:HHcontour} for a representation of the Hartle-Hawking and Maldacena contours. In terms of the $\rho$ coordinate the geometry becomes hyperbolic space in all-minus signature,
\beq
	ds^2 = - (d\rho^2 + \sinh^2(\rho)d\Omega_{d-1}^2)\,.
\eeq
The action along this contour reproduces what we described above. We show it in some detail with the future $\mathbb{S}^p\times\mathbb{S}^q$ problem in mind, where the important contributions we study at large anisotropy are all-minus geometries. To proceed we use that along the $\rho$ contour the volume element is 
\beq
	\sqrt{-g} = i^{d-1} \times \sinh^{d-1}(\rho)\,,
\eeq
and so
\beq
	S =i^{d-1} \frac{\text{vol}(\mathbb{S}^{d-1})}{16\pi G}2(d-1)\int d\rho \,\sinh^{d-1}(\rho) + (\text{bdy terms})\,.
\eeq
The contour for $\rho$ is that indicated in Fig.~\ref{F:HHcontour}, and the boundary terms are added on a $t =-\ln \varepsilon$ slice with $\varepsilon\to 0$. In even $d$ this result coincides with a contour where we omit the dip down to Lorentzian time, i.e.~from $\rho = 0$ to $\rho = -\ln \varepsilon$ with boundary terms placed there. This is the same thing as what we get from integrating the bulk action up to the cutoff, expanding at small $\varepsilon$, and discarding the divergent terms, giving
\beq
	S = i^{d-1} \frac{2(d-1)\text{vol}(\mathbb{S}^{d-1})}{16\pi G}\frac{\Gamma\left( \frac{d}{2}\right)\Gamma\left( \frac{1-d}{2}\right)}{2\sqrt{\pi}}  = -i\frac{S_0}{2}\times \left( i^d \text{sec}\left(\frac{\pi d}{2}\right)\right) = -i \frac{S_0}{2}\,,
\eeq
where in the second equality we used $\Gamma$ identities and in the last even $d$. 

In odd $d$ $\sqrt{-g}$ is real and so the integral over the real $\rho$ contour produces a real answer. The imaginary part of $S$ instead comes from the dip segment of the contour at large $|t|$. To evaluate it we use for odd $d$
\beq
\label{E:secularSphere}
	i^{d-1} \sinh^{d-1}(\rho) = (\text{divergent as }\rho\to\infty) + 2^{2-d} C\!\left(d-2,\frac{d-1}{2}\right) + O(e^{-2\rho})\,,
\eeq
with $C(a,b)=\frac{a!}{b!(a-b)!}$ the binomial coefficient. The integration over the real-$\rho$ and dip contours up to a late time $t = - \ln \varepsilon$ produces terms that are power-law divergent in $\frac{1}{\varepsilon}$, canceled off by ordinary counterterms, a real logarithmic divergence $\propto \ln \varepsilon$ canceled by a logarithmic counterterm analogous to what we encounter in holographic renormalization in odd-dimensional AdS$_d$, and a finite imaginary contribution from integrating the constant term in~\eqref{E:secularSphere} along the dip by $\frac{\pi}{2}$ in complex time. The real part is ambiguous, depending on how the logarithmic divergence is regulated, but the imaginary part is not, and we have
\beq
	S =(\text{real}) +  \left( -\frac{i \pi}{2}\right)\times \frac{2(d-1)\text{vol}(\mathbb{S}^{d-1})}{16\pi G}\times  2^{2-d}C\left( d-2,\frac{d-1}{2}\right)  = (\text{real}) - i \frac{S_0}{2}\,,
\eeq
where we have again used $\Gamma$ identities together with $d$ being odd.

%--------------------------------------
\subsection{$\mathbb{S}^p\times\mathbb{S}^q$ minisuperspace}
\label{S:SpSqsuperspace}
%--------------------------------------

In this work we consider the $\mathbb{S}^p\times\mathbb{S}^q$ wavefunction, where the conformal class of the metric on the late-time slice is
\beq
	\gamma_{ij}dx^idx^j = d\Omega_p^2 + \varphi^2 d\Omega_q^2\,.
\eeq
We are interested in bulk geometries that fill in these data and are invariant under an $SO(p+1)\times SO(q+1)$ isometry. The bulk metric can then be taken to be
\beq
	ds^2 = -dt^2 + A(t) d\Omega_p^2 + B(t)d\Omega_q^2\,.
\eeq
The gravitational action evaluated on such a metric is, up to boundary terms,
\begin{align}
	S &= \frac{\text{vol}(\mathbb{S}^p\times\mathbb{S}^q)}{16\pi G}\int dt\,L + (\text{bdy terms})\,,
	\\
	\nonumber
	L & = - A^{\frac{p}{2}}B^{\frac{q}{2}}\left( \frac{p(p-1)}{4}\frac{\dot{A}^2}{A^2} + \frac{q(q-1)}{4}\frac{\dot{B}^2}{B^2} + \frac{pq}{2}\frac{\dot{A}\dot{B}}{AB} - \frac{p(p-1)}{A} - \frac{q(q-1)}{B} + (p+q)(p+q-1)\right)\,.
\end{align}
The solutions we study are necessarily complex geometries with a late-time asymptotic region at large $t$. For general complex $A, B$, the phase of the volume element $A^{\frac{p}{2}}B^{\frac{q}{2}}$ is determined by starting in the late-time Lorentzian region, where it is positive, and following it back in complex time.  

The equations of motion are those that follow from variation with respect to $A$ and $B$ along with the Hamiltonian constraint
\begin{align}
\begin{split}
\label{E:SpSqEinstein}
	0 & = \frac{\ddot{A}}{2A} +\frac{p-2}{4}\frac{\dot{A}^2}{A^2} +\frac{q}{4}\frac{\dot{A}\dot{B}}{AB} + \frac{p-1}{A} -(p+q) \,,
	\\
	0 & = \frac{\ddot{B}}{2B} + \frac{q-2}{4}\frac{\dot{B}^2}{B^2} + \frac{p}{4}\frac{\dot{A}\dot{B}}{AB} + \frac{q-1}{B} - (p+q) \,,
	\\
	0 & = \frac{p(p-1)}{4}\frac{\dot{A}^2}{A^2} + \frac{q(q-1)}{4}\frac{\dot{B}^2}{B^2} + \frac{pq}{2}\frac{\dot{A}\dot{B}}{AB} + \frac{p(p-1)}{A} + \frac{q(q-1)}{B} - (p+q)(p+q-1) \,.
\end{split}
\end{align}
The ``early'' boundary condition is that the geometry caps off smoothly. This can happen by the $\mathbb{S}^p$ capping off smoothly with the $\mathbb{S}^q$ remaining finite size, or vice versa. The ``early'' boundary condition means 
\beq
	A = -\rho^2 ( 1 + O(\rho^2))\,, \qquad B = B_0( 1 + O(\rho^2))\,,
\eeq
or vice versa. It is straightforward to extend this to a perturbative solution in small $\rho$. The late-time boundary condition is that, for $t = \rho + i \Theta$ and some real $(t_0,\Theta)$,
\beq
	A = e^{2(t-t_0)}\left( 1 + O(e^{-2t})\right)\,, \qquad B =e^{2(t-t_0)} \left(  \varphi^2  + O(e^{-2t})\right)\,.
\eeq
This corresponds to a large-$\rho$ boundary condition for some real $\rho_0$,
\beq
	A = e^{2(\rho-\rho_0)+2i\Theta}\left( 1 + O(e^{-2\rho})\right)\,, \qquad B = e^{2(\rho-\rho_0)+2i\Theta}\left( \varphi^2 + O(e^{-2\rho})\right)\,,
\eeq
which is straightforward to implement numerically and which in turn allows us to extract $\Theta \in [0,\pi)$. In particular it amounts to the statement that $A$ and $B$ have the same phase at large $\rho$. This boundary condition can be extended to a perturbative solution in large $t$ or $\rho$. 

We wrap up with a short discussion of the holographic renormalization of the on-shell action. In $d>5$ the renormalized action includes the first few boundary terms
\beq
\label{E:RG1}
	S_{\rm CT} = \frac{1}{8\pi G}\int d^{d-1}x\sqrt{h}\left( d-2 - \frac{R_h}{2(d-3)} + \frac{1}{2(d-5)(d-3)^2} \left(R_{h ij}R_h^{ij} - \frac{d-1}{4(d-2)}R_h^2\right) + \hdots\right)\,,
\eeq
with $R_h$ the curvature of the boundary metric $h$. In $d=4$ the third term is absent, and in $d=5$ it is replaced by $-\frac{1}{8}\ln \varepsilon \left( R_{hij}R_h^{ij} - \frac{1}{3}R_h^2\right)$. The boundary terms are added on the $t=-\ln \varepsilon$ slice, the bulk action is integrated up to that time, and then the sum admits a smooth $\varepsilon\to 0$ limit which defines the renormalized action.

Within our minisuperspace treatment we can verify the coefficients of the first two terms in the following way. In $d>5$ solving the Einstein equations perturbatively at late time gives
\begin{align}
\begin{split}
	A & = e^{2t}\left( 1 + \frac{1}{n-2}\left(p-1- \frac{R_{\gamma}}{2(n-1)} \right) e^{-2t}+ O(e^{-4t})\right)\,,
	\\
	B & = e^{2t}\left( \varphi^2 + \frac{1}{n-2}\left( q-1 - \frac{\varphi^2R_{\gamma}}{2(n-1)}\right)e^{-2t} + O(e^{-4t})\right)\,,
\end{split}
\end{align}
with $n=p+q$ and
\beq
	R_{\gamma} = p(p-1) + \frac{q(q-1)}{\varphi^2}\,,
\eeq
the curvature of the metric $\gamma$. On-shell the bulk Lagrangian is $R - 2\Lambda= 2(d-1)=2n$  so that at large $t$ the action is
\begin{align}
	\nonumber
	S_{\rm EH}& = \frac{\text{vol}(\mathbb{S}^p\times\mathbb{S}^q)}{16\pi G} \int^{-\ln \varepsilon} dt\,A^{\frac{p}{2}}B^{\frac{q}{2}} \times 2n\,,
	\\
	S_{\rm GH} & =  - \frac{\text{vol}(\mathbb{S}^p\times\mathbb{S}^q)}{16\pi G} \left. A^{\frac{p}{2}}B^{\frac{q}{2}}\left(p \frac{\dot{A}}{A}+q\frac{\dot{B}}{B}\right)\right|_{t = -\ln \varepsilon} \,,
	\\
	\nonumber
	S_{\rm CT} & =\frac{\text{vol}(\mathbb{S}^p\times\mathbb{S}^q)}{16\pi G}\left. A^{\frac{p}{2}}B^{\frac{q}{2}}\left(  2(n-1) - \frac{1}{n-2}\left( \frac{p(p-1)}{A} + \frac{q(q-1)}{B}\right) + \cdots\right)\right|_{t = - \ln \varepsilon}\,,
\end{align} 
where we have only included the volume and curvature counterterms.

Plugging in the asymptotic forms of $A$ and $B$ we have
\begin{align}
\begin{split}
	\frac{16\pi G}{\text{vol}(\mathbb{S}^p\times\mathbb{S}^q)}S_{\rm EH,div} & = \frac{\varphi^q}{\varepsilon^{n}}\left(  2 + \frac{n}{2(n-1)(n-2)} R_{\gamma} \varepsilon^2 + O(\varepsilon^4)\right)\,,
	\\
	\frac{16\pi G}{\text{vol}(\mathbb{S}^p\times\mathbb{S}^q)}S_{\rm GH,div} & = \frac{\varphi^q}{\varepsilon^{n}}\left( -2n -\frac{n-2}{2(n-1)}R_{\gamma} \varepsilon^2 + O(\varepsilon^4)\right)\,,
	\\
	\frac{16\pi G}{\text{vol}(\mathbb{S}^p\times\mathbb{S}^q)}S_{\rm CT,div} & = \frac{\varphi^q}{\varepsilon^{n}}\left( 2 (n-1) + \frac{n-4}{2(n-2)}R_{\gamma} \varepsilon^2 + O(\varepsilon^4)\right)\,.
\end{split}
\end{align}
It is elementary to see that the leading $O(\varepsilon^{-n})$ and subleading $O(\varepsilon^{-n+2})$ divergences cancel on the nose in the total action $S = S_{\rm EH} + S_{\rm GH} + S_{\rm CT}$. 

We are now in a position to explain the numerical procedure we use in the next Sections. 
\begin{enumerate}
	\item Use the near-cap Einstein's equations to find near-cap solutions in terms of the complex initial condition $B_0$. For a given $B_0$, solve the equations numerically for real $\rho$ from a suitably small initial $\rho$ out to a large $\rho$ cutoff.
	\item Shoot in $B_0$ to determine the locus such that the geometry admits a late-time Lorentzian region. On this locus $\lim_{\rho\to\infty}B/A = \varphi^2 >0$. For any such solution, determine $\varphi$ and $\Theta$ from the late-time behaviour of $A$ and $B$, 
	\beq
	\label{E:largeRhoBC}
		A= e^{2(\rho-\rho_0)+2i \Theta} \left( 1+ O(e^{-2\rho})\right)\,, \qquad B =e^{2(\rho-\rho_0) + 2i\Theta}\left(\varphi^2 + O(e^{-2\rho})\right)\,,
	\eeq
	for real $\rho_0$ and $\Theta \in [0,\pi)$. 
	\item Evaluate the on-shell renormalized action of the solution. The tree-level wavefunction is  $\Psi_{\rm tree} \sim  \sum_{\alpha} e^{iS_{\alpha}}$ where $\alpha$ labels contributing saddles. The semiclassical approximation to $\Psi$ is dominated by that of largest $|e^{i S_{\alpha}}|=e^{-\text{Im}(S_{\alpha})}$.
\end{enumerate}
Our solutions are naturally generated along the Maldacena contour, with the dip to Lorentzian signature placed at large $|\rho|$. 

%--------------------------------------
\subsection{KSW}
\label{S:KSW}
%--------------------------------------

The Kontsevich--Segal--Witten (KSW) condition is a proposed restriction on the complex metrics entering the GPI \cite{Kontsevich:2021dmb, Witten:2021nzp}. In a real frame in which the metric is diagonal, with eigenvalues $\lambda_\mu$, it is
\beq
\label{E:KSW}
 \sum_{\mu=1}^d |\arg\lambda_\mu| < \pi\,.
\eeq
The arguments are chosen on the branch containing positive-definite Euclidean metrics. Lorentzian metrics lie on the boundary of this domain and are understood as limits of regulated complex metrics. For a saddle written using a complex time coordinate, the condition applies to the metric pulled back to a contour parameterized by real coordinates. Its validity can therefore depend on the time contour, even when the analytically continued solution is fixed.

We now review a few facts from the literature. As noted in \cite{Hertog:2024nbh}, a useful consequence of KSW is the weaker condition
\beq
\label{E:weakKSW}
 \text{Re}\sqrt{g_{\rm E}} \geq 0\,.
\eeq
The inequality is strict wherever~\eqref{E:KSW} is strict. On an Einstein solution with the de Sitter radius set to one, the bulk Euclidean action is $-(d-1)V^{\rm ket}_{\rm E}/(8\pi G)$~\cite{deHaro:2000vlm, Hertog:2024nbh, Cotler:2026flq}. The Gibbons-Hawking term has no finite real contribution in the late-time limit with real boundary data. Consequently,
\beq
\label{E:KSWvolume}
 \ln|e^{iS_\alpha}|^2
 = \frac{d-1}{4\pi G}\text{Re}V^{\rm ket}_{{\rm E},\alpha}
 = \frac{d-1}{8\pi G}\text{Re}V_{{\rm E},\alpha}\,.
\eeq
Here $V^{\rm ket}_{{\rm E},\alpha}$ is the complex Euclidean volume of one filling, and $V_{{\rm E},\alpha}=V^{\rm ket}_{{\rm E},\alpha}+\overline{V^{\rm ket}_{{\rm E},\alpha}}$ is the volume of the bra-ket geometry. These expressions refer to an individual saddle; interference can matter in the sum over saddles.

An allowable contour therefore has $\text{Re}V_{{\rm E},\alpha}>0$, and therefore has a positive semiclassical weight in~\eqref{E:KSWvolume}. This is a restriction on possible representatives of a saddle, and does not determine its coefficient in a chosen integration cycle. In particular, a positive real volume is only a necessary condition. The all-minus saddles considered below can have positive real volume while failing the full KSW condition.

There is a simple application in $d=3$. For saddles obtained from real asymptotically hyperbolic fillings with a minus AdS metric, and a genus-$g$ boundary, the anomaly gives $\ln|e^{iS}|^2=S_0\chi(\Sigma_g)/2=S_0(1-g)$. Such saddles have negative real Euclidean volume for $g>1$ and cannot admit a KSW-allowable contour. This conclusion concerns this family of fillings; it is not a classification of all complex saddles or a determination of the norm of the state. The torus has zero anomaly weight and requires a separate analysis of its cap.

KSW is most restrictive near a cap. Near where a $\mathbb{S}^p$ caps off smoothly along real $\rho$, the line element is
\beq
\label{E:KSWcap}
 ds^2\approx-\left(d\rho^2+\rho^2d\Omega_p^2\right)+B_0\,d\Omega_q^2\,.
\eeq
On a contour that leaves the cap in the direction $\rho=e^{i\alpha}s$, with $s$ real, the $p+1$ directions of the disk share the phase $\arg(-e^{2i\alpha})$, while the $q$ directions of the $\mathbb{S}^q$ have phase $\arg B_0$. The first phase can be removed by choosing the disk to be Euclidean, $\alpha=\pm\pi/2$. The second cannot. $B_0$ is fixed by the saddle, whatever the contour. So a saddle admits an allowable contour only if
\beq
\label{E:KSWwedge}
 q\,|\arg B_0|<\pi\,.
\eeq
This confines $B_0$ to a wedge of opening angle $2\pi/q$ around the positive real axis; for $q=2$ it is the half-plane $\text{Re}\,B_0>0$. In particular, every saddle with $B_0<0$ fails, and so does every saddle whose metric near the cap is minus a real Euclidean metric. When the $\mathbb{S}^q$ caps off instead, the same argument gives $p\,|\arg A_0|<\pi$, with $A_0$ the value of $A$ at the cap. The condition~\eqref{E:KSWwedge} is necessary but not sufficient. A contour that passes the test at the cap must still remain allowable all the way to the late-time region, and this can fail even when~\eqref{E:KSWwedge} holds.

KSW was proposed so that quantum fields on an allowable metric have convergent path integrals. For pure gravity around an all-minus saddle, the fluctuation integral can be made convergent directly. Along the part of the contour where the saddle is all minus, write its metric as $g=-g_{\rm a}$, with $g_{\rm a}$ a real asymptotically hyperbolic Einstein metric. Under $g\to-g$ the Ricci tensor is unchanged and $R$ changes sign, and so, for any metric $g=-(g_{\rm a}+h)$,
\beq
\label{E:minusAdSaction}
 iS_{\rm EH}\big[-(g_{\rm a}+h)\big]=i^{\,d}\,I_{\rm AdS}\big[g_{\rm a}+h\big]\,,
\eeq
where $I_{\rm AdS}$ is the Euclidean action of gravity with the same $G$ and cosmological constant $-\Lambda$, and the branch of the volume element is the one fixed by the late-time Lorentzian region. The quadratic action for metric fluctuations is therefore a constant phase times that of Euclidean AdS gravity around $g_{\rm a}$. Rotating every metric fluctuation by the same constant phase turns the Gaussian integral into the AdS one, with the usual rotation of the conformal mode, and the determinant changes only by a phase. So the one-loop determinant of gravity around an all-minus saddle is as well defined as that of the corresponding AdS problem. This works because the whole gravitational action maps to the AdS action. It need not work for matter. A scalar of mass $m$ maps to an AdS scalar of mass squared $-m^2$, which lies below the Breitenlohner-Freedman bound when $m>(d-1)/2$, and then no constant rotation makes its Gaussian integral converge.

%--------------------------------------
\subsection{Renormalized volume and $Q$-curvature}
\label{S:Qcurvature}
%--------------------------------------

The round-sphere calculation has a useful generalization to the all-minus saddles~\cite{Cotler:2026flq}, which we review below in slightly different notation. The generalization follows from adapting results in the mathematical literature on negative cosmological constant Einstein spacetimes \cite{Graham:2003GZ, Fefferman:2002QCurvaturePoincare, ChangQingYang:2008RenVol} to the current positive cosmological constant context. 

Suppose that a candidate filling $\alpha$ has a contour along which its metric is minus a real, positive-definite asymptotically hyperbolic Einstein metric $g_{\rm a}$, with $\text{Ric}(g_{\rm a}) = -(d-1)g_{\rm a}$ and conformal boundary $(\Sigma,[\gamma])$. The boundary dimension is $d-1$. The magnitude of its tree-level contribution can be expressed in terms of the integrated $Q$-curvature,
\beq
\label{E:chiQ}
 \chi_{\mathsf{Q},\alpha}[\gamma] = \frac{2}{\Gamma(d-1)\text{vol}(\mathbb S^{d-1})}\int_\Sigma d^{d-1}x\sqrt{\gamma}\, \mathsf{Q}_{d-1,\alpha}[\gamma]\,.
\eeq
For even $d-1$, $\mathsf{Q}_{d-1}$ is the local Branson $Q$-curvature of the boundary metric \cite{Branson:1985Conformal, Branson:1995SharpInequalities}. For odd $d-1$, it is defined by the scattering problem on the asymptotically hyperbolic filling \cite{Graham:2003GZ}, and depends on $\alpha$ as well as $\gamma$. In either case, its integral is invariant under a Weyl transformation of $\gamma$. We normalize $\mathsf{Q}_{d-1}$ so that $\mathsf{Q}_{d-1}[\mathbb S^{d-1}] = \Gamma(d-1)$, and hence $\chi_\mathsf{Q}[\mathbb S^{d-1}] = 2$ for the hyperbolic-ball filling~\cite{Chang:1997uniqueness}.

With the continuation and choice of volume-element branch fixed by the round-sphere calculation, the candidate saddle weight is
\beq
\label{E:Qweight}
 \ln |e^{iS_\alpha}|^2 = -2\text{Im}(S_\alpha) = \frac{S_0}{2}\chi_{\mathsf{Q},\alpha}[\gamma]\,.
\eeq
In odd $d$, this imaginary part comes from the logarithmic term in the asymptotic volume integrated along the downward leg of the Maldacena contour. Its coefficient is proportional to $\int_\Sigma \mathsf{Q}_{d-1}$, so~\eqref{E:Qweight} is another way to express the Weyl-anomaly calculation. This happens because $\mathsf{Q}_{\gamma}$ is proportional to the holographic Weyl anomaly $\langle T_{i}^{i}\rangle$ of Einstein gravity in AdS \cite{Henningson:1998gx, Skenderis:2002wp}, up to total derivative terms \cite{Graham:2003GZ, Graham:2007Qcurvature}. In even $d$, there is no logarithmic term. Instead the finite renormalized volume $V_{{\rm ren},\alpha}$ determines the magnitude,
\begin{align}
\label{E:Qvolume}
 \int_\Sigma d^{d-1}x\sqrt{\gamma}\,\mathsf{Q}_{d-1,\alpha} &= \frac{2^{d-1}\Gamma((d-1)/2)}{\Gamma(-(d-1)/2)}V_{{\rm ren},\alpha}\,,
 \\
\label{E:finiteVolumeWeight}
 -2\,\text{Im}(S_\alpha) &= (-1)^{d/2}\frac{d-1}{4\pi G}V_{{\rm ren},\alpha}\,, \qquad d\ \text{even}\,.
\end{align}
The distinction between the two cases is therefore geometric. In odd bulk dimension the magnitude is fixed locally by the boundary metric, whereas in even bulk dimension it retains information about the filling.

These statements use the reality of $g_{\rm a}$. For a genuinely complex asymptotically hyperbolic filling with even $d-1$, its finite volume $V_{{\rm fin},\alpha}$ can have an imaginary part, and the relation becomes
\beq
\label{E:complexQweight}
 -2\text{Im}(S_\alpha) = \frac{S_0}{2}\chi_\mathsf{Q}[\gamma] - (-1)^{(d-1)/2}\frac{d-1}{4\pi G}\text{Im}(V_{{\rm fin},\alpha})\,, \qquad d-1\ \text{even}\,.
\eeq
Complex Nariai branches need not therefore have the weight of a real AdS branch with the same conformal boundary. Nor does~\eqref{E:Qweight} determine whether an all-minus saddle has nonzero coefficient in the GPI.

For later use, in $d-1=4$ the integrated curvature takes a particularly simple form \cite{Maldacena:2011mk},
\beq
\label{E:chiQ4}
 \chi_\mathsf{Q}[\gamma] = \chi(\Sigma) - \frac{1}{32\pi^2}\int_\Sigma d^4x\sqrt{\gamma}\,W_{ijkl}W^{ijkl}\,, \qquad d=5\,,
\eeq
where $W$ is the Weyl tensor of $\gamma$. In $d-1=3$, the corresponding relation involves the bulk Weyl tensor instead,
\beq
\label{E:chiQ3}
 \chi_{\mathsf{Q},\alpha}[\gamma] = 2\chi(M_\alpha) - \frac{1}{16\pi^2}\int_{M_\alpha} d^4x\sqrt{g_{\rm a}}\,W_{\mu\nu\rho\sigma}W^{\mu\nu\rho\sigma}\,, \qquad d=4.
\eeq
Here $M_\alpha$ denotes the asymptotically hyperbolic filling, and the curvature conventions in both equations use the full tensor contraction for $W^2$ \cite{Anderson:2001AHE}.

%--------------------------------------
\section{$\mathbb{S}^1\times \mathbb{S}^q$ wavefunction}
\label{S:S1}
%--------------------------------------

In this Section we study the $\mathbb{S}^1\times\mathbb{S}^q$ wavefunction, with $q\geq2$ and $d=q+2$. The geometries and their actions can be found analytically, and their relation to AdS black holes gives us a toy model for deciding which saddles contribute. The cycle we study excludes the saddles whose weights run away at large anisotropy, including the black branes of toroidal slices, and we compare its saddle content with KSW. We then study loop effects in the large-$\mathbb{S}^1$ limit at general $q$ and the large-$\mathbb{S}^2$ limit in $d=4$.

%--------------------------------------
\subsection{Candidate saddles}
\label{S:S1saddles}
%--------------------------------------

In $d=4$ this is the familiar Kantowski-Sachs problem~\cite{Halliwell:1990tu,Conti:2014uda}. We follow the presentation of its complex no-boundary saddles in~\cite{Maldacena:2019cbz} and extend it to general $q$. The near-Nariai quantum problem was recently studied in~\cite{Turiaci:2025xwi}.

For $p=1$ it is convenient to replace the proper-time coordinate of Section~\ref{S:SpSqsuperspace} by the complex radius $z$ of the $\mathbb{S}^q$. The general solution with the required symmetry is
\beq
\label{E:S1metric}
 ds^2=-\frac{dz^2}{F(z)}+F(z)dx^2+z^2d\Omega_q^2\,,
 \qquad F(z)=z^2-1+\frac{\mu}{z^{q-1}}\,.
\eeq
Here $x$ has period $L$ and the de Sitter radius is unity. At large positive $z$ the conformal boundary metric is $dx^2+d\Omega_q^2$, so that
\beq
\label{E:S1period}
 L=\frac{2\pi}{\varphi}\,.
\eeq
Large $\varphi$ corresponds to a small circle, the analogue of high temperature in AdS. A complex contour in $z$ connects the cap to the late-time Lorentzian region.

Suppose that the circle caps off at $z=\zeta$. Then $F(\zeta)=0$ fixes $\mu$ and smoothness fixes $F'(\zeta)$ in terms of $\varphi$. On the sheet continued from the real late-time geometry we have
\beq
\label{E:S1smooth}
 \mu=\zeta^{q-1}(1-\zeta^2)\,,\qquad
 F'(\zeta)=\frac{(q+1)\zeta^2-(q-1)}{\zeta}=2i\varphi\,.
\eeq
Indeed, writing $z-\zeta=F'(\zeta)\rho^2/4$ and $x=\theta/\varphi$, with $\theta\sim\theta+2\pi$, gives $ds^2=-d\rho^2-\rho^2d\theta^2+\zeta^2d\Omega_q^2+\hdots$. The two solutions are
\beq
\label{E:S1roots}
 \zeta_\pm=\frac{i\varphi\pm\sqrt{q^2-1-\varphi^2}}{q+1}\,.
\eeq
For generic boundary data the mass parameter is complex.

To translate back to the variables of Section~\ref{S:SpSqsuperspace}, take
\beq
\label{E:S1translation}
 B=z^2\,,\qquad A=\frac{F(z)}{\varphi^2}\,,\qquad
 \frac{dz}{dt}=\sqrt{F(z)}\,.
\eeq
The square root is continued from $\sqrt F\sim z$ at late time, and $B_0=\zeta^2$. Define
\beq
\label{E:S1Nariaidata}
 r_{\rm N}^2=\frac{q-1}{q+1}\,,\qquad
 \varphi_{\rm c}=\sqrt{q^2-1}\,.
\eeq
For $0<\varphi<\varphi_{\rm c}$ the two branches trace a circle in the $B_0$ plane,
\beq
\label{E:S1B0circle}
 B_{0,\pm}=r_{\rm N}^2e^{\pm2i\vartheta}\,,\qquad
 \varphi=\varphi_{\rm c}\sin\vartheta\,,\qquad
 0<\vartheta<\frac{\pi}{2}\,.
\eeq
The point $\varphi = 0$, i.e. $\vartheta = 0$, is the Nariai limit of the extremal Schwarzschild-de Sitter exterior, with a $\mathrm{dS}_2\times\mathbb{S}^q$ throat, and we term it the ``Nariai endpoint.'' It is connected to the circle of radius $r_{\rm N}^2$ in the $B_0$ plane, which we call the ``Nariai branch.''

\begin{figure}[t]
\centering
    \includegraphics[width=3in]{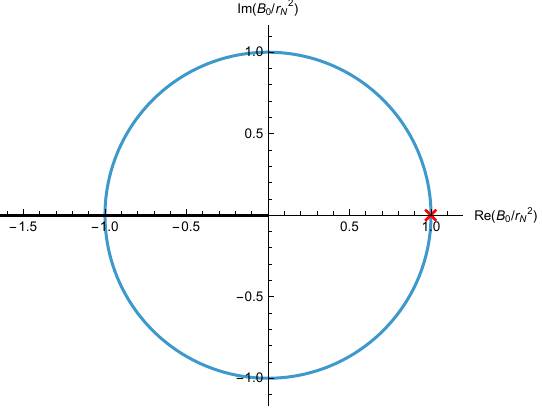}
\caption{\label{F:S1Sqphases} The various candidates of no-boundary geometries contributing to the $\mathbb{S}^1\times\mathbb{S}^q$ wavefunction. The Nariai branch (blue) begins at the Nariai endpoint at $B_0 = r_{\rm N}^2$ and follows a circle. The AdS branch exists (black) along the negative real axis, with (minus) small AdS black holes lying between $B_0 = -r_{\rm N}^2$ and $B_0=0$, and the (minus) large AdS black holes for $B_0<-r_{\rm N}^2$. These two meet where the Nariai branch intersects the negative real axis.}
\end{figure}

Above $\varphi_{\rm c}$ both roots are imaginary. Writing $\zeta=ir_h$ gives
\beq
\label{E:S1radii}
 r_{\rm s,l}=\frac{\varphi\mp\sqrt{\varphi^2-(q^2-1)}}{q+1}\,,\qquad
 L=\frac{4\pi r_h}{q-1+(q+1)r_h^2}\,.
\eeq
Along $z=ir$ the metric becomes
\beq
\label{E:S1minusAdS}
 ds^2=-\left(\frac{dr^2}{f(r)}+f(r)dx^2+r^2d\Omega_q^2\right)\,,
 \qquad
 f(r)=1+r^2-\frac{r_h^{q-1}(1+r_h^2)}{r^{q-1}}\,.
\eeq
These are minus the Euclidean small and large AdS-Schwarzschild black holes of the Hawking-Page problem. They have $\Theta=\pi/2$ in~\eqref{E:largeRhoBC} and form a line of solutions that we call the AdS branch. At large $\varphi$ the small black hole approaches $B_0=0$ from below and the large black hole has $B_0\to-\infty$, as shown in Fig.~\ref{F:S1Sqphases}.

In AdS the real black holes merge at $r_h=r_{\rm N}$, where $L$ is maximal. For larger $L$ their radii become complex, with $r_h=-i\zeta_\pm$. The Nariai branches are therefore minus the complexified AdS black holes of~\cite{Mahajan:2025bzo}. We will compare their contributions below.

There is also a filling where the $\mathbb{S}^q$ caps off. Setting $\mu=0$ gives minus thermal global AdS,
\beq
\label{E:S1thermal}
 ds^2=-\left(\frac{dr^2}{1+r^2}+(1+r^2)dx^2+r^2d\Omega_q^2\right)\,.
\eeq
The $\mathbb{S}^q$ shrinks at $r=0$ and the circle has arbitrary period. Its topology is $\mathbb{S}^1\times D^{q+1}$, while the black holes have topology $D^2\times\mathbb{S}^q$.

%--------------------------------------
\subsection{Tree-level wavefunction}
\label{S:S1tree}
%--------------------------------------

At tree level the wavefunction is a sum over saddles,
\beq
\label{E:S1saddlesum}
 \Psi_{\rm tree}(\varphi)=\sum_a n_a e^{iS_a}\,,
\eeq
where $a$ indexes the saddles of fixed $\varphi$ and the coefficients $n_a$ depend on the integration contour of the GPI. For now we include all candidates and compare their contributions. At a given $\varphi$ the dominant individual contribution has the largest $-\text{Im}S_a$, provided its coefficient $n_a$ is nonzero. We find the actions here and examine the coefficients $n_a$ in Subsection~\ref{S:S1contour}.

We evaluate the action directly on~\eqref{E:S1metric}, using the terms of Section~\ref{S:setup}. With a late-time cutoff at $z=Z$ the bulk and Gibbons--Hawking terms are
\begin{align}
\begin{split}
\label{E:S1regulatedaction}
 S_{\rm EH}&=\frac{L\Vol(\mathbb{S}^q)}{8\pi G}
 \left(Z^{q+1}-\zeta^{q+1}\right)\,,
 \\
 S_{\rm GH}&=-\frac{L\Vol(\mathbb{S}^q)}{8\pi G}
 \left((q+1)Z^{q+1}-qZ^{q-1}+\frac{q+1}{2}\mu\right)\,.
\end{split}
\end{align}
Here $R-2\Lambda=2(q+1)$ and the volume element is $z^q\,dz\,dx\,d\Omega_q$. The counterterms cancel the powers of $Z$ and contribute $qL\Vol(\mathbb{S}^q)\mu/(16\pi G)$ to the finite mass-dependent term. The renormalized action is
\begin{align}
\begin{split}
\label{E:S1directaction}
 S&=-\frac{\Vol(\mathbb{S}^q)}{8G\varphi }
 \left(2\zeta^{q+1}+\mu\right)+S_{\rm vac}=-\frac{\Vol(\mathbb{S}^q)}{8G\varphi}
 \zeta^{q-1}(1+\zeta^2)+S_{\rm vac}\,,
\end{split}
\end{align}
where $S_{\rm vac}\propto L$ is the real (and scheme-dependent in some $d$) action of minus thermal AdS where the circle has periodicity $2\pi /\varphi$.

This result has a thermodynamic interpretation. With $\kappa_q=\Vol(\mathbb{S}^q)/(4G)$, the mass and cap entropy are
\beq
\label{E:S1energyentropy}
 M=\frac{q\Vol(\mathbb{S}^q)}{16\pi G}\mu\,,\qquad
 \mathcal{S}_{\rm cap}=\frac{\Vol(\mathbb{S}^q)}{4G}\zeta^q\,.
\eeq
Using the condition that the geometry caps off smoothly~\eqref{E:S1smooth}, we can rewrite~\eqref{E:S1directaction} as
\begin{subequations}
\begin{align}
\label{E:S1offaction}
 iS&=\kappa_q\mathcal{E}_q+iS_{\rm vac}\,,
 \\
\label{E:S1exponent}
 \mathcal{E}_q&=\zeta^{q-1}\left( \zeta-\frac{iq}{2\varphi}(1-\zeta^2)\right)\,.
\end{align}
\end{subequations}
The saddle-dependent part is $\mathcal{S}_{\rm cap}-iL M$, relating the cap entropy to the energy conjugate to the boundary periodicity. In Subsection~\ref{S:S1contour} we will obtain this expression as an off-shell action where we allow for $\zeta$ to be a free parameter, that is for geometries that cap off with a conical singularity.

Treating $\zeta$ as a free parameter and extremizing~\eqref{E:S1exponent} with respect to it (and forbidding $\zeta = 0$) gives the condition~\eqref{E:S1smooth} that the geometry caps off smoothly. At any smooth cap saddle, $\mathcal{E}_q$ evaluates to
\beq
\label{E:S1onshell}
 \mathcal{E}_q  =\frac{\zeta^{q-1}}{q+1}\left(\zeta-\frac{iq}{\varphi}\right)\,,
 \qquad
 -\text{Im}S=\kappa_q\text{Re}\mathcal{E}_q\,.
\eeq
In evaluating this expression we must recall that $\zeta$ is in general complex.

The directly evaluated action~\eqref{E:S1directaction} also applies on the AdS branch, with $\zeta=ir_h$. To compare it with ordinary AdS gravity, define the Euclidean black hole action relative to thermal AdS,
\beq
\label{E:S1IE}
 I_{\rm E}=
 \frac{\Vol(\mathbb{S}^q)}{4G}
 \frac{r_h^q(1-r_h^2)}{q-1+(q+1)r_h^2}\,.
\eeq
The result for our saddle is
\beq
\label{E:S1AdSphase}
 iS=-i^q I_{\rm E}+iS_{\rm vac}\,.
\eeq
Along the imaginary-radius leg $z=ir$, the volume element obeys $z^q\,dz=i^{q+1}r^q\,dr$. Together with the $i$ in $e^{iS}$ this gives $-i^q$, with the sheet fixed by the Lorentzian late-time boundary condition. For odd $q$ the action is real, so $-\text{Im}S=0$ along the entire AdS branch. For even $q$,
\beq
\label{E:S1evenweight}
 -\text{Im}S=-(-1)^{q/2}I_{\rm E}\,.
\eeq
Minus thermal AdS has $\text{Im}S=0$ for all $q$.

At large anisotropy, the radii of the small and large black holes are
\beq
\label{E:S1largephi}
 r_{\rm s}=\frac{q-1}{2\varphi}+O(\varphi^{-3})\,,\qquad
 r_{\rm l}=\frac{2\varphi}{q+1}+O(\varphi^{-1})\,.
\eeq
For even $q$ the resulting actions for $\varphi\gg 1$ are
\begin{align}
\begin{split}
\label{E:S1weight}
 -\text{Im}S_{\rm s}
 &=(-1)^{q/2+1}\frac{\Vol(\mathbb{S}^q)(q-1)^{q-1}}{2^{q+2}G\varphi^q}
 \left(1+O(\varphi^{-2})\right)\,,
 \\
 -\text{Im}S_{\rm l}
 &=(-1)^{q/2}\frac{2^{q-2}\Vol(\mathbb{S}^q)}{(q+1)^{q+1}G}\varphi^q
 \left(1+O(\varphi^{-2})\right)\,.
\end{split}
\end{align}
We summarize the coefficients through $d=11$ in Table~\ref{T:S1weights}.

\begin{table}[H]
\centering
\renewcommand{\arraystretch}{1.55}
\begin{tabular}{c|c|c|c}
$q$ & $d$ & $-\text{Im}S_{\rm s}$ & $-\text{Im}S_{\rm l}$ \\ \hline
$2$ & $4$ & $\displaystyle\frac{\pi}{4G\varphi^2}$ &
 $\displaystyle-\frac{4\pi}{27G}\varphi^2$ \\
$4$ & $6$ & $\displaystyle-\frac{9\pi^2}{8G\varphi^4}$ &
 $\displaystyle\frac{32\pi^2}{9375G}\varphi^4$ \\
$6$ & $8$ & $\displaystyle\frac{625\pi^3}{48G\varphi^6}$ &
 $\displaystyle-\frac{256\pi^3}{15\cdot7^7G}\varphi^6$ \\
$8$ & $10$ & $\displaystyle-\frac{7^6\pi^4}{480G\varphi^8}$ &
 $\displaystyle\frac{2048\pi^4}{105\cdot9^9G}\varphi^8$ \\
$3,5,7,9$ & $5,7,9,11$ & $0$ & $0$
\end{tabular}
\caption{Leading large-$\varphi$ saddle weights on the AdS branch. These are logarithms of amplitudes; the corresponding logarithms of probabilities are twice these values. The zeros for odd $q$ are exact at tree level.}
\label{T:S1weights}
\end{table}

In $d=4$ the small black hole has the largest tree-level modulus at large $\varphi$, decreasing towards that of thermal AdS, while the large black hole is exponentially suppressed. These $d=4$ saddles were studied in~\cite{Anninos:2012ft,Banerjee:2013mca}, which consider both signs of $iS$. With the opposite sign the large black hole would run away even in $d=4$, but the same sign would give the round $\mathbb{S}^3$ the weight $e^{-\pi/(2G)}$ instead of the no-boundary weight $e^{\pi/(2G)}$. In $d=6$ the small black hole is mildly suppressed and the large black hole runs away. This runaway repeats in $d=10$ and motivates the contour analysis that follows.

%--------------------------------------
\subsection{A toy model for admissibility}
\label{S:S1contour}
%--------------------------------------

If the large black holes contribute for $q\equiv0\pmod4$, their weights give $|\Psi|^2\sim\exp(2c_q\varphi^q/G)$ with $c_q>0$, and hence a divergent contribution
\beq
\label{E:S1runawaynorm}
 \int^\infty\,d\varphi\,\mu(\varphi)
 \exp\!\left(\frac{2c_q}{G}\varphi^q\right)\,,
\eeq
to the no-boundary norm for $\mu(\varphi)$ the effective measure on the anisotropy. These smooth, weakly curved geometries lie within the regime of validity of gravitational EFT. Ordinary loop corrections remain smaller than the tree-level exponent wherever the expansion is controlled, and cannot remove this growth. 

So, assuming these saddles contribute, this runaway challenges the status of the no-boundary proposal as producing a state in semiclassical de Sitter quantum gravity. 

However, all is not lost. There is a useful analogy with AdS gravity. The complex black holes of~\cite{Mahajan:2025bzo} can have larger $e^{-I}$ than thermal AdS at low temperature, in AdS$_5$ and above, and still be absent from the canonical partition function. They argue that the integration contour of the GPI assigns them zero coefficient, $n_a = 0$. With this in mind, in this Subsection we study whether similar admissibility criteria eliminate the dangerous large black hole runaway.

It is not clear how seriously we should regard the KSW condition of Subsection~\ref{S:KSW}. For asymptotically hyperbolic geometries with $\mathbb{S}^1\times\mathbb{S}^q$ boundary, it allows the complex black holes just below the minimum black-hole temperature, which the integration contour of~\cite{Mahajan:2025bzo} excludes. Conversely, the sphere partition function of the minimal string comes from saddles that are minus a round sphere~\cite{Mahajan:2021maz}, and KSW would exclude these~\cite{Mahajan:2025bzo}. We proceed by constructing a toy model for the GPI, deducing the saddle coefficients by adapting the strategy of Appendix F of~\cite{Maldacena:2019cbz} and~\cite{Mahajan:2025bzo}.

The model is the following. Let us interpret the $\mathbb{S}^1\times\mathbb{S}^q$ wavefunction as an integral over an auxiliary parameter, which we take here to be the cap radius $\zeta$. On the AdS branch it is $i$ times the horizon radius, which~\cite{Mahajan:2025bzo} use as their integration variable. There is no smooth bulk solution at generic $(\varphi,\zeta)$. However, there is a near-solution, one which solves Einstein's equations everywhere apart from a conical singularity at the cap. We assign the off-shell integrand to be $e^{iS}$ for $S$ the off-shell action~\eqref{E:S1offaction} of this near-solution, so that the wavefunction is $\int d\zeta \,e^{iS}$ for some integration contour over $\zeta$. We study various candidate integration contours for $\zeta$, and see which can be deformed through the candidate bulk saddles of the last Subsection. 

Our toy integral is
\beq
\label{E:S1toy}
 \mathcal{I}_{\mathcal C}(\varphi)
 =e^{iS_{\rm vac}}
 \int_{\mathcal C}d\zeta\,\exp\left(\kappa_q\mathcal{E}_q\right)\,,
\eeq
with $\mathcal{E}_q$ given by~\eqref{E:S1exponent}. We posit a flat measure on $\zeta$, an acceptable approximation in the classical limit, and consider effects of the gravitational one-loop measure on normalizability later in the Section. The point $\zeta=0$ has the same action as thermal AdS. 

We could equally integrate over the vacuum-subtracted boundary energy $M$, retaining the Jacobian,
\beq
\label{E:S1energyintegral}
 \mathcal I_{\mathcal C}
 =e^{iS_{\rm vac}}\int_{\widetilde{\mathcal C}_M}dM\,
 \frac{d\zeta}{dM}
 \exp\!\left(\mathcal S_{\rm cap}(\zeta(M))-iLM\right)\,.
\eeq
Here $\widetilde{\mathcal C}_M$ follows the same caps as $\mathcal C$ on the branched inverse map $\zeta(M)$. Since $M\propto\zeta^{q-1}(1-\zeta^2)$ has a maximum $M_{\rm N}$ at $\zeta=r_{\rm N}$ on the positive real axis, the interval $0\leq\zeta\leq1$ maps to $0\to M_{\rm N}\to0$. The outward leg follows black hole caps and the return leg cosmological caps, with different entropies at the same energy. The Jacobian diverges integrably at the fold. Thus the energy contour goes out and comes back on different sheets, as observed in $d=4$ in Appendix F of~\cite{Maldacena:2019cbz}. The cap radius keeps these sheets explicit. In AdS the positive energy grows monotonically with the horizon radius, which is why~\cite{Mahajan:2025bzo} can pass freely between the two.

Choose a candidate cycle $\mathcal C_+$ running from the origin to the first convergent sector of the upper half-plane. One representative follows the positive real axis and turns at a finite large radius into that sector. The leading term in the effective action, $ib\zeta^{q+1}$, with $b=q/(2\varphi)>0$, puts its center at
\beq
\label{E:S1decaysector}
 \arg \zeta=\frac{\pi}{2(q+1)}\,.
\eeq
The turn is needed because the real exponent grows as $\zeta^q$ on the positive real axis. Since the integrand is entire, we can deform this curve to the straight central ray~\eqref{E:S1decaysector}, or to any ray within the same convergent sector. These are representatives of the same cycle, including the deformation used below to bound the large black hole contribution. This generalizes the $q=2$ prescription of~\cite{Maldacena:2019cbz}. In the energy plane the initial real segment continues past the return to $M=0$, through negative real $M$, and the central ray gives $M\to-i\infty$. Stopping at the return to zero would give the bounded cap contour $[0,1]$, whose extra endpoint we discuss below.

A saddle contributes when its upward cycle intersects $\mathcal C_+$. At real $G$, along that cycle $\text{Im}\mathcal E_q$ is constant and $\text{Re}\mathcal E_q$ increases, with flow $d\zeta/ds=\overline{\mathcal E'_q(\zeta)}$. Coincident saddle phases allow Stokes jumps in the saddle decomposition. We resolve them by following~\cite{Mahajan:2025bzo} and complexifying 
\beq
\label{E:S1lateral}
	\frac{1}{G}\to \frac{1}{|G|} (1\pm i \varepsilon)\,,\qquad \varepsilon\to0^+\,.
\eeq
Given a fixed contour and a fixed complexification of $G$, the toy integral is unambiguous. For a fixed convergent contour the two choices have the same exact limit. We average their thimble decompositions, the homology average of~\cite{Mahajan:2025bzo}, which we call the median prescription. It leaves the integral unchanged.

\begin{figure}[p]
\centering
\includegraphics[width=6in]{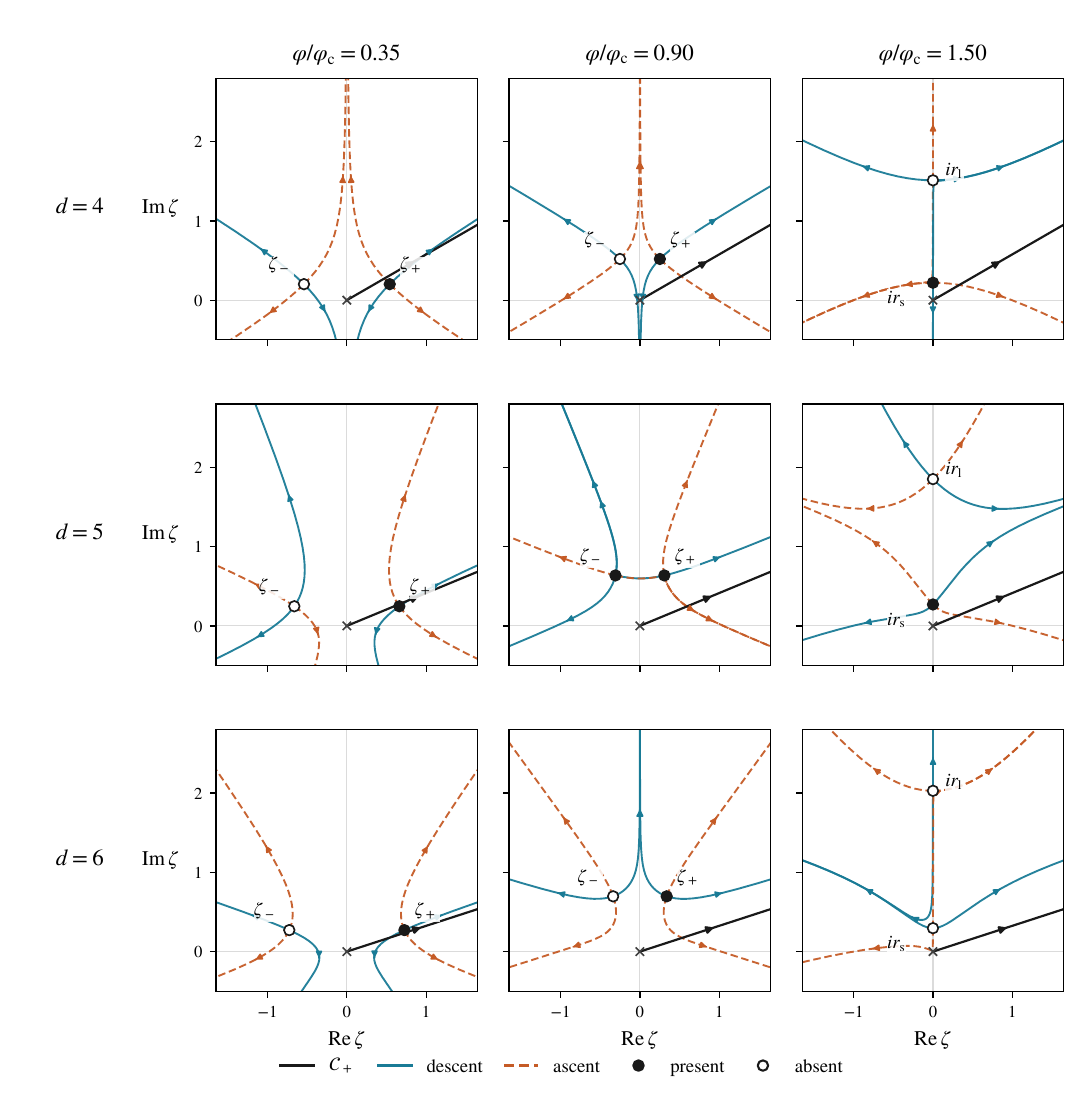}
\caption{Saddles and gradient flows of the cap-radius integral at three values of $\varphi/\varphi_{\rm c}$, with $\varphi_{\rm c}=\sqrt{q^2-1}$ and $q=d-2$. The black ray represents $\mathcal C_+$. Blue solid paths are descent cycles and orange dashed paths are ascent cycles, with arrows in the direction of flow. The cross marks the endpoint $\zeta=0$. Filled saddles have an upward intersection with the ray. We use $1/G\propto1+0.002i$ to separate Stokes connections. Changing the sign changes the Stokes-dependent entries of Table~\ref{T:S1contours}; the runaway large black hole in $d=6$ remains absent.}
\label{F:S1contours}
\end{figure}

Figure~\ref{F:S1contours} shows the smooth saddles and their flows in $d=4,5,6$. Filled points have an upward intersection with $\mathcal C_+$ for the indicated complexification of $G$. The paths make visible both the absence of the lower Nariai saddle at small $\varphi$ and the different saddle content after the merger.

Consider first $\varphi<\varphi_c$, below the merger. Taking $\zeta_+=r_{\rm N}e^{i\vartheta}$ we have
\beq
\label{E:S1generalNariaiexponent}
 \mathcal E_q(\zeta_+;\varphi)
 =\frac{r_{\rm N}^q}{q+1}
 \left(e^{iq\vartheta}
 -\frac{iq}{(q-1)\sin\vartheta}e^{i(q-1)\vartheta}\right)\,.
\eeq
The reflection identity
\beq
\label{E:S1reflection}
 \mathcal E_q(-\bar \zeta)=(-1)^q\overline{\mathcal E_q(\zeta)}
\eeq
gives the exponent at $\zeta_-=-\bar\zeta_+$. The two $B_0$ values are complex conjugates. The phases of $\mathcal{E}_q$ for these two values are opposite for even $q$ and coincide throughout the Nariai branch for odd $q$. Alignment with the thermal AdS endpoint at $\zeta= 0$ requires
\beq
\label{E:S1generalStokes}
 \text{Im}\mathcal E_q(\zeta_+)=0
 \quad\Longleftrightarrow\quad
 (q+1)\cos((q-1)\vartheta)
 +(q-1)\cos((q+1)\vartheta)=0\,.
\eeq
For even $q$ this also aligns the two smooth saddles. A Stokes jump requires a connecting flow, and several alignments can occur at larger $q$. On the AdS branch the exponents are $i^q$ times real numbers, so for even $q$ their phases coincide with the endpoint throughout. For odd $q$ a saddle aligns with the endpoint at $I_{\rm E}=0$, or $r_{\rm l}=1$ and $\varphi=q$ for the large black hole.

At large anisotropy we find that the large black hole does not contribute to this toy integral for odd $q$ and $q\equiv0\pmod4$. Since the integrand is entire, deform $\mathcal C_+$ within its convergent sector to the ray $\zeta=t e^{i\omega}$, with $t\geq0$ and
\beq
\label{E:S1boundedray}
 \frac{\pi}{2q}<\omega<\frac{\pi}{q+1}\,.
\eeq
The real parts of both the $\zeta^q$ and $ib\zeta^{q+1}$ terms in the effective action $\mathcal{E}_q$ are negative along this ray,
\beq
\label{E:S1raybound}
 \text{Re}\mathcal E_q(t e^{i\omega})
 =t^{q-1}\left(
 b\sin((q-1)\omega)
 +t\cos(q\omega)
 -bt^2\sin((q+1)\omega)\right)\,.
\eeq
The first coefficient in brackets is positive and the other two are negative, giving
\beq
\label{E:S1rayestimates}
 \sup_{t\geq0}\text{Re}\mathcal E_q(t e^{i\omega})\leq K_q b^q\,,
 \qquad
 \text{Re}\mathcal E_q\geq0
 \ \Longrightarrow\
 t=O(b)\,,\quad |\text{Im}\mathcal E_q|=O(b^q)\,,
\eeq
where $K_q>0$ is independent of $b$. Indeed, omitting the negative $t^2$ term shows that a nonnegative real exponent requires $t\leq b\sin((q-1)\omega)/|\cos(q\omega)|$, on which the full exponent is $O(b^q)$.

The exponent at the large black hole grows as $b^{-q}$,
\beq
\label{E:S1largesaddlebound}
 \mathcal E_q(ir_{\rm l})=i^q H_q\,,\qquad
 H_q=-\frac{I_{\rm E}(r_{\rm l})}{\kappa_q}
 \sim\frac{q^q}{(q+1)^{q+1}}b^{-q}>0\,.
\eeq
For $q\equiv0\pmod4$ the saddle's real exponent exceeds the maximum on the ray, so its upward cycle cannot intersect it. For odd $q$ an intersection would require a point with nonnegative real exponent and imaginary exponent $O(b^{-q})$. Equation~\eqref{E:S1rayestimates} bounds the latter by $O(b^q)$, excluding this case too. These separations persist under our prescription for complexifying $G$. The large black holes are therefore absent at sufficiently large $\varphi$ for odd $q$ and $q\equiv0\pmod4$. For $q\equiv2\pmod4$ their real exponent is large and negative, and this bound leaves their coefficients undetermined.

This is one of the main results of this Subsection. The contour choice $\mathcal C_+$ assigns zero coefficient to every runaway large black hole. So in this toy model the saddles that would spoil the normalizability of the no-boundary state simply do not contribute.

The same argument applies to toroidal slices. Let the late-time slice be a flat torus $\mathbb{T}^{q+1}$, with one circle of period $2\pi/\varphi$ and a transverse $\mathbb{T}^q$. When that circle caps off, the fillings are~\eqref{E:S1metric} with $d\Omega_q^2$ replaced by the flat metric on $\mathbb{T}^q$ and $F(z)=z^2+\mu/z^{q-1}$. The off-shell exponent is then the large-$\zeta$ part of~\eqref{E:S1exponent},
\beq
\label{E:S1planar}
 \mathcal{E}=\zeta^q+ib\zeta^{q+1}\,,
\eeq
with $\text{vol}(\mathbb{S}^q)$ replaced by $\text{vol}(\mathbb{T}^q)$ in $\kappa_q$. Its only smooth saddle is minus the Euclidean AdS black brane, at $\zeta=2i\varphi/(q+1)$, where $\mathcal{E}=i^q(2\varphi)^q/(q+1)^{q+1}$. For $q\equiv0\pmod4$ this is positive and grows as $\varphi^q$, a runaway, while for odd $q$ it is imaginary. On every ray~\eqref{E:S1boundedray} both terms in $\text{Re}\mathcal{E}$ are negative, so $\text{Re}\mathcal{E}<0$ there for all $t>0$ and every $\varphi$. For odd $q$ or $q \equiv 0\pmod 4$, the upward cycle of the brane therefore misses the ray. For $q\equiv0\pmod4$ it has $\text{Re}\mathcal{E}>0$ throughout, and for odd $q$ it has $\text{Re}\mathcal{E}\geq0$ and $\text{Im}\mathcal{E}\neq0$, while on the ray $\text{Re}\mathcal{E}\geq0$ only at $\zeta=0$, where $\mathcal{E}=0$. So $\mathcal{C}_{+}$ excludes the black brane at every $\varphi$ for odd $q$ or $q \equiv 0 \mod 4$. For $q \equiv 2 \mod 4$, this argument leaves its coefficient undetermined. This is the large-$\varphi$ limit of the black hole result: there the large black hole approaches the black brane, and the $\mathbb{S}^q$ curvature drops out of~\eqref{E:S1exponent}.

In $d=4$ this method recovers the results of Appendix F of~\cite{Maldacena:2019cbz}. Below $\varphi_{\rm c}=\sqrt3$ the upper Nariai saddle has an upward intersection and the lower saddle is absent. Above the merger the small black hole contributes. The large black hole's upward intersection changes depending on how we complexify $G$, making its coefficient Stokes-dependent. Its large-$\varphi$ weight is exponentially suppressed.

In $d=5$ the upper Nariai saddle contributes and the lower one undergoes an additional Stokes transition. Below the merger $\varphi<\varphi_{\rm c}=\sqrt8$ we have
\beq
\label{E:S1q3Nariaiexponents}
 \mathcal E_3(\zeta_\pm;\varphi)
 =\pm\frac{(8-\varphi^2)^{3/2}}{64}+iP(\varphi)\,,
 \qquad
 P(\varphi)=\frac{12\varphi-\varphi^3-24/\varphi}{64}\,.
\eeq
On the imaginary axis $\text{Im}\mathcal E_3(iy)=y^2(b(1+y^2)-y)\geq0$ for $b=3/(2\varphi)>1/2$, as holds below the merger. When $P<0$, the lower saddle's upward level curve cannot cross that axis to reach $\mathcal C_+$. The common saddle phase crosses the endpoint phase at
\beq
\label{E:S1q3Stokes}
 \varphi_{\rm S}=\sqrt{6-2\sqrt3}<\varphi_{\rm c}\,.
\eeq
For $\varphi_{\rm S}<\varphi<\varphi_{\rm c}$ an upward connection from the lower saddle to the upper one makes the lower saddle present with the plus sign in~\eqref{E:S1lateral} and absent with the minus sign. Its smaller weight leaves the leading near-Nariai wavefunction unchanged.

After the merger the small black hole contributes and the large one is absent. Their pure-phase actions conceal this distinction in the tree-level modulus. Following the level curves through the merger extends the large-$\varphi$ exclusion throughout the AdS branch.

In $d=6$ the upper Nariai saddle again contributes. Above the merger the large black hole has no upward intersection in either complexification of $G$, while the small black hole is Stokes-dependent. Table~\ref{T:S1contours} summarizes the results in $d=4,5,6$.

\begin{table}[H]
\centering
\renewcommand{\arraystretch}{1.3}
\begin{tabular}{c|c|c|c}
$d$ & Nariai branches & Small AdS black hole & Large AdS black hole \\ \hline
$4$ & upper only & present & Stokes-dependent  \\ \hline
$5$ & \shortstack{upper; lower\\Stokes-dependent$^\dagger$} & present & absent \\ \hline
$6$ & upper only & Stokes-dependent & absent
\end{tabular}
\caption{Smooth saddle content of the unit-measure toy integral on $\mathcal C_+$. The Nariai entries refer to $\varphi<\varphi_{\rm c}$ and the black hole entries to $\varphi>\varphi_{\rm c}$; the merger itself requires a uniform approximation. Stokes-dependent entries require the lateral or median prescription. $^\dagger$The lower $d=5$ Nariai saddle is absent for $\varphi<\varphi_{\rm S}$ and Stokes-dependent for $\varphi_{\rm S}<\varphi<\varphi_{\rm c}$.}
\label{T:S1contours}
\end{table}

It is instructive to compare with the canonical AdS contour, the positive real $r$ axis. Its off-shell exponent is
\beq
\label{E:S1canonical}
 \mathcal{E}_{\rm AdS}(r)=r^q-b r^{q-1}(1+r^2)\,,\qquad
 \mathcal{E}_q(ir)=i^q\mathcal{E}_{\rm AdS}(r)\,.
\eeq
This contour is convergent. Complex black holes, geometries of the form $(-\text{Nariai branch})$, do not contribute, while at high temperature the large real black hole contributes and the unstable small black hole is absent when averaging over complexifications of $G$~\cite{Mahajan:2025bzo}. Setting $\zeta=ir$ maps the canonical AdS contour to the positive imaginary axis and multiplies the exponent by $i^q$. For $q=2$ it reverses the exponent, while for $q=4$ it preserves it. The different cycles explain why a large black hole favored in thermal AdS can be absent from the wavefunction, and why complex black holes can contribute as Nariai saddles. For $q=3$ our transition occurs where $\text{Re}\mathcal E_{\rm AdS}=-P(\varphi)$ crosses zero. The complex black holes remain absent from the canonical contour on both sides.

There are several assumptions in this toy model. One is the simple off-shell form, and another is the integration contour. Already in $d=4$ integrating $\zeta$ over a regulated real-line contour includes both Nariai saddles, while the interval from zero to one adds another endpoint~\cite{Maldacena:2019cbz}. Thus, requiring that the physically sensible Nariai endpoint contributes does not fix the candidate cycle for $\zeta$. 

The comparison with KSW is also instructive. The AdS branch, and likewise the black brane, has $B_0<0$, so it fails the KSW cap test~\eqref{E:KSWwedge}. The cap test cannot tell the two Nariai branches apart, since $B_{0,-}=\overline{B_{0,+}}$, but the full condition can.

The necessary cap condition can be made quantitative on the upper Nariai branch. With $\zeta_+=r_{\rm N}e^{i\vartheta}$, the $q$ sphere directions each have phase $2\vartheta$. Even if the two directions of the disk that caps off the circle are chosen optimally, KSW requires $2q\vartheta<\pi$. Hence
\beq
\label{E:S1KSWthreshold}
 \varphi < \varphi_{\rm KSW}
 = \sqrt{q^2-1}\sin\left(\frac{\pi}{2q}\right)
 < \varphi_{\rm c}\,.
\eeq
For $q=2,3,4$ this gives $\varphi_{\rm KSW}=\sqrt{3/2},\sqrt{2},\sqrt{15}\sin(\pi/8)$, respectively.

For the upper branch, below this bound, one can construct a sequence of strictly allowable finite-cutoff contours approaching the late-time saddle. An explicit contour is
\beq
\label{E:S1KSWcontour}
 z(s)=\left(\zeta_+^{q+1}+s\left(Z^{q+1}-\zeta_+^{q+1}\right)\right)^{1/(q+1)}\,,
 \qquad 0\leq s\leq1\,,
\eeq
with the branch continued from the cap to a large real cutoff $Z$. At finite $Z$ the cap and the circle normalization must be shifted to impose exactly real boundary data; the corresponding contour is given in Appendix~\ref{S:KSWappendix}. It obeys the strict inequality in the interior for fixed $0<\varphi<\varphi_{\rm KSW}$ and sufficiently large $Z$. The limiting representative saturates KSW throughout, with a Lorentzian metric in the future. Admissibility here therefore refers to this regulated sequence. At $\varphi=\varphi_{\rm KSW}$ the cap condition itself saturates, and the construction does not establish the same strict statement.

Below $\varphi_{\rm KSW}$ the lower Nariai cap passes the same test, but no allowable contour that respects the $SO(q+1)\times U(1)$ symmetry connects it to the late-time region. Along any such contour, the $\mathbb{S}^q$ eigenvalue is $z(s)^2$, so KSW requires $q|\arg z^2|<\pi$ throughout. This confines $z$ to one of two disjoint sectors, $|\arg z|<\pi/(2q)$ or $|\arg z-\pi|<\pi/(2q)$. The lower cap $\zeta_-=-\bar\zeta_+$ lies in the second and the late-time cutoff $z=Z$ in the first, so every contour between them crosses the imaginary $z$ axis, where the $\mathbb{S}^q$ has negative squared radius. The contour cannot avoid this by passing through $z=0$, where the geometry is singular. Applying the reflection and conjugation $z\to -\bar{z}$ to the upper-branch contour gives an allowable contour from $\zeta_-$ to $z = -Z$. 
In the coordinate $-z$ it describes the complex conjugate of the upper saddle, which contributes to $\Psi^*$ rather than to $\Psi$.

This sharpens the comparison with the toy integral. In the dimensions analyzed above, KSW and the contour model agree on two points. Both exclude the runaway large black holes and black branes. Both also exclude the lower Nariai saddle, KSW at least for symmetric contours and the contour model apart from the Stokes-dependent range in $d=5$. The only smooth saddle that KSW allows, the upper Nariai saddle below $\varphi_{\rm KSW}$, appears on $\mathcal{C}_+$. Beyond this the contour model is more permissive. The upper Nariai saddle contributes beyond $\varphi_{\rm KSW}$ and up to the merger, although no contour can satisfy the KSW cap bound there, and in $d=4,5$ the small all-minus black hole contributes although it fails KSW at the cap. The endpoint $\zeta=0$, which has the action of minus thermal AdS, also contributes, although minus thermal AdS fails KSW. Its circle has negative squared length where the $\mathbb{S}^q$ caps off.

We have tried, and found it challenging to extend this analysis to $p,q>1$. The difficulty is finding a family of off-shell candidate bulk geometries.\footnote{One could use gravitational constrained instantons~\cite{Cotler:2020lxj}, performing the saddle approximation at fixed constraint and then integrating over its value. A suitable gauge can also give forced saddles, the $\lambda$-solutions of~\cite{Cotler:2021cqa}, with auxiliary fields sourcing Einstein's equations. These methods parameterize AdS wormhole amplitudes~\cite{Cotler:2021cqa, Cotler:2022rud} and may give an analogous model for these product-sphere fillings.}

%--------------------------------------
\subsection{Large $\mathbb{S}^1$ limit}
\label{S:S1nearNariai}
%--------------------------------------

We now return to the upper Nariai branch as $\varphi\to0$. The boundary circle becomes large and the bulk develops a long nearly $\mathrm{dS}_2\times\mathbb{S}^q$ throat. Dimensional reduction extends the leading quantum calculation of~\cite{Turiaci:2025xwi} from $q=2$ to general $q$.

Reducing on the $\mathbb{S}^q$, write $ds_d^2=\bar g_{ab}dx^a dx^b+\Phi^{2/q}d\Omega_q^2$ and take $G_2=G/\Vol(\mathbb{S}^q)$. The reduced action is
\beq
\label{E:S1reduction}
 \begin{split}
 S_2&=\frac{1}{16\pi G_2}\int d^2x\sqrt{-\bar g}
\left( \Phi\bar R+\frac{q-1}{q}\frac{(\bar\nabla\Phi)^2}{\Phi}+q(q-1)\Phi^{1-2/q}-q(q+1)\Phi\right)
 +(\text{bdy terms})\,.
 \end{split}
\eeq
A 2d Weyl transformation removes the dilaton kinetic term. We choose its constant part so that the Nariai $\mathrm{dS}_2$ radius is unity,
\beq
\label{E:S1Weyl}
 \bar g_{ab}=\frac{1}{q+1}
 \left(\frac{\Phi}{\Phi_0}\right)^{-(q-1)/q}(g_2)_{ab}\,,
 \qquad \Phi_0=r_{\rm N}^q\,.
\eeq
The action becomes
\begin{align}
\begin{split}
\label{E:S1dilaton}
 S_2&=\frac{1}{16\pi G_2}\int d^2x\sqrt{-g_2}
 \left(\Phi R_2-\mathcal U(\Phi)\right)+(\text{bdy terms})\,,
 \\
 \mathcal U(\Phi)&=q\Phi_0\left(
 \left(\frac{\Phi}{\Phi_0}\right)^{1/q}
 -\left(\frac{\Phi}{\Phi_0}\right)^{-1/q}\right)\,.
\end{split}
\end{align}
In these conventions
\beq
\label{E:S1potentialexpansion}
 \mathcal U(\Phi_0)=0\,,\qquad
 \mathcal U'(\Phi_0)=2\,,\qquad
 \mathcal U''(\Phi_0)=-\frac{2}{\Phi_0}\,,
 \qquad
 \mathcal U(\Phi_0+\delta\Phi)
 =2\delta\Phi-\frac{\delta\Phi^2}{\Phi_0}+\cdots\,.
\eeq
The constant-dilaton solution, the Nariai endpoint dS$_2\times\mathbb{S}^q$, has $R_2=\mathcal U'(\Phi_0)=2$, as required.

The quadratic deformation controls the departure from Nariai, as in~\cite{Ivo:2025yek, Cotler:2026jdz}. To leading order the throat and dilaton perturbation are
\beq
\label{E:S1throat}
 ds_2^2=-d\tau^2+\cosh^2\tau\,d\theta^2\,,\qquad
 \delta\Phi=\eta\sinh\tau\,,\qquad
 \eta=\frac{q r_{\rm N}^{q-1}}{q+1}\varphi\,,
 \qquad \theta\sim\theta+2\pi\,.
\eeq
The cap value $\zeta_+^q=\Phi_0+i\eta+O(\varphi^2)$ fixes $\eta$. On the Euclidean hemisphere $\tau=iu$ the perturbation becomes $\delta\Phi=i\eta\sin u$. The quadratic deformation gives
\beq
\label{E:S1quadraticcap}
-\text{Im}(S_2)_{\rm quadratic}
 =-\frac{\eta^2}{24G_2\Phi_0}\,,
\eeq
which is consistent with the general result in deformed dS JT gravity in~\cite{Cotler:2026jdz}. The imaginary cap perturbation turns the breathing instability~\cite{Ivo:2025yek,Shi:2025xxx} of the Euclidean product space $\mathbb{S}^2\times\mathbb{S}^q$ into a suppressing Gaussian when evaluating the wavefunction. We will encounter a similar feature numerically for the $\mathbb{S}^2\times\mathbb{S}^2$ wavefunction in the next Section.

Expanding the full action~\eqref{E:S1onshell} gives the same result,
\begin{align}
\begin{split}
\label{E:S1Nariaigaussian}
 -\text{Im}S(\varphi)
 &=\frac{\Vol(\mathbb{S}^q) r_{\rm N}^q}{4G}
 -\frac{u_q}{2G}\varphi^2+O(\varphi^4/G)\,,
 \\
 u_q&=\frac{\Vol(\mathbb{S}^q) q^2 r_{\rm N}^{q-2}}{12(q+1)^2}>0\,.
 \end{split}
\end{align}
Define $\mathcal S_{\rm N}=\Vol(\mathbb{S}^q)r_{\rm N}^q/(2G)$, the logarithm of $|e^{iS_{\rm N}}|^2$, the tree-level probability distribution at the Nariai endpoint, which is also minus the Euclidean action of the $\mathbb{S}^2\times\mathbb{S}^q$ saddle. For $q=2$, $\mathcal S_{\rm N}=2\pi/(3G)$ and $u_2=4\pi/27$.

The constant tree-level probability gives a logarithmic divergence when integrated against the measure $d\varphi/\varphi$. At one loop we expect that the throat's Schwarzian modes supply a $\varphi^{3/2}$ prefactor to the wavefunction, and that each generator of the rotational $SO(q+1)$ sector supplies $\varphi^{1/2}$. Together they give $\varphi^{a_q}$, with
\beq
\label{E:S1softpower}
 n_{\rm rot}=\frac{q(q+1)}{2}\,,\qquad
 a_q=\frac32+\frac{n_{\rm rot}}2\,.
\eeq
These powers count pairs of boundary Fourier modes. The Schwarzian has $n\geq2$, with regularized count $\sum_{n\geq2}1=-3/2$, and each gauge generator has $n\geq1$, with count $-1/2$. Their Gaussian determinants give the stated powers.

With the covariant fluctuation measure normalized by the $1/G$ in the quadratic action, Gaussian factors of $G$ cancel mode by mode. At fixed subtraction scale and renormalized couplings in de Sitter units, this gives
\beq
\label{E:S1Nariaiwave}
 \Psi_{\rm N}(\varphi)\simeq
 \mathcal N_q\,\varphi^{a_q}
 \exp\left(\frac{\mathcal S_{\rm N}}2
 -\frac{u_q}{2G}\varphi^2+i\mathcal P_q(\varphi)\right)\,,
\eeq
where $\mathcal P_q$ is real, and where there are higher-order corrections in small $\varphi$. The prefactor $\mathcal N_q$ includes the finite massive-mode determinant and late-time measure normalization, and is $G$-independent at one loop in this convention. Higher corrections come from further terms in the cap action and interactions of the soft modes, controlled for $G\ll\varphi\ll1$. The wavefunction also depends on other fluctuations of the boundary metric. In these conventions, integrating over those fluctuations in the norm using the ultralocal measure on fluctuations of the boundary metric ought to produce a $G$- and $\varphi$-independent determinant. The Gaussian selects $\varphi\sim\sqrt G$ within this regime, giving schematically
\beq
\label{E:S1NariaiNorm}
 \|\Psi\|_{\rm N}^2\simeq
 e^{\mathcal S_{\rm N}}
 \int_0^\infty\frac{d\varphi}{\varphi}
 \varphi^{2a_q}e^{-u_q\varphi^2/G}
 \propto  G^{a_q}e^{\mathcal S_{\rm N}}\,.
\eeq

The exponential and power $G^{a_q}$ agree with the schematic form of the Euclidean gravity amplitude on $\mathbb{S}^2\times\mathbb{S}^q$. Its residual isometry group $SO(3)\times SO(q+1)$ has dimension $3+n_{\rm rot}=2a_q$, and division by the properly normalized volume of its isometry group produces the same power of $G$. The Lorentzian norm and Euclidean amplitude thus agree at tree and logarithmic order. We expect that their finite prefactors differ on account of examples of one-loop differences between the Lorentzian inner product and Euclidean amplitude discussed in~\cite{Cotler:2026jdz}.

%--------------------------------------
\subsection{$\mathbb{S}^1\times\mathbb{S}^2$ at large $\mathbb{S}^2$}
\label{S:S1largeS2}
%--------------------------------------

We can also proceed analytically at large $\varphi$ in $d=4$, where the $\mathbb{S}^2$ becomes large. In our toy model for admissibility the small black hole contributes on $\mathcal C_+$, with
\beq
\label{E:S1small4}
 r_{\rm s}=\frac{1}{2\varphi}+O(\varphi^{-3})\,,\qquad
 -\text{Im}S_{\rm s}
 =\frac{\pi}{4G\varphi^2}+O\!\left(\frac{1}{G\varphi^4}\right)\,.
\eeq
The tree-level amplitude tends to a constant, so loop effects determine whether or not the wavefunction is normalizable at large anisotropy. In this case we can give an approximate result for the one-loop prefactor of the wavefunction, from the shrinking black hole and the graviton gas surrounding it.

Near the cap the geometry approaches minus the flat space Euclidean Schwarzschild geometry. Its growing curvature requires
\beq
\label{E:S1capcontrol}
 \frac{G}{r_{\rm s}^2}\sim G\varphi^2\ll1\,.
\eeq
The cap becomes Planckian at $\varphi\sim G^{-1/2}$, although the backreaction of the graviton gas will become important for smaller $\varphi$.

In the corresponding AdS geometry the small black hole is a heavy compact object near the center, with $r_{\rm s}\ll1$ and $LM_{\rm AdS}\gg1$. Flat-space Schwarzschild describes its local fluctuations. The AdS potential lifts its translation modes, and the exterior supplies the thermal graviton gas. These factors are evaluated on the rotated fluctuation contours of our all-minus saddle.

Gross, Perry, and Yaffe~\cite{Gross:1982cv} calculated the local Schwarzschild determinant after dividing out the asymptotic thermal gas,
\beq
\label{E:S1GPYflat}
 |\mathcal D_{\rm flat}|
 \propto(\mu_{\rm R}L)^{212/45}
 \left(\frac{M_{\rm AdS}}{2\pi L}\right)^{3/2}
 \int d^3X\,,
 \qquad M_{\rm AdS}\simeq\frac{r_{\rm s}}{2G}\,.
\eeq
Here $M_{\rm AdS}$ is the positive mass of the corresponding AdS black hole, $\mu_{\rm R}$ is a fixed renormalization scale, $X$ are the three translation collective coordinates, and $212/45$ is the one-loop gravitational scaling anomaly. 

Global AdS localizes the black hole near its center, lifting the translations. Its energy there is $M_{\rm AdS}\sqrt{1+X^2}\simeq M_{\rm AdS}+M_{\rm AdS}X^2/2$, so each position integral gives $(LM_{\rm AdS})^{-1/2}$. The rotated steepest-descent contours of our saddle give the same modulus. In this way the small black hole contribution to the determinant is
\beq
\label{E:S1GPYAdS}
 |\mathcal D_{\rm local}|
 \propto(\mu_{\rm R}L)^{212/45}L^{-3}
 \propto\varphi^{-77/45}\,.
\eeq
The factors of $M_{\rm AdS}$, including their powers of $G$, cancel. The AdS potential replaces the flat-space position volume, with localization controlled by $LM_{\rm AdS}\sim1/(G\varphi^2)\gg1$. This determinant already includes the radius-instability Gaussian of the toy integral.

To leading order in $r_{\rm s}$ the remaining determinant is the thermal global AdS graviton gas, with one-particle partition function and Bose-Einstein sum~\cite{Giombi:2014yra}
\begin{align}
\begin{split}
\label{E:S1gravitoncharacter}
 z_{\rm grav}(x)&=\frac{5x^3-3x^4}{(1-x)^3}\,,
 \\
 \ln Z_{\rm gas}(L)
 &=\sum_{n=1}^\infty\frac1n z_{\rm grav}(e^{-nL})
 =\frac{\pi^4}{45L^3}+O(L^{-1})
 =\frac{\pi}{360}\varphi^3+O(\varphi)\,.
 \end{split}
\end{align}
The leading term counts two graviton polarizations in global AdS. The all-minus geometry reverses the $d=4$ classical action, while its Gaussian determinants have the same modulus on the rotated contours. The gas therefore enhances the wavefunction. The small black hole changes its exponent by $O(r_{\rm s}\varphi^3)=O(\varphi^2)$.

Putting these pieces together gives
\beq
\label{E:S1smallwave}
 |\Psi_{\rm s}(\varphi)|\simeq
 \mathcal N_{\rm s}\,\varphi^{-77/45}Z_{\rm gas}(2\pi/\varphi)
 \exp\left(\frac{\pi}{4G\varphi^2}\right)\,,
\eeq
at one loop and leading order at large $\varphi$, with $\mathcal N_{\rm s}$ independent of $G$ in these conventions. The local power alone gives a square-integrable tail against $d\varphi/\varphi$, although the effective measure at large $\varphi$ is presumably no longer $1/\varphi$ on account of the integration over the directions perpendicular to $\varphi$ in the distribution $|\Psi|^2$. The gas eventually makes the distribution grow.

The logarithm of the probability is
\beq
\label{E:S1largeprob}
 \begin{split}
 \mathcal F(\varphi)=\ln|\Psi_{\rm s}|^2
 &=\frac{\pi}{2G\varphi^2}
 +\frac{\pi}{180}\varphi^3-\frac{154}{45}\ln\varphi+2\ln\mathcal N_{\rm s}\\
 &\quad+O(\varphi^2)+O\!\left(\frac{1}{G\varphi^4}\right)+O(G\varphi^7)\,.
 \end{split}
\eeq
Gas backreaction first contributes $O(G\varphi^7)$, smaller than the free-gas exponent by $G\varphi^4$. The decreasing tree-level term and increasing gas term balance at
\beq
\label{E:S1minimum}
 \varphi_*=\left(\frac{60}{G}\right)^{1/5}(1+o(1))\,,
 \qquad
 \mathcal F(\varphi_*)=
 \frac{\pi}{72}\left(\frac{60}{G}\right)^{3/5}
 +O(G^{-2/5})+O(\ln G)\,.
\eeq
This is a local minimum of $|\Psi_{\rm s}|^2$, whose leading location is insensitive to power-law prefactors. A minimum of the fully integrated anisotropy distribution also depends on the transverse determinant. Both $G\varphi_*^2\sim G^{3/5}$ and $G\varphi_*^4\sim G^{1/5}$ vanish as $G\to0$, so the cap and gas expansions are controlled. Thermal AdS shares the leading gas factor but is subdominant throughout relative to the small black hole.

Beyond the minimum the gas drives the distribution upwards. Its central temperature is $O(L^{-1})$ and stress tensor $O(L^{-4})$, making backreaction important at
\beq
\label{E:S1gasbreakdown}
 G\varphi^4\sim1\,,\qquad
 \varphi_{\rm gas}\sim G^{-1/4}\,.
\eeq
This precedes the Planckian cap, with energies and curvature still below the Planck scale. Continuing further requires making sense of higher loop effects.

To estimate the controlled tail's contribution to the norm, take a cutoff $\varphi_{\rm cut}=\epsilon G^{-1/4}$ with fixed $\epsilon\ll1$. The upper endpoint dominates the one-loop integral from the minimum to the cutoff,
\beq
\label{E:S1cutnorm}
 \ln\!\left(\int_{\varphi_*}^{\varphi_{\rm cut}}
 \frac{d\varphi}{\varphi}|\Psi_{\rm s}(\varphi)|^2\right)
 =\frac{\pi\epsilon^3}{180}G^{-3/4}
 +O(G^{-1/2})+O(\ln G)\,.
\eeq
Backreaction changes the leading coefficient by a small relative amount $O(\epsilon^4)$. Comparing with the near-Nariai norm $\sim G^3e^{2\pi/(3G)}$ gives $\|\Psi\|_{\rm tail}^2/\|\Psi\|^2_{\rm N} \sim e^{-2\pi/(3G)+O(G^{-3/4})}$, so within this homogeneous estimate the semiclassical part of the large-$\varphi$ tail contributes an exponentially smaller amount than the Nariai regime to the no-boundary norm.

The circle has given us analytic control over the candidate saddles and their actions, a toy contour excluding the runaway saddles, and two limits where loop effects matter. For general $p,q>1$ we must find the geometries numerically. Determining their contributions still requires an off-shell model for the GPI and a complete integration contour, which is presently beyond us. So in what follows we will continue to distinguish candidate weights from the question of which saddles contribute to the no-boundary wavefunction.

%--------------------------------------
\section{Case study: $\mathbb{S}^2\times\mathbb{S}^2$ wavefunction}
\label{S:S2S2}
%--------------------------------------

In this Section we study the simplest example of a $\mathbb{S}^p\times\mathbb{S}^q$ wavefunction where both factors have $p,q>1$, namely the $p=q=2$ wavefunction in $5d$. Unlike for the $\mathbb{S}^1\times\mathbb{S}^q$ wavefunction, all of the bulk geometries must be obtained numerically, and to do so we follow the algorithm described at the end of Section~\ref{S:setup}. We first present our solutions, with a rather complex ``phase diagram,'' and then use those solutions to present the tree-level wavefunction $e^{iS}$.

%--------------------------------------
\subsection{Candidate saddles}
\label{S:S2S2solutions}
%--------------------------------------

In our numerical algorithm we begin with solutions that, for $t = \rho - \rho_0 + i \Theta$ with some $(\rho_0,\Theta)$, smoothly cap off at $\rho = 0$, and then numerically solve for them along the $\rho>0$ axis up to some large value of $\rho$ where they obey~\eqref{E:largeRhoBC}, which we reprise here:
\begin{equation*}
	A = e^{2(\rho-\rho_0)+2i\Theta}(1 + O(e^{-2\rho}))\,, \qquad B = e^{2(\rho-\rho_0)+2i\Theta}(\varphi^2+O(e^{-2\rho}))\,,
\end{equation*}
These are labeled by a single complex constant $B_0$ labeling the solution with
\begin{align}
\begin{split}
\label{E:nearCapSol}
	A(\rho) & = -\rho^2 \left( 1 + \frac{2B_0+1}{9B_0} \rho^2 + \frac{2(92B_0^2-70B_0+23)}{2025B_0^2}\rho^4 + O(\rho^6)\right)\,,
	\\
	B(\rho) &= B_0\left( 1 + \frac{4B_0-1}{3B_0}\rho^2 + \frac{(4B_0-1)(16B_0-1)}{135B_0^2}\rho^4 + O(\rho^6)\right)\,.
\end{split}
\end{align}
We use this approximate solution (extended to several more orders) to initialize our numerics at a value of $\rho$, call it $\rho_i>0$, sufficiently small that higher-order terms are negligible, but sufficiently large that the smallness of $A$ does not spoil the numerical evolution. In practice this can be achieved with $\rho_i \sim 10^{-5}$ unless $|B_0|$ is sufficiently small, in which case we must ensure that $\rho_i \ll \sqrt{|B_0|}$. 

It is instructive to compute the Kretschmann scalar, a measure of the curvature, near the cap,
\begin{align}
\begin{split}
	K(\rho)& = R_{\mu\nu\rho\sigma}R^{\mu\nu\rho\sigma} = \frac{8(6B_0^2-2B_0+1)}{B_0^2} -\frac{64(B_0-1)^2(4B_0-1)}{9B_0^3}\rho^2 
	\\
	& \qquad \qquad \qquad \qquad + \frac{256(B_0-1)^2(4B_0-1)(121B_0-34)}{2025B_0^4}\rho^4 + O(\rho^6)\,.
\end{split}
\end{align}
In particular  at the cap it is
\beq
	K(0) = 48 - \frac{16}{B_0} + \frac{8}{B_0^2}\,,
\eeq
which becomes strong when $|B_0|$ is small.

There is a special initial condition $B_0 = \frac{1}{4}$ such that the geometry can be determined analytically,
\beq
	ds^2 = -\left( d\rho^2 + \frac{1}{2}\sinh^2(\sqrt{2}\rho)d\Omega_2^2\right) + \frac{1}{4}d\widetilde{\Omega}_2^2\,,
\eeq
which becomes Lorentzian dS$_3\times \mathbb{S}^2$ under the continuation $\sqrt{2}\rho = t - \frac{\pi i}{2}$. We refer to this solution as the ``Nariai point'' and at it the dS$_3$ has radius $\frac{1}{\sqrt{2}}$ and the $\mathbb{S}^2$ has radius $\frac{1}{2}$. This geometry in turn is the continuation of the Euclidean $\mathbb{S}^3\times\mathbb{S}^2$ saddle.

\begin{figure}[t]
\begin{center}
	\includegraphics[width=4in]{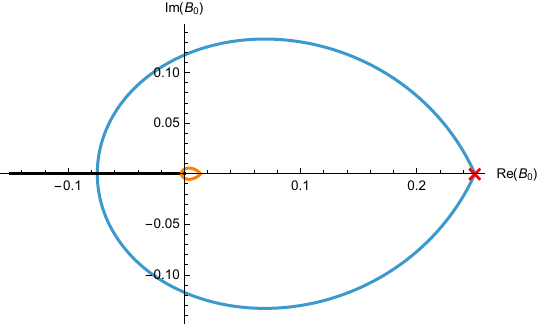}
	\caption{\label{F:S2S2phases} The various branches of no-boundary geometries that can contribute to the $\mathbb{S}^2\times\mathbb{S}^2$ wavefunction. These include the Nariai branch (blue), the AdS branch (black), and the first Russian doll branch (orange). The Nariai point $B_0=1/4$ is indicated as a red cross.}
\end{center}
\end{figure}

Shooting in $B_0$ until we achieve the late-time boundary condition $\lim_{\rho\to\infty}B/A=\varphi^2>0$, we find three types of solutions:
\begin{enumerate}
	\item The ``Nariai branch,'' topologically a circle in the complex $B_0$ plane, emanating from the Nariai point $B_0$ and encircling the origin, intersecting the negative real axis at $B_0 \approx -0.075$.
	\item The ``AdS branch,'' solutions with $B_0<0$. These solutions have $A,B<0$ along the Maldacena contour. They are the all-minus version of asymptotically hyperbolic geometries with $\mathbb{S}^2\times\mathbb{S}^2$ boundary.
	\item A ``Russian doll'' of nested complex branches with $|B_0|\to 0$. Each of these is a topological circle, encircling the origin. The first lies inside the Nariai branch, the second lies inside the first, and so on. 
\end{enumerate}
We plot these results in Fig.~\ref{F:S2S2phases} and comment on the solutions in a little more detail below. 

The reduced Einstein equations and regularity conditions have real coefficients and are invariant under complex conjugation. Consequently, if $\{A(\rho),B(\rho)\}$ is a solution for some $B_0=b$ and $\varphi$, then $\{\overline{A(\rho)},\overline{B(\rho)}\}$ is a solution with $B_0=\bar b$ and the same $\varphi$. This explains the symmetry of Fig.~\ref{F:S2S2phases} under complex conjugation of $B_0$. The corresponding asymptotic phase transforms as $\Theta\to \pi-\Theta$.

The map from $B_0$ to the boundary anisotropy $\varphi$ is not one-to-one. For some regimes in $\varphi$ there are multiple ways of filling in the same fixed boundary geometry. Each such filling is a candidate saddle contribution to the wavefunction. To find which saddle dominates we must determine which has the largest $-\text{Im}(S)$. We discuss both the anisotropy $\varphi$ and renormalized action along these branches in the remainder of this Section.

As we show in Section~\ref{S:S1}, the basic structure consisting of a Nariai point connected through conjugate complex branches to an AdS branch also appears in the exactly solvable $\mathbb S^1\times\mathbb S^q$ problem. The infinite sequence of nested branches accumulating at $B_0=0$, however, has no analogue there.

In the solutions described above, it is always the $A$-sphere that caps off. Since $p=q$, exchanging the two spheres maps an $A$-capping solution with anisotropy $\varphi$ to a $B$-capping solution with anisotropy $1/\varphi$. The latter therefore provide no independent solutions.

%--------------------------------------
\subsubsection{Nariai branch} 
%--------------------------------------

\begin{figure}[t]
\centering
	\begin{minipage}{0.45\textwidth}
        \centering
        \includegraphics[width=3in]{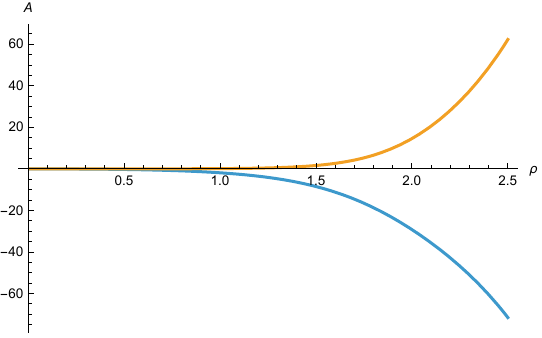} 
	\end{minipage}\hfill
 	\begin{minipage}{0.45\textwidth}
        \includegraphics[width=3in]{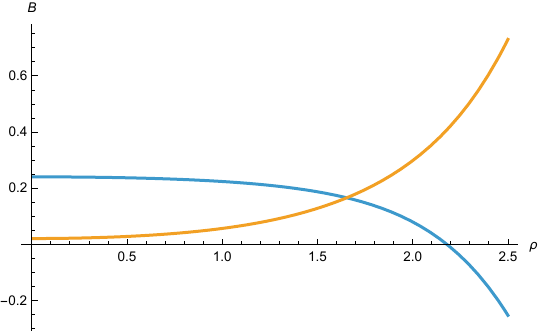} 
	\end{minipage}
	\caption{\label{F:nearNariai} The real (blue) and imaginary (orange) parts of the warpfactors $A$ (left) and $B$ (right) for a solution on the Nariai branch with $B_0 = 0.2406072 + 0.0210020 i$. This is a ``domain wall'' geometry interpolating between a perturbation of $(-\mathbb{H}^3)\times\mathbb{S}^2$ at small $\rho$, and an (analytic continuation of an) asymptotically dS$_5$ space  at large $\rho$.}
        \centering
\end{figure}

The Nariai point $B_0 = \frac{1}{4}$ is connected to a family of solutions, the blue curve in Fig.~\ref{F:S2S2phases}. Apart from the Nariai point and where the curve intersects the AdS branch at $B_0 \approx - 0.075$, the solutions for $A$ and $B$ are complex. The solution with $\text{Im}(B_0)<0$ is the complex conjugate of the solution with $\text{Im}(B_0)>0$. The anisotropy $\varphi$ tends to 0 as $B_0\to \frac{1}{4}$ along the branch. We plot the warpfactors $A$ and $B$ for a representative solution near the Nariai point with $\varphi = 0.1$ and $\Theta = 1.12754$ in Fig.~\ref{F:nearNariai}. We must shoot to identify solutions on the Nariai branch, say fixing an angle in the complex-$B_0$ plane and shooting in $|B_0|$ until locating a solution with $B/A>0$ as $\rho\to\infty$.

%--------------------------------------
\subsubsection{AdS branch} 
%--------------------------------------

When $B_0<0$, the perturbative solution for $A$ and $B$ near $\rho=0$ is negative. Numerically, this remains true for all $\rho$, in which case solutions with negative $B_0$ obey the late-time boundary condition with $\Theta = \frac{\pi}{2}$. As a result these metrics describe Euclidean AdS$_5$ geometries in all minus signature, ones with $\mathbb{S}^2\times\mathbb{S}^2$ conformal boundary. In the context of AdS/CFT these have recently been studied on their own terms in~\cite{Aharony:2019vgs,Kiritsis:2020bds}. Since $B_0<0$, they fail the KSW cap test~\eqref{E:KSWwedge}.

\begin{figure}[t]
\centering
	\begin{minipage}{0.45\textwidth}
        \centering
        \includegraphics[width=3in]{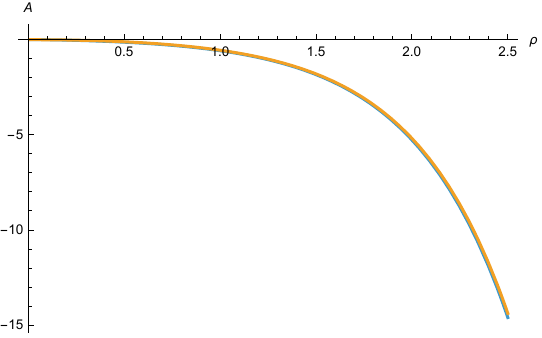} 
	\end{minipage}\hfill
 	\begin{minipage}{0.45\textwidth}
        \includegraphics[width=3in]{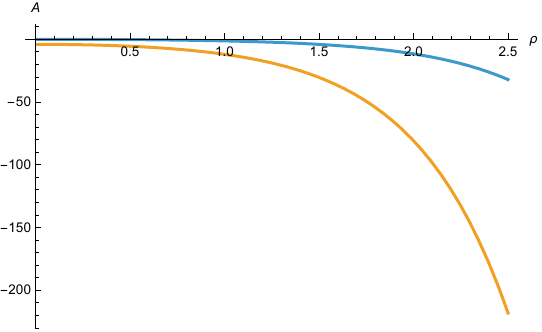} 
	\end{minipage}
	\caption{\label{F:AdS} Plots of $A$ (blue) and $B$ (orange) for points on the AdS branch with $B_0 = -10^{-2}$ (left) and $B_0=-4$ (right). The first has $\varphi = 0.9929$ and the second $\varphi = 2.590$.}
        \centering
\end{figure}

Along the AdS branch $\varphi$ oscillates around $1$ as $B_0\to 0$ and tends to $\infty$ as $B_0\to -\infty$. We plot two representatives in Fig.~\ref{F:AdS}.

%--------------------------------------
\subsubsection{The Russian doll}
%--------------------------------------

Finally we find a successive tower of nested branches looping around $B_0=0$. Indeed, the Nariai branch is in a sense the outermost layer of the doll, although we refer to the largest one inside it, the orange branch in Fig.~\ref{F:S2S2phases}, as the first branch. The branches of the doll appear at successively smaller values of $B_0$, accumulating as $B_0\to 0$. 

While we can locate these geometries by shooting at very small $|B_0|$, it is more instructive to derive them semianalytically by working perturbatively in $|B_0|$. 

This semianalytic solution has two regimes.
\begin{enumerate}
	\item A ``core'' regime near the cap with $\rho/\sqrt{|B_0|}$ fixed.
	\item An asymptotic regime for $\rho \gg \sqrt{|B_0|}$. 
\end{enumerate}
These regimes overlap, and by matching we build a complete solution from cap to asymptotia. We describe the details in Appendix~\ref{A:russianDoll}. 

Using that semianalytic solution we can predict the location of the Russian doll branches in the $B_0$ plane, as well as the generational separation between the branches. The $k^{\rm th}$ branch traces a path $r_k(\theta)$ in the $B_0$ plane, and in the $B_0\to 0$ limit, $\frac{r_k(\theta)}{r_{k+1}(\theta)} = e^{4\pi/\sqrt{15}}\approx 25.652$, so that the generational separation is $\sim 25.7$. Our solution predicts the location of the branches to a relative error that scales as $|B_0|$.

%--------------------------------------
\subsection{The wavefunction}
%--------------------------------------

Having located the bulk saddles that fill in the late-time data, and without knowing a priori which saddles contribute to the GPI and which do not, we write the tree-level wavefunction as
\beq
	\Psi_{\rm tree}(\varphi) \sim \sum_a n_ae^{iS_a}\,,
\eeq
where $a$ indexes the possible saddles at fixed $\varphi$ and $S$ is the renormalized action and $n_a$ indexes the contribution of the saddle to the wavefunction. The semiclassical approximation comes from the saddle with the largest $-\text{Im}(S)$ that contributes, giving
\beq
	|\Psi_{\rm tree}(\varphi)| \sim e^{-\text{Im}(S_{\rm dom})}\,.
\eeq

In $d=5$, the renormalized action of these saddles discussed in Subsection~\ref{S:SpSqsuperspace} is
\begin{align}
\begin{split}
\label{E:S2S2action}
	S &=\frac{2\pi}{G} \lim_{\rho_c\to\infty}\left(\bar{S}_{\rm bulk} + \bar{S}_{\rm bdy} + \bar{S}_{\rm finite}\right)\,,
	\\
	\bar{S}_{\rm bulk}&=4 \int_0^{\rho_c}d\rho\,AB  \,,
	\\
	\bar{S}_{\rm bdy} &= \left. \left( - B \partial_{\rho}A - A \partial_{\rho} B+ 3AB- \frac{1}{2}(A+B)  +\frac{\rho-\rho_0}{12}\left(\varphi^2 + \varphi^{-2} -4\right)  \right)\right|_{\rho = \rho_c}\,,
	\\
	\bar{S}_{\rm finite} &=  i\frac{\Theta}{12} \left( \varphi^2 + \varphi^{-2}-4\right)\,.
\end{split}
\end{align}
The imaginary term $\bar{S}_{\rm finite}$ comes from a bulk integral on the downward segment from the line of fixed $\rho$, to the line of fixed $t$ at asymptotia. It is fixed by the ``Weyl anomaly'' of the wavefunction in a way that we explain in more detail in Subsection~\ref{S:odd}. 

We find $S$ numerically in the following way. Given a numerical solution for $A$ and $B$ along the $\rho>0$ axis; determine $\Theta$ and $\varphi$ from a large-$\rho$ fit; numerically integrate up to a large cutoff, place the boundary terms there; and extrapolate final answers by adjusting the cutoff. For the near-Nariai limit we adopt a more refined procedure we describe below.

\begin{figure}[t]
\begin{center}
	\includegraphics[width=6in]{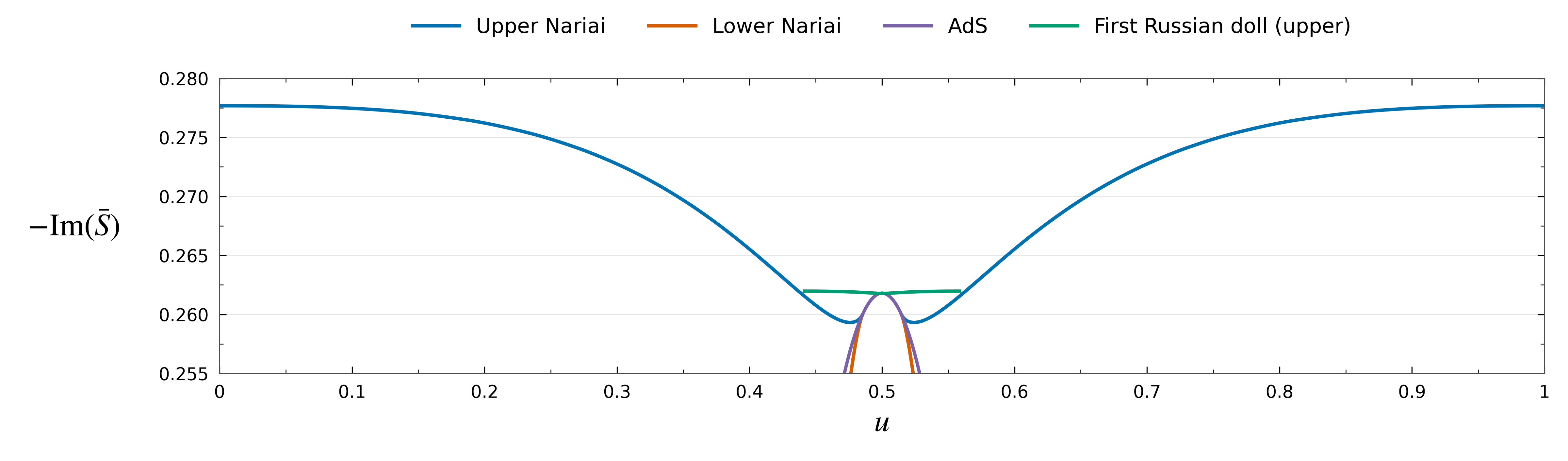}
	\caption{\label{F:S2S2wavefunction} Here we plot $-\text{Im}(\bar{S}) = -\frac{G}{2\pi}\text{Im}(S)$ as a function of $u = \frac{\varphi}{1+\varphi}$ for the various saddles that can contribute to the $\mathbb{S}^2\times\mathbb{S}^2$ wavefunction. Note that any saddle with some $(\varphi,\text{Im}(S))$ where the $A$-sphere caps off implies a corresponding saddle with $(1/\varphi,\text{Im}(S))$ where the $B$-sphere caps off, which explains the symmetry of the plot under $\varphi \leftrightarrow 1/\varphi$, i.e. $u\to 1-u$. Note also that the weight is globally maximized near the Nariai endpoints at $\varphi = 0,\infty$.}
\end{center}
\end{figure}

Because there are many saddles, this results in a somewhat complicated picture for the wavefunction. We present our results for $-\text{Im}(S)$ in Fig.~\ref{F:S2S2wavefunction}, and describe some more details in the remainder of this Subsection.

%--------------------------------------
\subsubsection{Near-Nariai limit}
%--------------------------------------

Near the Nariai point, i.e. for small $\varphi$, we must be a little more clever in order to extract the action to high accuracy. There are two factors to contend with. The first is that the bulk geometry is a long ``domain wall'' between an approximate (continuation of an) dS$_3\times\mathbb{S}^2$ region at $\rho$ of $O(1)$, and a (continuation of an) dS$_5$ region at large $\rho$. The second is that, at least along the upper half of the Nariai branch, there is a large cancellation in Im$(S)$ between the finite anomaly counterterm $\bar{S}_{\rm finite}$ in~\eqref{E:S2S2action}, and the rest of the terms in the renormalized action so that the result is $O(1)$ as $\varphi\to 0$.

In the presentation~\eqref{E:S2S2action}, there are cancellations between the integral and boundary terms involving terms growing as $e^{4\rho_c}$, which in a numerical implementation becomes delicate at large $\rho_c$. For that reason we reorganize the numerical evolution of both the geometry and the action. Define
\begin{align}
\begin{split}
	\mathcal F(\rho)&=4\int_0^\rho d\rho'\,A(\rho')B(\rho')
	+\left.\left(-(AB)'+3AB-\frac{A+B}{2}\right)\right|_{\rho}\,,
	\\
	H&=\frac14\left(\frac{A'}A+\frac{B'}B\right)\,,
	\qquad
	\sigma=\frac14\left(\frac{A'}A-\frac{B'}B\right)\,,
	\\
	D&=AB\sigma=\frac14\left(BA'-AB'\right)\,.
\end{split}
\end{align}
The Einstein equations and Hamiltonian constraint give a closed system for $A$, $B$, $D$, and $\mathcal F$,
\begin{align}
\begin{split}
	A'&=2A(H+\sigma)\,,
	\qquad
	B'=2B(H-\sigma)\,,
	\qquad
	D'=\frac{A-B}{2}\,,
	\\
	\mathcal F'&=\frac{4AB\sigma^2}{H+1}
	-\frac{(A+B)H(H-1)}{H+1}-(A-B)\sigma\,,
\end{split}
\end{align}
where $\sigma$ and $H$ are determined algebraically at each step by
\beq
	\sigma=\frac{D}{AB}\,,
	\qquad
	H=\sqrt{1+\frac{\sigma^2}{3}
	-\frac16\left(\frac1A+\frac1B\right)}\,.
\eeq
We choose the square-root sheet connected to $H\sim1/(2\rho)$ near the cap and $H\to 1$ at large $\rho$. In evaluating $\mathcal F'$, we use
\beq
	H-1=\frac{2\sigma^2-\frac1A-\frac1B}{6(H+1)}\,.
\eeq
We initialize all four evolution variables at a small positive $\rho_i$ using the near-cap series. In particular, $\mathcal F(0)=-B_0/2$. Evolving $D$ avoids obtaining the small late-time shear $\sigma$ by subtracting nearly equal logarithmic derivatives of $A$ and $B$.

We impose the late-time boundary condition by shooting in $B_0$ for a positive real limit of
\beq
	\frac{B'}{A'}=\frac{B(H-\sigma)}{A(H+\sigma)}
	\longrightarrow\varphi^2\,.
\eeq
This ratio eliminates the constant terms in the large-$\rho$ expansions of $A$ and $B$, and therefore converges faster than $B/A$. We extract $\Theta$ from the common asymptotic phase as before.

At fixed $\varphi>0$ and large $\rho$, $A,B$ are $O(e^{2\rho})$ while $\sigma$ and $H-1$ are $O(e^{-2\rho})$, so each term in the evolution equation for $\mathcal F$ has a finite late-time limit. Their sum obeys $\mathcal F'\to-Q/12$, where $Q=\varphi^2+\varphi^{-2}-4$. Writing
\beq
	\mathcal C=\lim_{\rho\to\infty}
	\left(\mathcal F(\rho)+\frac{\rho Q}{12}\right)\,,
\eeq
the imaginary part of~\eqref{E:S2S2action} is
\beq
	-\text{Im}(S)=-\frac{2\pi}{G}
	\left(\text{Im}(\mathcal C)+\frac{\Theta Q}{12}\right)\,.
\eeq
We check convergence by increasing the cutoff and numerical precision, and by independently integrating the original second-order equations together with the bulk integral. The near-Nariai limit is taken after the renormalized large-$\rho$ limit.

This formulation removes the exponentially large cutoff cancellations, although the cancellation with the finite anomaly term remains substantial at small $\varphi$. For example, at $\varphi=10^{-2}$ the two contributions to $-\frac{G}{2\pi}\text{Im}(S)$ on the upper branch are approximately $926.322354438669$ and $-926.044674844068$, while their sum is $0.277679594601$, very close to the value at the Nariai point
\beq
	-\frac{G}{2\pi}\text{Im}(S_{\rm Nariai})
	=\frac{\pi}{8\sqrt{2}}\approx0.2776801836\,.
\eeq

\begin{figure}[t]
\begin{center}
	\includegraphics[width=3in]{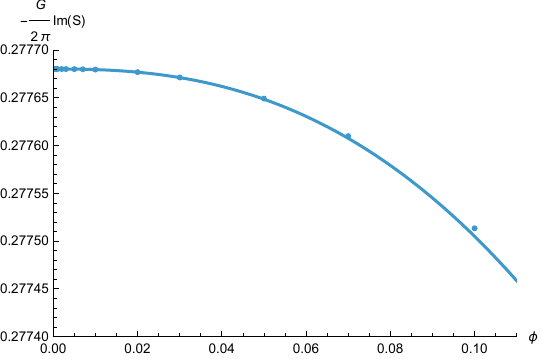}
	\caption{\label{F:nearNariaiS} The quantity
	$-\mathrm{Im}(GS/(2\pi))$ on the upper half of the Nariai branch
	at several small values of $\varphi$ (dots), as well as its
	fit~\eqref{E:nearNariaiFit} (solid blue).}
\end{center}
\end{figure}

On the upper Nariai branch, our numerical results give for small $\varphi$
\begin{align}
\begin{split}
\label{E:nearNariaiFit}
	-\mathrm{Im}\left(\frac{GS}{2\pi}\right)
	&=\frac{\pi}{8\sqrt2}-0.05197256\,\varphi^{2\kappa}+\hdots\,,
	\\
	\Theta&=\frac{\pi}{2\sqrt2}
	+0.28998119\,\varphi^\kappa
	-0.02239\,\varphi^{2\kappa}+\hdots\,,
\end{split}
\end{align}
where
\beq
	\kappa=\sqrt5-1\,.
\eeq
The fractional power follows from perturbation theory about the dS$_3\times\mathbb S^2$ Nariai endpoint. Writing $B=\frac14+(B_0-\frac14)f(\rho)+\cdots$, the regular linearized mode behaves at large (but not too large so that the perturbation backreacts) $\rho$ as
\beq
	f(\rho)=\frac{\sinh(\sqrt{10}\rho)}
	{\sqrt5\,\sinh(\sqrt2\rho)}
	\sim\frac1{\sqrt5}e^{\sqrt2(\sqrt5-1)\rho}\,.
\eeq
Matching the point where this mode becomes order one to the asymptotically dS$_5$ region gives $|B_0-\frac14|\simeq0.402746576\,\varphi^\kappa$. The leading change in the imaginary action is quadratic in the cap deformation $|B_0-1/4|$, explaining the power $2\kappa$ in the first fit. Its negative coefficient in $-\mathrm{Im}(S)$ shows that the Nariai endpoint is a local maximum of the tree-level wavefunction. We plot this behaviour in Fig.~\ref{F:nearNariaiS}.

The Nariai branch has upper and lower halves related by complex conjugation of the metric functions. At fixed boundary anisotropy $\varphi$, denote their actions by $S_+$ and $S_-$. Their bulk and boundary contributions are complex conjugates apart from the finite term $\bar{S}_{\rm finite}$ in~\eqref{E:S2S2action}, since if the point on the upper branch has some $\Theta$, then the conjugate point has $\pi-\Theta$. As a result
\beq
\label{E:S2S2sumrule}
	\text{Im}(S_++S_-)=\frac{2\pi}{G}\frac{\pi Q}{12}\,.
\eeq
Since $Q\sim\varphi^{-2}$ and $\text{Im}(S_+)$ is finite as $\varphi\to0$, this implies that
\beq
	-\text{Im}(S_-)\sim-\frac{\pi^2}{6G}Q
	\sim-\frac{\pi^2}{6G}\frac{1}{\varphi^2}\,,
\eeq
when approaching the Nariai point along the lower branch, and its contribution to the wavefunction is exponentially suppressed relative to the upper branch. From Fig.~\ref{F:S2S2wavefunction} we see that this suppression continues along the rest of the lower branch.

%--------------------------------------
\subsubsection{Other branches}
%--------------------------------------

The weight of the AdS branch is rather simple. Because the AdS branch has $A,B$ everywhere negative, we immediately infer $\Theta = \pi/2$, and the only contribution to $\text{Im}(S)$ comes from what we call $\bar{S}_{\rm finite}$ in~\eqref{E:S2S2action}. So, along the AdS branch, we simply have
\beq
	-\text{Im}(S) = -\frac{\pi^2}{12G} (\varphi^2 + \varphi^{-2} - 4)\,,
\eeq
for those values of $\varphi$ that arise on the branch. As we comment in the preamble to the next Section, this result is ultimately due to the Weyl anomaly of the tree-level wavefunction. 

There is a direct conformal-geometric derivation of this result, using what we reviewed in Subsection~\ref{S:Qcurvature}. For the boundary metric $\gamma = d\Omega_2^2 + \varphi^2 d\Omega_2^2$, the two sectional curvatures are $1$ and $\varphi^{-2}$, giving
\beq
\label{E:S2S2Weyl}
 W_{ijkl}W^{ijkl} = \frac{4}{3}(1+\varphi^{-2})^2\,, \qquad \text{vol}(\Sigma,\gamma) = 16\pi^2\varphi^2\,.
\eeq
Using $\chi(\mathbb S^2\times\mathbb S^2)=4$ in~\eqref{E:chiQ4}, we obtain
\beq
\label{E:S2S2chiQ}
 \chi_Q = \frac{2}{3}(4-\varphi^2-\varphi^{-2})\,.
\eeq
Since the $d=5$ static patch entropy is $S_0 = \pi^2/(2G)$,~\eqref{E:Qweight} immediately gives
\beq
\label{E:S2S2Qaction}
 -\text{Im}(S_{\rm AdS}) = \frac{S_0}{4}\chi_Q = -\frac{\pi^2}{12G}(\varphi^2+\varphi^{-2}-4)\,.
\eeq
Thus the complete imaginary action along the real AdS branch, and not only its large-anisotropy limit, follows from a local conformal invariant of the boundary. The fact that the complex Nariai and Russian doll branches can have different magnitudes at the same $\varphi$ is consistent with the extra finite contribution for genuinely complex fillings described in Section~\ref{S:Qcurvature}.

In particular, the limit $-B_0 \gg 1$ generates solutions with $\varphi\gg 1$, which generates an exponentially suppressed contribution $|\Psi| \sim e^{-\frac{\pi^2}{12G}\varphi^2}$. There is a ``mirror'' AdS branch in which the $B$-sphere caps off which has the same action with $\varphi \to 1/\varphi$, so that $-B_0\gg 1$ also corresponds to geometries with $\varphi \to 0$. So in this regime the AdS and Nariai branches both give contributions to the wavefunction, but since the Nariai branch has $-\text{Im}(S)$ tending to a constant and the AdS branch has $-\text{Im}(S) \propto -\frac{1}{G\varphi^2}$, the Nariai branch dominates. 

The Russian doll branches have $\varphi \approx 1$, and given that they intersect the AdS branch, it is not surprising that they have an action whose imaginary part is approximately that of the AdS branch in the $\varphi\to 1$ limit, $-\text{Im}(S) \approx \frac{\pi^2}{6G}$. From the plot in Fig.~\ref{F:S2S2wavefunction} we see that the upper half dominates the AdS branch. It is guaranteed by the sum rule~\eqref{E:S2S2sumrule} that either the upper or lower dominates, since by it the average of $-\text{Im}(S)$ of the upper and lower Russian doll branch is that of the AdS branch.

%--------------------------------------
\section{$\mathbb{S}^p\times \mathbb{S}^q$ wavefunction}
\label{S:anisotropy}
%--------------------------------------

In the last Section we investigated the $\mathbb{S}^2\times\mathbb{S}^2$ wavefunction in detail. Among the candidate saddles we found an AdS branch, along which the metric is real and negative-definite. Its contribution to $-\text{Im}(S)$ is fixed by the constant term in the asymptotic bulk Lagrangian. Integrating that term along the downward leg of the complex time contour gives the finite piece $\bar S_{\rm finite}$ in~\eqref{E:S2S2action}. At large $\varphi$ the saddle weight is therefore
\beq
\label{E:SpSqfiveweight}
 |e^{iS_{\rm AdS}}|\sim\exp\left(-\frac{\pi^2}{12G}\varphi^2\right)\,.
\eeq
Whether this saddle appears in $\Psi$ is a separate question.

Here we find the corresponding large-anisotropy weights for $p,q\geq2$ and $d=p+q+1\leq11$. The large negative-$B_0$ solutions have a simple scaling limit, analogous to the large black holes of Section~\ref{S:S1}. In odd $d$ an anomaly calculation determines their weights analytically. In even $d$ we evaluate a finite renormalized action numerically. The weights grow exponentially when the noncontracting sphere has $q=0,3$ mod $4$. In view of Section~\ref{S:S1contour}, this growth makes the question of which saddles contribute essential.

%--------------------------------------
\subsection{Solutions at large anisotropy}
\label{S:SpSqscaling}
%--------------------------------------

Suppose the $\mathbb{S}^p$ caps off and take $B_0=-\delta^{-2}$ with $0<\delta\ll1$. Along the real-$\rho$ portion of the AdS branch we parameterize the line element as
\beq
\label{E:SpSqcoremetric}
 ds^2=-\left(d\rho^2+a(\rho)d\Omega_p^2+\delta^{-2}b(\rho)d\Omega_q^2\right)\,,
 \qquad A=-a\,,\quad B=-\delta^{-2}b\,.
\eeq
The functions $a,b$ are non-negative along the AdS branch. The curvature of the $\mathbb{S}^q$ enters Einstein's equations through $(q-1)\delta^2/b$, which vanishes as $\delta\to0$. Locally the large sphere becomes flat. Its volume supplies an overall factor $\delta^{-q}$, while its radial expansion remains coupled to that of the contracting sphere.

Set $n=p+q=d-1$. Dropping the $\delta^2$ term in Einstein's equations~\eqref{E:SpSqEinstein} gives
\begin{align}
\begin{split}
\label{E:SpSqcoreeom}
 0&=\frac{a''}{2a}+\frac{p-2}{4}\frac{a'^2}{a^2}
       +\frac{q}{4}\frac{a'b'}{ab}-\frac{p-1}{a}-n\,,\\
 0&=\frac{b''}{2b}+\frac{q-2}{4}\frac{b'^2}{b^2}
       +\frac{p}{4}\frac{a'b'}{ab}-n\,,\\
 0&=\frac{p(p-1)}{4}\frac{a'^2}{a^2}
       +\frac{q(q-1)}{4}\frac{b'^2}{b^2}
       +\frac{pq}{2}\frac{a'b'}{ab}-\frac{p(p-1)}{a}-n(n-1)\,.
\end{split}
\end{align}
Primes denote $\rho$ derivatives. The choice $B_0=-\delta^{-2}$ fixes $b(0)=1$. Smoothness at the cap $\rho = 0$ then gives
\beq
\label{E:SpSqcorecap}
 a=\rho^2\left(1+\frac{n(p+1-q)}{3p(p+1)}\rho^2+O(\rho^4)\right)\,,
 \qquad b=1+\frac{n}{p+1}\rho^2+O(\rho^4)\,.
\eeq
There is thus one limiting solution for fixed $p$, $q$ with the $\mathbb{S}^p$ capping off. At large $\rho$ it has
\beq
\label{E:SpSqcoreboundary}
 a=\alpha e^{2\rho}(1+O(e^{-2\rho}))\,,\qquad
 b=\beta e^{2\rho}(1+O(e^{-2\rho}))\,,\qquad
 \widetilde\varphi^2=\frac{\beta}{\alpha}=\lim_{\rho\to\infty}\frac{b}{a}\,.
\eeq
Matching to~\eqref{E:largeRhoBC} gives $\Theta=\pi/2$ and $\rho_0=-\frac12\ln\alpha$, and so the boundary anisotropy
\beq
\label{E:SpSqcoretranslation}
 \varphi=\frac{\widetilde\varphi}{\delta}\,(1+O(\delta^2))\,.
\eeq
The distinction between $\varphi$ and $\widetilde\varphi$ matters when translating the action of the limiting solution back to the original boundary data.

Let us discuss the evaluation of the renormalized action in the limit of large anisotropy and in a larger dimension than we discussed in Subsection~\ref{S:SpSqsuperspace}. Define the radial volume density and extrinsic curvature of the positive-definite metric inside the parentheses of~\eqref{E:SpSqcoremetric},
\beq
\label{E:SpSqVK}
 V=a^{p/2}b^{q/2}\,,\qquad K=\frac{p a'}{2a}+\frac{q b'}{2b}\,.
\eeq
Along real $\rho$, direct evaluation of the action gives
\beq
\label{E:SpSqcoreaction}
 S_{\rm real\,\rho}
 =i^n\frac{\Vol(\mathbb{S}^p\times\mathbb{S}^q)}{8\pi G}\delta^{-q}
 \left(n\int_0^{\rho_c}d\rho\,V+V(-K+u(a))\big|_{\rho_c}\right)\,.
\eeq
The phase $i^n$ comes from the all-minus volume element, with its branch fixed by the late-time Lorentzian metric, while $u$ parameterizes the local counterterms. We have,
\begin{align}
\begin{split}
\label{E:SpSquct}
 u(a)&=\sum_k\frac{u_k}{a^k}\,,\quad u_0=n-1\,,
 \quad
 u_1=\frac{p(p-1)}{2(n-2)}\,,
 \quad
 u_2=\frac{p(p-1)^2(4(n-1)-np)}{8(n-4)(n-1)(n-2)^2}\,.
\end{split}
\end{align}
where $u_0$, $u_1$, and $u_2$ come from the counterterms~\eqref{E:RG1} dropping the terms with inverse powers of $B$. Only terms with $2k<n$ are power counterterms. When $n$ is even the next term is logarithmic, and we must also include the downward leg of the time contour. These two features account for the different calculations in odd and even $d$.

By phrasing ``radial'' evolution in $\rho$ as a Hamilton-Jacobi problem, we can easily construct higher counterterms. With $x=1/a$ and parameterizing $u=u(x)$, the local radial Hamilton-Jacobi equation is
\begin{align}
\begin{split}
\label{E:SpSqHJ}
 k_p&=\frac{p u+2(q-1)x\partial_xu}{p(n-1)}\,,\qquad
 k_q=\frac{u-2x\partial_xu}{n-1}\,,\\
 0 & = p(p-1)k_p^2+q(q-1)k_q^2+2pqk_pk_q- n(n-1) - p(p-1)x\,.
\end{split}
\end{align}
Here $k_p,k_q$ are the extrinsic-curvature eigenvalues inferred from the local counterterm functional. Expanding this equation fixes $u_k$ successively. It reproduces~\eqref{E:SpSquct} and supplies all counterterms needed below.

Our goal is the dimensionless coefficient $C_{p,q}$ in
\beq
\label{E:SpSqCdefinition}
 -\text{Im}(S_{\rm AdS})
 =\frac{\Vol(\mathbb{S}^p\times\mathbb{S}^q)}{8\pi G}
 C_{p,q}\varphi^q\,(1+o(1))\,,\qquad\varphi\to\infty\,,
\eeq
with $C_{p,q}>0$ corresponding to a runaway direction and $C_{p,q}<0$ to a suppressed one.

%--------------------------------------
\subsection{Results in odd $d$}
\label{S:odd}
%--------------------------------------

For odd $d$, $n$ is even and $i^n=(-1)^{n/2}$ is real. Thus the real-$\rho$ integral and its boundary terms have real action. The imaginary part comes from the downward leg $\Delta\rho=-i\pi/2$. All power divergences on that leg cancel against boundary terms, leaving the constant $v_{\star}$ in the asymptotic form of $V$,
\beq
\label{E:SpSqvolconstant}
 V=\sum_{k<n/2}v_k e^{(n-2k)\rho}+v_\star+o(1)\,.
\eeq
Its integral along real $\rho$ is secular, $n v_\star\rho_c$, and is removed by the logarithmic counterterm. Its integral along the downward leg remains finite and imaginary. Consequently
\beq
\label{E:SpSqoddC}
 C_{p,q}=(-1)^{n/2}\frac{\pi n}{2}
 \frac{v_\star}{\widetilde\varphi^q}\,.
\eeq
The ratio $v_\star/\widetilde\varphi^q$ is fixed by the boundary curvature. It is independent of the normalizable data and hence of the details of the cap.

For example, in $d=5$ with $p=q=2$, use $u=\rho-\rho_0$ and the positive squared radii $\mathcal A=-A$, $\mathcal B=-B$ of the full solution. Their first asymptotic terms are
\beq
\label{E:SpSqfiveasymptotic}
 \mathcal A=e^{2u}+\frac{\varphi^{-2}-2}{6}+O(ue^{-2u})\,,\qquad
 \mathcal B=\varphi^2e^{2u}+\frac{\varphi^2-2}{6}+O(ue^{-2u})\,.
\eeq
Solving through the next order gives
\beq
\label{E:SpSqfivevolume}
 \mathcal A\mathcal B
 =\varphi^2e^{4u}-\frac{\varphi^2+1}{6}e^{2u}
 -\frac{\varphi^2+\varphi^{-2}-4}{48}+O(ue^{-2u})\,.
\eeq
The constant term in the on-shell bulk Lagrangian is four times the last displayed term. Using $\Vol(\mathbb{S}^2\times\mathbb{S}^2)/(8\pi G)=2\pi/G$ and integrating by $-i\pi/2$ gives
\beq
\label{E:SpSqfiveanomaly}
 -\text{Im}(S_{\rm AdS})
 =-\frac{\pi^2}{12G}(\varphi^2+\varphi^{-2}-4)\,.
\eeq
This reproduces~\eqref{E:S2S2action} at $\Theta=\pi/2$, including its finite term, and gives $C_{2,2}=-\pi/24$.

The same calculation is the familiar holographic Weyl anomaly~\cite{Henningson:1998gx}. Under a constant boundary Weyl rescaling $\gamma\to e^{2\sigma}\gamma$, the logarithmic term produces
\beq
\label{E:SpSqWeylshift}
 S_{\rm ren}[e^{2\sigma}\gamma]-S_{\rm ren}[\gamma]
 =i^n\frac{\Vol(\mathbb{S}^p\times\mathbb{S}^q)}{8\pi G}
 \delta^{-q} n v_\star\sigma\,.
\eeq
The shift $\sigma=-i\pi/2$ turns the negative-definite asymptotic metric into the positive-definite late-time metric. Its anomalous finite action is precisely that of the downward leg. This explains why the boundary expansion alone determines $-\text{Im}(S)$ in odd $d$.

In $d=7$ the calculation requires one more asymptotic order. For $p=2,q=4$, the constant in the full volume density $\mathcal A\mathcal B^2$ is
\beq
\label{E:SpSqsevenconstant}
 -\frac{\varphi^4-7\varphi^2+33-9\varphi^{-2}}{800}\,.
\eeq
The factor $i^6=-1$ reverses its contribution relative to $d=5$, and we obtain
\beq
\label{E:SpSqsevenanomaly}
 -\text{Im}(S_{\rm AdS})
 =\frac{\pi^3}{200G}(\varphi^4-7\varphi^2+33-9\varphi^{-2})\,.
\eeq
Thus $C_{2,4}=3\pi/800>0$. At large $\varphi$ the candidate contribution to the wavefunction grows as $\exp(\pi^3\varphi^4/(200G))$, even though its bulk curvature remains of order the de Sitter scale.

The same anomalous contribution is expressed geometrically by the integrated $\mathsf{Q}_{d-1}$-curvature in~\eqref{E:Qweight}. For even $d-1$, this integral is fixed locally by the boundary conformal class. The $\mathbb{S}^2\times\mathbb{S}^2$ calculation in~\eqref{E:S2S2chiQ} is the $d-1=4$ example. Thus, the anomaly and $\mathsf{Q}$-curvature calculations determine the same saddle weight; the Hamilton--Jacobi equation provides a direct way to evaluate it for the higher-dimensional products considered here.

For higher odd $d$ it is economical to extract the anomaly from~\eqref{E:SpSqHJ}. Set $m=n/2$ and substitute $u_0,\ldots,u_{m-1}$ into that equation. If $r_m$ is the coefficient of $x^m$ in the last line of~\eqref{E:SpSqHJ}, the vanishing coefficient of $u_m$ is the logarithmic obstruction, with
\beq
\label{E:SpSqobstruction}
 \frac{n v_\star}{\widetilde\varphi^q}=\frac{r_m}{2}\,,\qquad
 C_{p,q}=(-1)^{n/2}\frac{\pi}{4}r_m\,.
\eeq
This supplies the exact coefficients through $d=11$, listed in Table~\ref{T:SpSqresults} in the Summary. 

%--------------------------------------
\subsection{Results in even $d$}
\label{S:SpSqeven}
%--------------------------------------

In even $d$ there is no constant term in the asymptotic volume density and no logarithmic counterterm. The downward leg adds no finite imaginary action. Instead $i^n$ is imaginary, and the finite part of the real-$\rho$ action determines the weight. Denote the renormalized expression in brackets in~\eqref{E:SpSqcoreaction} by
\beq
\label{E:SpSqFren}
 F_{p,q}=\lim_{\rho_c\to\infty}
 \left(n\int_0^{\rho_c}d\rho\,V+V(-K+u(a))\big|_{\rho_c}\right)\,.
\eeq
Then the translation~\eqref{E:SpSqcoretranslation} gives
\beq
\label{E:SpSqevenC}
 C_{p,q}=(-1)^{d/2}\frac{F_{p,q}}{\widetilde\varphi^q}\,.
\eeq

For odd $d-1$, the scattering definition of $\mathsf{Q}_{d-1}$ instead depends on the bulk filling. Its integral is proportional to the renormalized volume, as described in Subsection~\ref{S:Qcurvature}. Consequently, the geometrical formula~\eqref{E:Qweight} packages the same finite interior information as $F_{p,q}$; it does not determine that information from local boundary curvature alone. This is why the even-$d$ weights require the bulk calculation.

The first cases are $(p,q)=(2,3)$ and $(3,2)$ in $d=6$. Starting with~\eqref{E:SpSqcorecap}, numerical integration gives
\beq
\label{E:SpSqsixtilde}
 \widetilde\varphi_{2,3}^2\simeq2.12630\,,\qquad
 \widetilde\varphi_{3,2}^2\simeq1.34402\,.
\eeq
To obtain the renormalized action we require the counterterms \beq
\label{E:SpSqsixtct}
 u_{2,3}=4+\frac{1}{3a}+\frac{1}{24a^2}\,,\qquad
 u_{3,2}=4+\frac{1}{a}+\frac{1}{24a^2}\,.
\eeq
We find $F_{2,3}\simeq-0.0664807$ and $F_{3,2}\simeq0.156866$. Since $(-1)^{d/2}=-1$, the corresponding coefficients have opposite signs,
\beq
\label{E:SpSqsixC}
 C_{2,3}\simeq0.0214417\,,\qquad C_{3,2}\simeq-0.116714\,.
\eeq
For the original $\mathbb{S}^2\times\mathbb{S}^3$ boundary this gives the two limits
\beq
\label{E:SpSqsixweights}
 -\text{Im}(S_{\rm AdS})\sim\frac{\pi^2}{G}
 \begin{cases}
  0.0214417\,\varphi^3\,, & \varphi\to\infty\,,\quad\mathbb{S}^2\text{ caps off}\,,\\
  -0.116714\,\varphi^{-2}\,, & \varphi\to0\,,\quad\mathbb{S}^3\text{ caps off}\,.
 \end{cases}
\eeq
Thus the first general product-sphere runaway already occurs in $d=6$.

The subtraction in~\eqref{E:SpSqFren} requires some care. Its separate terms grow as $e^{n\rho_c}$ while their sum tends to a constant. We subtract the asymptotic divergent terms before taking the finite part, and check stability under changes of the cutoff and of the near-cap starting point. The Hamiltonian constraint in~\eqref{E:SpSqcoreeom} provides a separate check on the geometry. The result can also be checked by solving at small finite $\delta$ and comparing $-\text{Im}(S)/\varphi^q$ with~\eqref{E:SpSqCdefinition}.

The same procedure in $d=8,10$ requires counterterms through $a^{-3},a^{-4}$ respectively, obtained from~\eqref{E:SpSqHJ}. We collect the numerical coefficients with the anomaly results in Table~\ref{T:SpSqresults}.

\newpage
$$$$
\begin{table}[H]
\centering
\begingroup
\small
\definecolor{SpSqrunaway}{RGB}{182,50,29}
\definecolor{SpSqsuppressed}{RGB}{31,93,122}
\renewcommand{\arraystretch}{1.18}
\setlength{\tabcolsep}{12pt}
\begin{tabular}{c|c|c|r|c}
 $d$ & $p$ & $q$ & $C_{p,q}$ & R/S \\ \hline
 $5$ & $2$ & $2$ & $-\pi/24$ & \textcolor{SpSqsuppressed}{S} \\
\hline
 $6$ & $2$ & $3$ & $0.0214417$ & \textcolor{SpSqrunaway}{R} \\
  & $3$ & $2$ & $-0.116714$ & \textcolor{SpSqsuppressed}{S} \\
\hline
 $7$ & $2$ & $4$ & $3\pi/800$ & \textcolor{SpSqrunaway}{R} \\
  & $3$ & $3$ & $3\pi/400$ & \textcolor{SpSqrunaway}{R} \\
  & $4$ & $2$ & $-27\pi/800$ & \textcolor{SpSqsuppressed}{S} \\
\hline
 $8$ & $2$ & $5$ & $-0.00239465$ & \textcolor{SpSqsuppressed}{S} \\
  & $3$ & $4$ & $0.0112786$ & \textcolor{SpSqrunaway}{R} \\
  & $4$ & $3$ & $0.0222466$ & \textcolor{SpSqrunaway}{R} \\
  & $5$ & $2$ & $-0.0977405$ & \textcolor{SpSqsuppressed}{S} \\
\hline
 $9$ & $2$ & $6$ & $-95\pi/395136$ & \textcolor{SpSqsuppressed}{S} \\
  & $3$ & $5$ & $-10\pi/9261$ & \textcolor{SpSqsuppressed}{S} \\
  & $4$ & $4$ & $69\pi/21952$ & \textcolor{SpSqrunaway}{R} \\
  & $5$ & $3$ & $20\pi/3087$ & \textcolor{SpSqrunaway}{R} \\
  & $6$ & $2$ & $-34375\pi/1185408$ & \textcolor{SpSqsuppressed}{S} \\
\hline
 $10$ & $2$ & $7$ & $0.000161248$ & \textcolor{SpSqrunaway}{R} \\
  & $3$ & $6$ & $-0.000926929$ & \textcolor{SpSqsuppressed}{S} \\
  & $4$ & $5$ & $-0.00342384$ & \textcolor{SpSqsuppressed}{S} \\
  & $5$ & $4$ & $0.00853156$ & \textcolor{SpSqrunaway}{R} \\
  & $6$ & $3$ & $0.0185254$ & \textcolor{SpSqrunaway}{R} \\
  & $7$ & $2$ & $-0.0856389$ & \textcolor{SpSqsuppressed}{S} \\
\hline
 $11$ & $2$ & $8$ & $35\pi/2985984$ & \textcolor{SpSqrunaway}{R} \\
  & $3$ & $7$ & $35\pi/331776$ & \textcolor{SpSqrunaway}{R} \\
  & $4$ & $6$ & $-245\pi/884736$ & \textcolor{SpSqsuppressed}{S} \\
  & $5$ & $5$ & $-35\pi/34992$ & \textcolor{SpSqsuppressed}{S} \\
  & $6$ & $4$ & $3125\pi/1327104$ & \textcolor{SpSqrunaway}{R} \\
  & $7$ & $3$ & $595\pi/110592$ & \textcolor{SpSqrunaway}{R} \\
  & $8$ & $2$ & $-924385\pi/35831808$ & \textcolor{SpSqsuppressed}{S} \\
\hline
\end{tabular}
\caption{Large-anisotropy coefficients defined in~\eqref{E:SpSqCdefinition}, with the $\mathbb{S}^p$ capping off and the $\mathbb{S}^q$ large. Odd-$d$ entries are exact and even-$d$ entries are numerical. \textcolor{SpSqrunaway}{R} denotes runaway and \textcolor{SpSqsuppressed}{S} exponential suppression of the candidate saddle weight.}
\label{T:SpSqresults}
\endgroup
\end{table}

%--------------------------------------
\newpage

%--------------------------------------
\subsection{Summary}
%--------------------------------------

Table~\ref{T:SpSqresults} summarizes the large-anisotropy coefficients for all ordered pairs $p,q\geq2$ with $d\leq11$. Their signs follow the empirical pattern
\beq
\label{E:SpSqsignpattern}
 \text{sgn}(C_{p,q})=(-1)^{q(q+1)/2}
 =\begin{cases}
 +1\,,&q=0,3\pmod4\,,\\
 -1\,,&q=1,2\pmod4\,.
 \end{cases}
\eeq
We do not have an analytic explanation for this pattern, and have only verified it up to $d=11$. Our results identify at least one runaway direction in every bulk dimension $6\leq d\leq11$. Exchanging the two sphere factors gives a complementary tail $C_{q,p}\varphi^{-p}$ as $\varphi\to0$.

These saddles remain weakly curved as $\delta\to0$, so their runaway is within the regime of validity of gravitational EFT. If a saddle with $C_{p,q}>0$ has nonzero coefficient in the GPI, its contribution grows as $\exp(2c\varphi^q/G)$ in $|\Psi|^2$, with $c>0$, and implies a divergent norm under the ordinary semiclassical measure. 

Recall that for the $\mathbb{S}^1\times \mathbb{S}^q$ wavefunction, there is an analogue of these runaways produced by large minus AdS black holes with $q\equiv0\pmod4$. In Subsection~\ref{S:S1contour} we presented a toy model for the GPI that excludes those contributions, resolving the corresponding tree-level obstruction to the norm within that model. We do not have an analogous off-shell family and integration cycle for $p,q>1$ to give a verdict on these runaways, and therefore do not know if they contribute to the wavefunction or not. KSW, if imposed, would exclude them: they all have $B_0<0$, and so fail the cap test~\eqref{E:KSWwedge}.

\section{Other geometries with enhanced saddle weights}
\label{S:otherGeometries}
%--------------------------------------

The all-minus product-sphere fillings are not the only geometries for which the candidate no-boundary weight can become large. The relation to $Q$-curvature lets us distinguish enhancement from a continuous boundary modulus, enhancement from the topology of the boundary, and enhancement from the topology of the filling at fixed boundary. We discuss examples of each. As before, a large tree-level weight does not determine the coefficient of a saddle in the GPI. 

%--------------------------------------
\subsection{Products of hyperbolic surfaces}
\label{S:hyperbolicProducts}
%--------------------------------------

Consider $d=5$ and a boundary $\Sigma_g\times\Sigma_h$, where $\Sigma_g$ and $\Sigma_h$ are compact hyperbolic surfaces of genera $g,h\geq2$. Normalize each surface to sectional curvature $-1$, and introduce the relative scale $\varphi$ through
\beq
\label{E:hyperbolicProductMetric}
 \gamma = d\Sigma_g^2 + \varphi^2d\Sigma_h^2\,.
\eeq
Gauss-Bonnet fixes their areas to $4\pi(g-1)$ and $4\pi(h-1)$. The boundary Euler characteristic and Weyl invariant are
\begin{align}
\begin{split}
\label{E:hyperbolicProductInvariants}
 \chi(\Sigma_g\times\Sigma_h) &= 4(g-1)(h-1)\,,
 \\
 W_{ijkl}W^{ijkl} &= \frac{4}{3}(1+\varphi^{-2})^2\,, \qquad \text{vol}(\Sigma,\gamma) = 16\pi^2(g-1)(h-1)\varphi^2\,.
\end{split}
\end{align}
It follows from \eqref{E:chiQ4} that
\beq
\label{E:hyperbolicProductQ}
 \chi_Q[\gamma] = \frac{2}{3}(g-1)(h-1)(4-\varphi^2-\varphi^{-2})\,.
\eeq
This has precisely the $\varphi$ dependence of~\eqref{E:S2S2chiQ}, multiplied by $(g-1)(h-1)$. At fixed genera, it is maximized at $\varphi=1$ and tends to $-\infty$ at either extreme anisotropy. At equal curvature radii,
\beq
\label{E:hyperbolicProductWeight}
 \chi_Q = \frac{4}{3}(g-1)(h-1)\,, \qquad \ln|e^{iS_\alpha}|^2 = \frac{\pi^2}{3G}(g-1)(h-1)\,.
\eeq
The second equality holds if a smooth real asymptotically hyperbolic Einstein filling $\alpha$ exists and is continued with the branch used in~\eqref{E:Qweight}. Its candidate weight exceeds that of the round $\mathbb S^4$ when $(g-1)(h-1)>3/2$; the simplest example is $\Sigma_2\times\Sigma_3$. The enhancement is unbounded as the genera grow, even though the local curvature of the boundary remains fixed.

Unlike the product-sphere and black-brane examples, this calculation does not construct the requisite bulk fillings. A higher-genus surface cannot cap off in the same smooth minisuperspace manner as a sphere. Equation~\eqref{E:hyperbolicProductQ} is an exact boundary calculation, but its implication for the no-boundary state is conditional on the existence of suitable fillings and their inclusion in the integration cycle. It identifies a possible enhancement from boundary topology, rather than a large-anisotropy runaway at fixed topology.

%--------------------------------------
\subsection{Toroidal black branes}
\label{S:blackBranes}
%--------------------------------------

For toroidal boundaries, the regular fillings can be written explicitly. Take a flat $\mathbb T^{d-1}$ with one circle of length $\beta$. The positive-definite asymptotically hyperbolic metric is
\beq
\label{E:blackBraneMetric}
 ds_{\rm a}^2 = \frac{dr^2}{f(r)} + f(r)d\chi^2 + r^2d\mathbb T_{d-2}^2\,, \qquad f(r)=r^2-\frac{r_{\rm h}^{d-1}}{r^{d-3}}\,, \qquad \chi\sim\chi+\beta\,.
\eeq
The $\chi$ circle caps off at $r=r_{\rm h}$, and regularity fixes $r_{\rm h}=4\pi/((d-1)\beta)$. The conformal boundary metric is $d\chi^2+d\mathbb T_{d-2}^2$; write its total volume as $V_{d-1}=\text{vol}(\mathbb T^{d-1})$. For odd $d-1$, its renormalized bulk volume is
\beq
\label{E:blackBraneVolume}
 V_{\rm ren} = -\frac{V_{d-1}}{2(d-1)}\left(\frac{4\pi}{(d-1)\beta}\right)^{d-1}\,.
\eeq
Indeed, the bare volume is $V_{d-1}(r_c^{d-1}-r_{\rm h}^{d-1})/(d-1)$, while the boundary volume is $V_{d-1}(r_c^{d-1}-r_{\rm h}^{d-1}/2+o(1))$. Subtracting the leading volume counterterm divided by $d-1$ gives~\eqref{E:blackBraneVolume}; the higher intrinsic-curvature counterterms vanish because the boundary is flat.

The continued all-minus filling therefore has
\beq
\label{E:blackBraneWeight}
 \ln|e^{iS_{\rm bb}}|^2 = (-1)^{(d-2)/2}\frac{V_{d-1}}{8\pi G}\left(\frac{4\pi}{(d-1)\beta}\right)^{d-1}\,, \qquad d-1\ \text{odd}\,.
\eeq
It is enhanced when $d-1\equiv1\pmod4$ and suppressed when $d-1\equiv3\pmod4$. In particular, for $d=6$,
\beq
\label{E:T5weight}
 \chi_Q = \frac{384\pi^2}{3125}\frac{V_5}{\beta^5}\,, \qquad \ln|e^{iS_{\rm bb}}|^2 = \frac{128\pi^4}{3125G}\frac{V_5}{\beta^5}\,.
\eeq
Holding the other four circle lengths fixed, the exponent grows as $\beta^{-4}$ as $\beta\to0$. The bulk curvatures stay bounded in de Sitter units: increasing $r_{\rm h}$ rescales the planar black brane rather than producing a region of large curvature. This is an explicit weakly curved candidate runaway in a toroidal component. It is represented by an all-minus metric and violates KSW, and the toy contour $\mathcal{C}_+$ of Subsection~\ref{S:S1contour} excludes it.

%--------------------------------------
\subsection{Hyperbolic fillings at fixed boundary}
\label{S:hyperbolicFillings}
%--------------------------------------

There is also a construction in which the conformal boundary is held fixed and the topology of the filling changes. This leads to a pathology similar to the old problem \cite{Myers:1988sp, Klebanov:1988eh, Halliwell:1989pu} regarding the sums over chains of Euclidean wormholes, where increasing the number of repetitions makes the on-shell action increasingly negative. See \cite{Aguilar-Gutierrez:2023ril, Blommaert:2025bgd, Ivo:2026ijv, Blommaert:2026jrh} for a modern discussion. This is also a similar feature of the ``chain-of-spheres'' saddles that motivated the KSW criterion \cite{Witten:2021nzp} in the first place.

Let $X$ be a compact hyperbolic $d$-manifold with a single totally geodesic boundary $\Sigma_0$, so that $K_{ij}=0$. The induced metric on $\Sigma_0$ has sectional curvature $-1$. Attach the end
\beq
\label{E:FuchsianEnd}
 ds_{\rm a}^2 = d\rho^2 + \cosh^2(\rho)d\Sigma_0^2\,, \qquad \rho\geq0\,.
\eeq
Because both the induced metric and its normal derivative match at $\rho=0$, the union $M_X=X\cup_{\Sigma_0}([0,\infty)\times\Sigma_0)$ is a smooth complete hyperbolic filling with conformal boundary $\Sigma_0$. Taking minus this metric and continuing the end to $\rho=t-\pi i/2$ gives a real Lorentzian late-time region, $ds^2=-dt^2+\sinh^2(t)d\Sigma_0^2$.

\begin{figure}
    \centering
    \includegraphics[width=\linewidth]{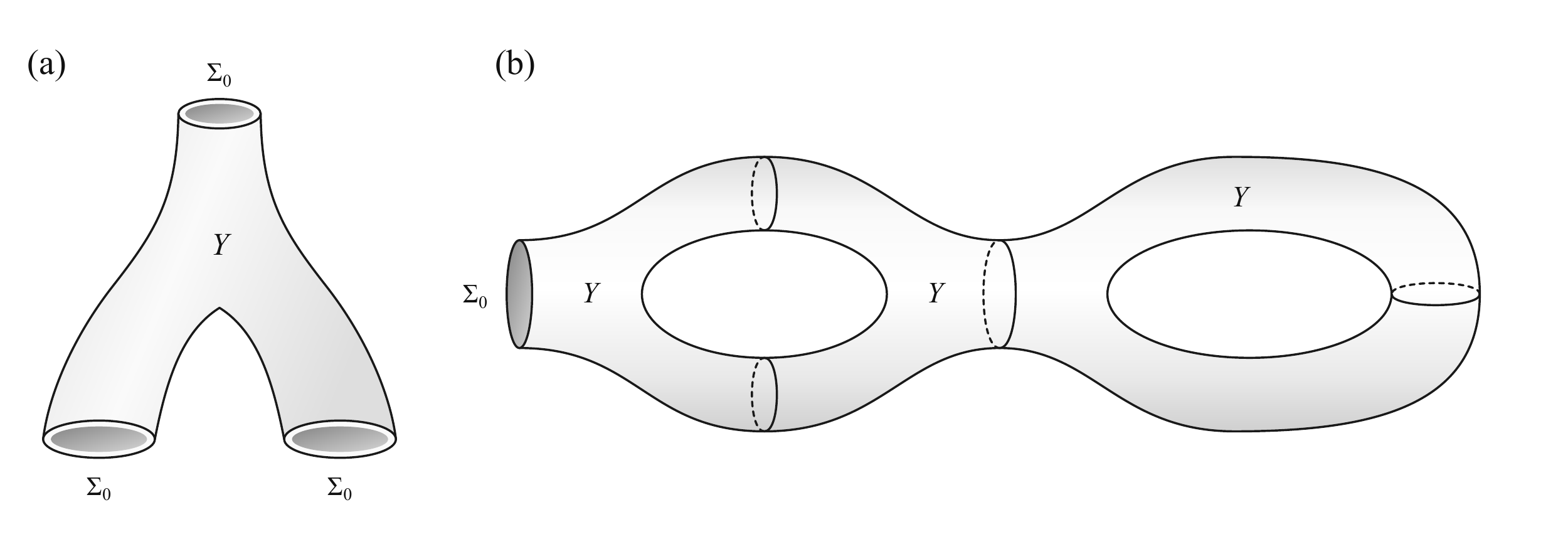}
    \caption{Hyperbolic building blocks and their gluings, depicted schematically as two-dimensional surfaces. (a) A four-dimensional hyperbolic ‘pair of pants’ $Y$ with three isometric, totally geodesic boundary components $\Sigma_{0}$. (b) Gluing three copies along four pairs of boundary components produces a connected hyperbolic manifold with a single remaining boundary $\Sigma_{0}$. The marked circles indicate the gluing interfaces. Repeating this construction gives increasingly complicated interiors while keeping the boundary geometry fixed.}
    \label{hyppant}
\end{figure}

For even $d$, the Weyl tensor vanishes throughout $M_X$. The renormalized Gauss-Bonnet relation reduces to
\beq
\label{E:hyperbolicFillingVolume}
 V_{{\rm ren},X} = (-1)^{d/2}\frac{\text{vol}(\mathbb S^d)}{2}\chi(X)\,,
\eeq
and hence
\beq
\label{E:hyperbolicFillingWeight}
 \ln|e^{iS_X}|^2 = S_0\chi(X)\,.
\eeq
The topology of $X$ thus controls the candidate magnitude at one and the same boundary conformal class.

In $d=4$, compact orientable hyperbolic building blocks $Y$ exist with three isometric totally geodesic boundaries $\Sigma_0$ (see Figure \ref{hyppant}).\footnote{In~\cite{Ratcliffe:2000nn}, the authors construct such a four-dimensional pair of pants and prove superexponential growth of the number of compact hyperbolic fillings for a special three-dimensional hyperbolic boundary. Their original application is to negative cosmological constant; here we use their four-dimensional hyperbolic geometry and continue its metric to the positive-cosmological-constant saddle.} Glue an odd number $N$ of copies along orientation-reversing boundary isometries, leaving exactly one copy of $\Sigma_0$ unglued. Suitable connected gluings give smooth compact fillings $X_N$. Since a closed odd-dimensional manifold has zero Euler characteristic,
\beq
\label{E:pantsEuler}
 \chi(X_N)=N\chi(Y)\,, \qquad \chi(Y)=\frac{3}{4\pi^2}\text{vol}(Y)>0\,,
\eeq
so their candidate weights obey $\ln|e^{iS_{X_N}}|^2=N S_0\chi(Y)$. These are actual smooth weakly curved geometries, with arbitrarily large weights and a prescribed conformal boundary. The outstanding question is their coefficient in the gravitational integration cycle. Their all-minus representation violates KSW, and the increasing weight of individual saddles does not by itself establish divergence of the full topology sum, whose phases, determinants, and saddle coefficients are additional data.

The same construction gives enhancement in higher bulk dimensions $d\equiv0\pmod4$ whenever an appropriate compact hyperbolic building block is available. For $d\equiv2\pmod4$, Gauss-Bonnet instead fixes $\chi(X)<0$, and~\eqref{E:hyperbolicFillingWeight} suppresses these constant-curvature fillings. The required building blocks cannot be assumed to exist in every dimension.

There is a related boundary diagnostic in odd bulk dimension. A closed hyperbolic four-manifold $\Sigma$ has $W=0$ and
\beq
\label{E:hyperbolicBoundaryWeight}
 \chi_Q[\Sigma]=\chi(\Sigma)=\frac{3}{4\pi^2}\text{vol}(\Sigma)\,, \qquad \ln|e^{iS_\alpha}|^2=\frac{S_0}{2}\chi(\Sigma)\,, \qquad d=5\,.
\eeq
Gluing an even number of the same four-dimensional building blocks along all their boundaries produces closed hyperbolic boundaries with unbounded positive Euler characteristic. As with $\Sigma_g\times\Sigma_h$, the weight formula in~\eqref{E:hyperbolicBoundaryWeight} is conditional on finding a smooth real asymptotically hyperbolic Einstein filling. The four-dimensional gluing construction establishes the boundary geometries, not their five-dimensional fillings.

%--------------------------------------
\subsection{Lens spaces}
\phantomsection\label{S:lens}
%--------------------------------------

In $d=4$, the Taub-Bolt fillings of squashed lens spaces give a family of saddles whose weights grow without bound at large squashing. Unlike the families above, they are not all-minus geometries, and KSW can be tested on them directly. We construct KSW-allowable contours in a bounded-weight range and exclude symmetry-preserving allowable contours in the runaway regime. Complex no-boundary lens-space bolts were considered qualitatively at large volume in \cite{Jensen:1990xc}. Here we evaluate their finite late-time weights and extend the squashed-sphere KSW analysis of \cite{Hertog:2024nbh} to lens spaces.

Take the late-time slice to be the lens space $L(p,1)$ with the squashed metric
\beq
\label{E:lensboundary}
\gamma = 4\nu_0^2(d\psi+\cos\theta\,d\phi)^2+d\Omega_2^2\,,\qquad \psi\sim\psi+\frac{4\pi}{p}\,.
\eeq
Here $2\nu_0$ measures the squashing, and $2\nu_0=1$ gives the round metric. The Taub-Bolt saddles are
\begin{align}
\label{E:lensmetric}
ds^2 &= 4\nu^2 f(t)(d\psi+\cos\theta\,d\phi)^2-\frac{dt^2}{f(t)}+(t^2+\nu^2)d\Omega_2^2\,,\nn\\
f(t) &= \frac{t^4+(6\nu^2-1)t^2+\nu^2(1-3\nu^2)+2mt}{t^2+\nu^2}\,.
\end{align}
The fiber circle caps off smoothly at a bolt $t=t_h$, where $f(t_h)=0$ and $\nu f'(t_h)=ip/2$. When the cutoff is removed, $\nu\to\nu_0$, $t_h\to ir_0$ and $m\to i\mu_0$, and the bolt conditions become
\beq
\label{E:lensbolt}
 3r_0^2-\frac{p\,r_0}{2\nu_0}+1-3\nu_0^2=0\,,\qquad
 r_\pm=\frac{p\pm\sqrt{p^2-48\nu_0^2+144\nu_0^4}}{12\nu_0}\,,
\eeq
with $\mu_0$ given in~\eqref{E:lensmass}. The two-sphere at the bolt has squared radius $\nu_0^2-r_0^2$. For $p\geq3$, the larger root has $r_+>\nu_0$, so this squared radius is negative, and the saddle violates KSW at the bolt. We use the smaller root $r_0=r_-$. It lies in $-\nu_0<r_0<\nu_0$, and it changes sign at $\nu_0^2=1/3$: $r_0>0$ below and $r_0<0$ above. For $p=2$ the corresponding root is $r_0=1/(3\nu_0)-\nu_0$, a genuine bolt for $\nu_0^2>1/6$.\footnote{For $p=1$ the larger root also gives a genuine bolt, with $0<r_+<\nu_0$, when $\nu_0^2>(2+\sqrt3)/12$. The argument below does not exclude it, but its weight is suppressed at large squashing.}

The saddle has
\begin{align}
\label{E:lensaction}
W_{\rm E} \equiv 2\,\text{Im}\,S
=\frac{S_0}{108p\nu_0^2}\left(p^3-72p\nu_0^2-\left(p^2-48\nu_0^2+144\nu_0^4\right)^{3/2}\right)\,,\qquad S_0=\frac{\pi}{G}\,.
\end{align}
For every $\nu_0$, $W_{\rm E}<0$, so each saddle is enhanced, $|e^{iS}|^2=e^{-W_{\rm E}}>1$. For $\nu_0^2<1/3$ the weight is bounded,$-\frac23S_0<W_{\rm E}<0$, and stays below that of the round three-sphere (see Appendix~\ref{VI:A:lens}). For $\nu_0^2>1/3$ it is not. At large squashing $W_{\rm E}\simeq-16\nu_0^4S_0/p$, a runaway. The runaway saddles remain weakly curved. Their Kretschmann scalar is
\beq
\label{E:lensKretschmann}
R_{\mu\nu\rho\sigma}R^{\mu\nu\rho\sigma}=24+24\left(\Psi_+^2+\Psi_-^2\right)\,,\qquad
 \Psi_\pm=\frac{m\pm i\nu(1-4\nu^2)}{(t\pm i\nu)^3}\,.
\eeq
At fixed $p$, the curvature remains bounded as $\nu_0\to\infty$. At the bolt,
\beq
 R_{\mu\nu\rho\sigma}R^{\mu\nu\rho\sigma}
 \longrightarrow
 48+96\frac{(p-1)^2}{(p+2)^2}\,,
\eeq
which gives $63.36$ for $p=3$, compared with $24$ for de Sitter space. The curvature also remains bounded along the contour running from the bolt up the imaginary $t$ axis to $t=0$ and then along the positive real axis. At $t=0$, the Kretschmann scalar tends to $1560$. Thus, the large-squashing runaway comes from a regime of finite curvature.

For $p\geq3$ and $\nu_0^2<1/3$, we construct a strictly KSW-allowable contour at any sufficiently large finite cutoff in Appendix~\ref{VI:A:lens}, and the same holds for $p=2$ with $1/6<\nu_0^2<1/3$. For $\nu_0^2>1/3$ there is no such contour, at least among contours that respect the $SU(2)\times U(1)$ symmetry. To see this, write $t=ir$ and parameterize such a contour by $r(s)$, with $s$ real and increasing from the bolt to the cutoff. The radial and fiber eigenvalues of the metric along it multiply to
\beq
\label{eqargnu1}
 \lambda_{\rm radial}\lambda_{\rm fiber}
 =\frac{r'^2}{f(ir)}\,4\nu^2f(ir)
 =4(\nu r')^2\,.
\eeq
KSW requires $|\arg\lambda_{\rm radial}|+|\arg\lambda_{\rm fiber}|<\pi$, so this product cannot be zero or negative real. Hence $\text{Re}(\nu r')$ never vanishes, and it keeps one sign along the whole contour. The sign is fixed at the bolt. Smoothness there requires $r-r_h=c\rho^2+\cdots$ with $\rho$ real, and the two eigenvalues of the disk are then $-8\nu c/p$ and $-2p\nu c\rho^2$, both with phase $\arg(-\nu c)$. KSW requires $\text{Re}(\nu c)<0$, so $\text{Re}(\nu r')<0$ everywhere on the contour. Since $\nu$ is constant, integrating gives
\beq
\label{eqargnu2}
 \text{Re}\!\left[\nu(r_c-r_h)\right]<0\,.
\eeq
The cutoff conditions of Appendix~\ref{VI:A:lens} instead give
\beq
\label{eqargnu3}
 \text{Re}\!\left[\nu(r_c-r_h)\right]
 =-\nu_0r_0+O(\epsilon^2)\,.
\eeq
For $\nu_0^2>1/3$ we have $r_0<0$, so this is positive at small enough $\epsilon$, a contradiction. For $\nu_0^2<1/3$ it is negative, consistent with the contour of Appendix~\ref{VI:A:lens}.

Thus KSW admits the strict ranges established above and excludes symmetry-preserving allowable contours for $\nu_{0}^{2}>\frac{1}{3}$. Our contour model has nothing to say here, since we have no off-shell family. For the squashed three-sphere, $p=1$, a minisuperspace study of the lapse integral finds that Stokes phenomena remove the bolt saddles that would otherwise dominate~\cite{Lehners:2024kus}. Finally, bounded weights for the allowed saddles do not by themselves make the sum over lens-space topologies converge; that also depends on the coefficients of the saddles and on the one-loop measure.

%--------------------------------------
\section{Comments on coupling to an inflaton}
\label{S:inflaton}
%--------------------------------------

In pure gravity, the magnitude of the no-boundary wavefunction depends only on the conformal class of the late-time metric. Coupling to a slowly rolling inflaton makes the overall scale physical when the wavefunction is evaluated on a slice of fixed inflaton $\chi = \chi_r$. We write the reheating metric as $R_r^2(d\Omega_p^2 + \varphi^2 d\Omega_q^2)$ and parameterize homogeneous changes of scale by $\zeta_0 = \delta\ln R_r$, at fixed anisotropy $\varphi$.

As shown in recent work~\cite{Cotler:2026flq}, the leading semiclassical dependence on $\zeta_0$ is
\beq
\label{E:inflatonZeroMode}
\ln |\Psi_\alpha(\zeta_0,\varphi)|^2 = \mathcal B_\alpha(\varphi) \left[1 - (d-2)\epsilon_* \zeta_0\right] + \cdots\,, \qquad \mathcal B_\alpha(\varphi) = \frac{d-1}{8\pi G H_*^{d-2}} \text{Re} V_{E,\alpha}(\varphi)\,.
\eeq
Here $\alpha$ labels a gravitational saddle, $V_{E,\alpha}$ is the dimensionless Euclidean volume of the bra-ket geometry obtained by gluing the saddle to its complex conjugate, and $H_*$ and $\epsilon_*$ are the Hubble scale and slow-roll parameter evaluated near the early cap. The KSW criterion implies $\text{Re} \,V_{E,\alpha} > 0$ for an admissible filling. Consequently, every KSW-admissible saddle individually favors a smaller reheating surface. Thus the familiar $\ell = 0$ problem of the round sphere is a more general feature of KSW-admissible no-boundary geometries. This expansion is valid to leading order in slow roll for $\epsilon_* |\zeta_0| \ll 1$.

The volume entering this expression is the bra-ket volume defined in~\eqref{E:KSWvolume}. Thus the sign statement used here follows from the local inequality~\eqref{E:weakKSW}, together with the absence of a finite real Gibbons--Hawking contribution at late times. This argument constrains each saddle separately and does not determine which saddles occur in the GPI.

For the $\mathbb S^2 \times \mathbb S^2$ wavefunction in $d = 5$, the Nariai endpoint has $\mathcal B_{\rm Nariai} = \pi^2/(2\sqrt{2} G H_*^3) > 0$ and exhibits the preference for a smaller reheating volume. In $d > 5$, the large-anisotropy runaway saddles have $\mathcal B_\alpha \sim c\varphi^q/(G H_*^{d-2})$ with $c > 0$, so their preference for smaller reheating volume is parametrically enhanced. However, the all-minus geometries representing these saddles violate KSW, and it remains unclear whether the corresponding saddles contribute to the gravitational path integral. Our conclusions hold at leading order in slow roll, provided the inflaton evolution remains perturbative along the saddle. Near the Nariai endpoint this requires $\epsilon_* \ln(1/\varphi) \ll 1$.

%--------------------------------------
\section{Discussion}
\label{S:discuss}
%--------------------------------------

We have found several families of smooth, weakly curved saddles whose weights grow without bound. If these saddles contribute to the no-boundary wavefunction, their unbounded weights threaten the normalizability of the state and, more generally, the mathematical integrity of the no-boundary framework. Large-anisotropy runaways occur for product-sphere boundaries in every bulk dimension $6 \leq d \leq 11$ that we have studied. The simplest examples are the all-minus continuation of a large AdS-Schwarzschild black hole with $\mathbb S^1 \times \mathbb S^4$ boundary and the filling of $\mathbb S^2 \times \mathbb S^3$ in which the two-sphere caps off. Since their exponential growth persists within the regime of validity of gravitational EFT, it cannot be dismissed as a consequence of large curvature or repaired by ordinary loop corrections.

The product-sphere families belong to a broader bestiary of examples with dangerously large weights. Toroidal black branes exhibit explicit runaways in a boundary modulus, the Taub-Bolt fillings of squashed lens spaces in $d=4$ run away at large squashing, and hyperbolic fillings in $d = 4$ have arbitrarily large individual weights at a fixed boundary conformal class, obtained by varying the topology of the bulk filling. For products of hyperbolic surfaces and closed hyperbolic four-manifolds in $d = 5$, the boundary conformal invariants likewise predict arbitrarily large saddle weights, provided suitable bulk fillings exist.

Whether any of these potentially problematic saddles contribute to the no-boundary wavefunction remains an open and important question. A smooth solution with the prescribed boundary conditions may have a vanishing coefficient in the gravitational integration cycle, and KSW does not determine that coefficient, as it is a local restriction on complex metrics, rather than a prescription for the integration cycle. Most of the dangerous geometries considered here arise naturally as minus asymptotically hyperbolic fillings, whose all-minus representatives violate KSW; for the product-sphere and black-brane examples, the noncontracting directions at the cap already obstruct an allowable time contour. In pure gravity their one-loop determinants are nevertheless as well defined as those of the corresponding AdS problem, but perturbative regularity does not determine whether they contribute. There are three broad possibilities for the no-boundary proposal.

If KSW is a fundamental restriction on the GPI, it excludes saddles without an allowable representative, including the product-sphere and toroidal runaway families. The restriction would, however, also exclude saddles with bounded weights that contribute to our contour model. Some nonspherical topologies remain compatible with KSW. In particular the lens-space Taub-Bolt saddles admit regulated KSW contours in a range of squashing, where their individual weights remain smaller than that of the round sphere. At larger squashing, where their weights grow without bound, they admit no allowable symmetric contour. What remains unclear is how rich a state survives the restriction, and whether the surviving contributions define a normalizable state once all boundary components and bulk topologies are included.

A second possibility is that the integration cycle retains a product-sphere or toroidal runaway in its asymptotic regime, in which case the exponential tail gives a divergent norm under the ordinary semiclassical measure unless the full saddle sum cancels its leading growth. This would challenge the no-boundary proposal as a prescription for a state in de Sitter quantum gravity already within gravitational EFT; because these families fail KSW at their caps, their inclusion would also challenge KSW as a universal restriction on the GPI.

A third possibility is that the gravitational integration cycle excludes the dangerous saddles because the corresponding thimbles have zero coefficient in the chosen integration cycle, without imposing KSW as a separate restriction. In a contour model for the $\mathbb S^1 \times \mathbb S^q$ wavefunction, we adapt the arguments of~\cite{Maldacena:2019cbz, Mahajan:2025bzo} by using the cap radius to parameterize an off-shell family with a conical singularity and studying its saddle decomposition on a chosen convergent contour. In the dimensions analyzed, this contour retains the upper Nariai saddle and excludes the lower one, apart from a Stokes-dependent range in $d=5$, while the large-anisotropy bound excludes the runaway large black holes and black branes for $q \equiv 0 \pmod 4$. On the canonical AdS contour, by contrast, the large real black hole contributes at high temperature, while the complex black holes are absent. Analytically continuing an on-shell solution and its action therefore does not determine whether the corresponding saddle contributes to the no-boundary wavefunction, since its coefficient must be determined from the integration cycle.

Where the two can be compared, the contour model and KSW agree on the runaway large black holes and on the lower Nariai saddle, the latter at least for symmetric contours, and every saddle that KSW allows appears on the contour. The contour also keeps saddles that KSW excludes: the upper Nariai saddle beyond the KSW bound and, in $d = 4,5$, the small all-minus black holes. An integration cycle can thus remove the dangerous runaway while retaining contributions that KSW would exclude. We have not derived this contour from the full GPI. Both the off-shell family and the contour are assumptions, and requiring a particular saddle to contribute does not, by itself, determine the contour uniquely.

A related example in which the integration cycle excludes exponentially enhanced saddles arises in the cosmological necklace problem. Gluing copies of a periodic Euclidean solution into progressively longer necklaces produces saddles with increasingly negative actions~\cite{Blommaert:2026jrh}. There the observable is a closed no-boundary density-matrix trace, and the repetitions are covers that need not change the abstract topology; our hyperbolic construction instead fixes a conformal boundary and varies the bulk topology. In their three-dimensional minisuperspace model, a regulated Lorentzian lapse contour excludes the higher necklaces even though their real Euclidean metrics satisfy KSW, providing another example in which the integration cycle removes an unbounded sequence of saddle weights independently of KSW.

The existence of large saddle weights does not necessarily preclude a sensible Lorentzian interpretation of the corresponding geometries. For example, our hyperbolic fillings have an exactly real Lorentzian region,
\beq
 ds^2=-dt^2+\sinh^2(t)d\Sigma_0^2\,,\qquad t>0\,.
\eeq
Although the fillings have arbitrarily large weights, they share this locally de Sitter region, in which a late-time observer can live. Whether these different fillings prepare distinguishable states of fluctuations, and whether their correlators could reveal the topology of the bulk filling to a late-time observer, remain open questions.

The occurrence of both local maxima and runaways also raises the question of how extrema of the no-boundary distribution are related to Lorentzian classicality. For an individual saddle contribution $\Psi_\alpha[h]\sim e^{iS_\alpha[h]}$, the connection between probability extrema and real extrinsic curvature follows from the leading semiclassical relation
\beq
 \frac{\delta\ln|\Psi_\alpha[h]|^2}{\delta h_{ij}(x)}
 =-2\,\text{Im}\,\pi_\alpha^{ij}(x)\,,
 \qquad \pi_\alpha^{ij}(x)=\frac{\delta S_\alpha[h]}{\delta h_{ij}(x)}\,.
\eeq
For the real induced metric $h_{ij}$ at a finite cutoff, stationarity under all independent metric variations therefore implies real extrinsic curvature in pure gravity, although it does not distinguish a maximum from a minimum~\cite{Ivo:2024ill}. In the presence of matter, the matter fields and their conjugate momenta must also be real. Stationarity with respect to the anisotropy constrains only its conjugate momentum; it says nothing about saddles with different bulk topologies, which cannot be related by a continuous variation of the boundary metric.

More generally, an extremum of the probability distribution is not necessary for WKB classicality, since the phase need only vary much more rapidly than the logarithm of the amplitude~\cite{Hartle:2008ng}. For the circle-cap solutions, at fixed boundary anisotropy the relative imaginary correction to the late-time metric is $O(e^{-(d-1)t})$, although this approximation need not be uniform along the runaway. Establishing a semiclassical observer interpretation still requires control of the fluctuations.

The anisotropic runaways are not the only potential difficulty for the no-boundary proposal. Coupling to a slowly rolling inflaton introduces a dependence on the overall size of the reheating surface, which is absent from the magnitude of the pure-gravity wavefunction. Within the controlled slow-roll expansion, the positive real Euclidean volume of a KSW-admissible filling makes its individual contribution favor a smaller reheating surface, extending the familiar homogeneous-mode problem of the round sphere to other allowable geometries. If the runaway saddles contribute, the same large factor that enhances their anisotropic weights also amplifies their preference for smaller reheating volumes. Removing the anisotropic runaways would not, by itself, resolve the preference for small reheating volumes. Whether that preference produces a divergent integral over the overall scale depends on the inflaton potential, the range of validity of the expansion, and the measure.

Ultimately, resolving the status of the runaway saddles requires a better understanding of the gravitational integration cycle. When a circle caps off, its radius provides a natural parameter for an off-shell family, but we do not have an equally controlled construction for $p,q > 1$ or for fillings whose topology changes. Constrained instantons or $\lambda$-saddles may offer a route~\cite{Cotler:2020lxj,Cotler:2021cqa,Cotler:2022rud}, but we do not yet know a suitable constraint and its integration contour. Such a model should reproduce the smooth saddles and their actions and determine their coefficients from the integration cycle. The larger question is whether de Sitter quantum gravity admits an integration cycle that yields a sufficiently rich, normalizable no-boundary state. Our examples make this question difficult to avoid, but do not settle it.

%----------------------------------------------------
\subsection*{Acknowledgements}
%----------------------------------------------------

We would like to thank A.~Blommaert, J.~Kudler-Flam, J.~Maldacena, B.~M\"{u}hlmann, X.~Shi, G.~Turiaci, and E.~Witten for insightful discussions.
We acknowledge the use of ChatGPT and Claude as research aids to help identify errors, work through intermediate steps of calculations and derivations, assist with coding and numerics, and suggest edits to the text. All AI-assisted calculations, derivations, and numerics were independently checked by the authors, and all results and arguments were written up and verified by the authors.
JC is supported by the Simons Collaboration on Celestial Holography, as well as a Fellowship from the Alfred P. Sloan Foundation. PD, WH, and KJ are supported in part by an NSERC Discovery Grant.

\begin{appendix}

%--------------------------------------
\section{The semianalytic Russian doll}
\label{A:russianDoll}
%--------------------------------------

In this Appendix we construct the Russian doll branches of the $\mathbb{S}^2\times\mathbb{S}^2$ wavefunction of Subsection~\ref{S:S2S2solutions} at leading order in $|B_0|$. The result is the closed formula~\eqref{E:dollPhi} for $\varphi$ as a function of $B_0$. It gives the generational factor $e^{4\pi/\sqrt{15}}$ exactly, and it locates each branch with a relative error of order $|B_0|$.

Let $z=-B_0$ with $|z|\ll1$. Near the cap the geometry varies on the scale $\rho\sim\sqrt{|z|}$, and far from the cap on the scale $\rho\sim1$. We solve in each region and match the two solutions where $\sqrt{|z|}\ll\rho\ll1$.

%--------------------------------------
\subsection{Core}
%--------------------------------------

Set $\rho=\sqrt{z}\,x$, $A=z\,\alpha(x)$ and $B=z\,\beta(x)$. At leading order in $z$ the cosmological constant drops out of~\eqref{E:SpSqEinstein}, and the equations become
\beq
\label{E:capEqns}
 \alpha''+\frac{\alpha'\beta'}{\beta}+2=0\,,\qquad
 \beta''+\frac{\alpha'\beta'}{\alpha}+2=0\,.
\eeq
The near-cap solution~\eqref{E:nearCapSol} fixes
\beq
 \alpha=-x^2\left(1-\frac{x^2}{9}+\frac{46x^4}{2025}+O(x^6)\right)\,,\qquad
 \beta=-\left(1+\frac{x^2}{3}+\frac{x^4}{135}+O(x^6)\right)\,.
\eeq
Up to an overall sign this is a Ricci-flat metric in which one two-sphere caps off smoothly. We integrate~\eqref{E:capEqns} numerically. At large $x$ the solution approaches the Ricci-flat cone over $\mathbb{S}^2\times\mathbb{S}^2$, $\alpha=\beta=-y^2/3$ with $y=x+c$. The approach is controlled by the squashing mode of the cone, $v=\frac12\ln(\alpha/\beta)$, which at linear order obeys
\beq
 v''+\frac{4}{y}\,v'+\frac{6}{y^2}\,v=0\,,\qquad v\propto y^{\sigma_\pm}\,,\qquad \sigma_\pm=-\frac32\pm\frac{\sqrt{15}}{2}\,i\,.
\eeq
So at large $x$
\beq
\label{E:coreAsymptotics}
 \frac12\ln\frac{\alpha}{\beta}=\frac12\left(D_+y^{\sigma_+}+D_-y^{\sigma_-}\right)+O\big(y^{-9/2}\big)\,,
\eeq
and fitting the numerical solution gives
\beq
 c=0.86125\,,\qquad D_\pm=-1.22014\pm0.02193\,i\,.
\eeq
These are the exponents Kol found for the squashing mode of the Ricci-flat cone over $\mathbb{S}^2\times\mathbb{S}^2$~\cite{Kol:2002xz}. For the cone over $\mathbb{S}^p\times\mathbb{S}^q$ they are $\sigma_\pm=-\frac{d-2}{2}\pm\frac12\sqrt{(d-2)(d-10)}$, complex for $d<10$. The same fact underlies B\"ohm's infinite sequences of cohomogeneity-one Einstein metrics~\cite{bohm1998inhomogeneous,Gibbons:2002th}. Here, the doll branches cross the positive real $B_0$ axis at the B\"ohm metrics on $\mathbb{S}^3\times\mathbb{S}^2$, continued to Lorentzian signature from their totally geodesic midpoint. Indeed, the squares of the values for $b_0$ listed in~\cite{Gibbons:2002th} for the metrics $(2,2)_3$ and $(2,2)_5$ are the values $B_0=0.013875$ and $5.5559\times10^{-4}$ of Table~\ref{T:doll}, and $(2,2)_1$ is the Nariai point $B_0=1/4$.

%--------------------------------------
\subsection{Outer region}
%--------------------------------------

For $\rho\gg\sqrt{|z|}$ the geometry is close to
\beq
 A=B=-\frac{\sinh^2 s}{3}\,,\qquad s=\rho+\sqrt{z}\,c\,.
\eeq
This is an exact solution of~\eqref{E:SpSqEinstein}, cosmological constant included. It is minus the hyperbolic cone over $\mathbb{S}^2\times\mathbb{S}^2$, with a singular tip at $s=0$. At small $s$ it reduces to the Ricci-flat cone of the core. Write
\beq
 A=-\frac{\sinh^2 s}{3}\,e^{U+V}\,,\qquad B=-\frac{\sinh^2 s}{3}\,e^{U-V}\,.
\eeq
Exchanging the two spheres sends $V\to-V$ and leaves $U$ fixed. The equations are invariant under the exchange, so the equation for $V$ is odd in $V$:
\beq
\label{E:outerV}
 V''+4\coth(s)\,V'+\frac{6}{\sinh^2 s}\,V=O\big(V^3,UV\big)\,.
\eeq
Its linear solutions are
\beq
 f_\pm=\tanh^{\sigma_\pm}(s)\;{}_2F_1\!\left(\frac{\sigma_\pm}{2},\frac{\sigma_\pm+1}{2};\sigma_\pm+\frac52;\tanh^2 s\right)\,,
\eeq
with $f_\pm\to s^{\sigma_\pm}$ as $s\to0$ and
\beq
\label{E:Fpm}
 F_\pm\equiv\lim_{s\to\infty}f_\pm=\frac{\Gamma\!\left(\sigma_\pm+\frac52\right)}{\Gamma\!\left(\frac{\sigma_\pm}{2}+\frac52\right)\Gamma\!\left(\frac{\sigma_\pm}{2}+2\right)}\,.
\eeq
Apart from a shift of $s$, the function $U$ is even in $V$ and starts at second order. It changes only the overall scale of $A$ and $B$, so it does not enter the anisotropy, which at late times is $\varphi^2=\lim B/A=e^{-2V(\infty)}$.

%--------------------------------------
\subsection{Matching and the doll branches}
%--------------------------------------

In the overlap $\sqrt{|z|}\ll s\ll1$, \eqref{E:coreAsymptotics} gives $V=\frac12\sum_\pm D_\pm z^{-\sigma_\pm/2}s^{\sigma_\pm}$. The outer solution with this behavior at small $s$ is $V=\frac12\sum_\pm D_\pm z^{-\sigma_\pm/2}f_\pm(s)$, and so
\beq
\label{E:dollPhi}
 \ln\varphi^2=-D_+F_+\,z^{-\sigma_+/2}-D_-F_-\,z^{-\sigma_-/2}\,,\qquad D_+F_+=-0.54367+0.10202\,i\,.
\eeq
Here $z^{-\sigma_\pm/2}$ is continued from $z>0$ through $\text{Im}\,B_0>0$. The lower half plane follows by complex conjugation. We draw three conclusions.

First, on the AdS branch $z>0$ the right side of~\eqref{E:dollPhi} is real, and
\beq
 \varphi\approx1-|D_+F_+|\,|B_0|^{3/4}\cos\!\left(\frac{\sqrt{15}}{4}\ln|B_0|-\arg(D_+F_+)\right)\,.
\eeq
So $\varphi$ oscillates about one, with an amplitude that falls as $|B_0|^{3/4}$.

Second, a saddle requires $\varphi^2>0$, so the branches are the curves on which $\text{Im}\ln\varphi^2=0$. The positive real $z$ axis is one of these curves; it is the AdS branch. The others are the dolls, and they cross the AdS branch where $\varphi$ is extremal along it. Under $z\to\lambda z$ with $\lambda=e^{4\pi/\sqrt{15}}$, both terms on the right side of~\eqref{E:dollPhi} are multiplied by $-\lambda^{3/4}$. So the family of curves is invariant under this rescaling. At every angle successive branches obey $r_k(\theta)/r_{k+1}(\theta)=e^{4\pi/\sqrt{15}}\approx25.65$, and $\varphi-1$ changes sign and shrinks by $\lambda^{3/4}\approx11.4$ from one branch to the next. The generational factor depends only on $\sigma_\pm$. The core enters only through the single number $D_+F_+$, which fixes where the branches lie and the size of $\varphi-1$ on them.

Third, the error is under control. The cosmological constant enters the core at relative order $z$, and so shifts $D_\pm$ by a relative amount of order $z$. Nonlinear terms enter $\ln\varphi^2$ only at cubic order. So~\eqref{E:dollPhi} has a relative error of order $|B_0|$, as Table~\ref{T:doll} shows. The formula even reproduces the outermost layer, the Nariai branch, to within about 15\% in $|B_0|$ where $|B_0|\lesssim0.2$.

\begin{table}[h]
\centering
\renewcommand{\arraystretch}{1.2}
\begin{tabular}{c|c|c|c|c|c}
$\arg B_0$ & branch & $|B_0|$, numerics & $|B_0|$, \eqref{E:dollPhi} & $\varphi$, numerics & $\varphi$, \eqref{E:dollPhi} \\ \hline
$\pi$   & 1 & $2.4960\times10^{-3}$ & $2.4784\times10^{-3}$ & $1.004886$ & $1.004869$ \\
$\pi$   & 2 & $9.6643\times10^{-5}$ & $9.6615\times10^{-5}$ & $0.999574$ & $0.999574$ \\
$\pi/2$ & 1 & $4.3841\times10^{-3}$ & $4.3815\times10^{-3}$ & $1.020834$ & $1.021011$ \\
$\pi/2$ & 2 & $1.70806\times10^{-4}$ & $1.70804\times10^{-4}$ & $0.998178$ & $0.998177$ \\
$\pi/4$ & 1 & $7.6729\times10^{-3}$ & $7.7520\times10^{-3}$ & $1.071226$ & $1.072570$ \\
$\pi/4$ & 2 & $3.0206\times10^{-4}$ & $3.0220\times10^{-4}$ & $0.993877$ & $0.993873$ \\
$0$     & 1 & $1.3875\times10^{-2}$ & $1.4268\times10^{-2}$ & $1.265115$ & $1.270156$ \\
$0$     & 2 & $5.5559\times10^{-4}$ & $5.5623\times10^{-4}$ & $0.979254$ & $0.979238$
\end{tabular}
\caption{The first two doll branches at fixed $\arg B_0$, from direct numerical solution and from the leading-order formula~\eqref{E:dollPhi}. At $\arg B_0=\pi$ the entries are the extrema of $\varphi$ along the AdS branch, where the dolls attach to it. At $\arg B_0=0$ the numerical entries are time-symmetric solutions, found on the Euclidean section $\rho=i\tau$.}
\label{T:doll}
\end{table}

%--------------------------------------
\section{KSW contours for $\mathbb{S}^1\times\mathbb{S}^q$}
\label{S:KSWappendix}
%--------------------------------------

We construct an allowable contour for the upper Nariai branch at fixed $0<\varphi<\varphi_{\rm KSW}$. It is important to impose the boundary conditions at a finite cutoff before taking the late-time limit. Put the boundary at $z=Z>0$ and write
\beq
\label{E:KSWfiniteMetric}
 ds^2=-\frac{dz^2}{F(z)}+a_Z^2F(z)d\chi^2+z^2d\Omega_q^2\,,\qquad
 \chi\sim\chi+2\pi\,,\qquad
 F(z)=z^2-1-\frac{\zeta_Z^{q-1}(\zeta_Z^2-1)}{z^{q-1}}\,.
\eeq
Here $\zeta_Z$ and $a_Z$ obey
\beq
\label{E:KSWfiniteData}
 a_Z=\frac{Z}{\varphi\sqrt{F(Z)}}\,,\qquad
 a_ZF'(\zeta_Z)=2i\,,\qquad
 \zeta_Z\longrightarrow\zeta_+\,,\quad a_Z\longrightarrow\varphi^{-1}\,.
\eeq
The square root is continued from the large real-$z$ sheet. These equations make the boundary metric exactly $Z^2(\varphi^{-2}d\chi^2+d\Omega_q^2)$. The finite root is a smooth perturbation of $\zeta_+$, since we work below the merger.

Using $d-1=q+1$ in this appendix, define
\beq
\label{E:KSWfiniteContour}
 D=Z^{d-1}-\zeta_Z^{d-1}\,,\qquad
 z(s)^{d-1}=\zeta_Z^{d-1}+sD\,,\qquad 0\leq s\leq1\,.
\eeq
The continuous root has positive real and imaginary parts in the interior. Write
\beq
 \theta(s)=\arg z(s)\,,\qquad
 \eta(s)=\arg\bigl(a_Z^2F(z(s))\bigr)\,,\qquad
 \epsilon_Z=-\arg(a_ZD)\,.
\eeq
These are continuous arguments, with $\eta(1)=\theta(1)=0$. Since $z'=D/((d-1)z^{d-2})$, the metric eigenvalues on the real $s$ contour have phases
\beq
\label{E:KSWphases}
 \arg\lambda_s=\pi-2\epsilon_Z-2q\theta-\eta\,,\qquad
 \arg\lambda_\chi=\eta\,,\qquad
 \arg\lambda_\Omega=2\theta\,.
\eeq
We will show that all three are positive in the interior. Their absolute values then sum to $\pi-2\epsilon_Z<\pi$.

Let $\vartheta_Z=\arg\zeta_Z$ and $\zeta_+=r_{\rm N}e^{i\vartheta}$. At fixed $\varphi$ below~\eqref{E:S1KSWthreshold}, $q\vartheta<\pi/2$ by a finite margin. For large $Z$, the boundary conditions give
\beq
\label{E:KSWregulator}
 \epsilon_Z=\frac{\text{Im}\bigl(\zeta_+^{d-3}+\zeta_+^{d-1}\bigr)}{2Z^{d-1}}
 +o(Z^{-(d-1)})>0\,,\qquad
 q\vartheta_Z+\epsilon_Z<\frac{\pi}{2}\,.
\eeq
Note that the last inequality follows from assuming that KSW holds at the horizon as we take the asymptotic limit, which is the condition on $\varphi$ that we are assuming, and that the small finite time effect on $\epsilon$ does not change it. Indeed, both $(d-3)\vartheta$ and $(d-1)\vartheta$ lie between zero and $\pi$. The strict margin is why we do not include the threshold itself in this construction. Since $\text{Im}z(s)^{d-1}=(1-s)\text{Im}\zeta_Z^{d-1}>0$, we also have $\theta(s)>0$ for $0<s<1$.

To prove positivity of the radial phase, define $c=(d-3)/(d-1)$ and
\beq
 \mathcal A(z)=z^{d-1}-\zeta_Z^{d-1}-(z^{d-3}-\zeta_Z^{d-3})=z^{d-3}F(z)\,.
\eeq
The contour gives the exact identity
\beq
\label{E:KSWradialIdentity}
 \text{Im}\frac{\mathcal A(z(s))z(s)^{d-1}}{D^2}
 =\int_0^s dv\,\text{Im}\left[
 \frac{z(v)^{d-1}}{D}\left(1-\frac{c}{z(v)^2}\right)
 +(s-v)\left(1-\frac{c}{z(v)^2}\right)\right]\,.
\eeq
The second term is positive for $v<s$, because $\text{Im}(z^{-2})<0$. For the first, set
\beq
 \mathcal F(v)=\text{Im}\left[
 \frac{z(v)^{d-1}}{D}\left(1-\frac{c}{z(v)^2}\right)\right]\,.
\eeq
Then $\mathcal F'(v)=-c^2\text{Im}(z(v)^{-2})>0$, and the smooth-cap equation implies
\beq
 \mathcal F(0)=\text{Im}\frac{2i\zeta_Z^{d-2}}{(d-1) a_ZD}>0\,.
\eeq
The last inequality follows from $q\vartheta_Z+\epsilon_Z<\pi/2$. Thus the left side of~\eqref{E:KSWradialIdentity} is positive. Its argument is $\eta+2\epsilon_Z+2q\theta$, modulo $2\pi$. At the future endpoint it is $2\epsilon_Z$, which fixes the continuous branch throughout the contour. It follows that
\beq
\label{E:KSWradialBound}
 0<\eta+2\epsilon_Z+2q\theta<\pi\,.
\eeq
This proves $0<\arg\lambda_s<\pi$.

It remains to show that $\eta>0$. Near the cap, the exact equation $a_ZF'(\zeta_Z)=2i$ gives
\beq
 \eta(0^+)=\frac{\pi}{2}-\epsilon_Z-q\vartheta_Z>0\,.
\eeq
Away from the cap we use two overlapping late-time expansions. First, keep $x=sZ^{d-1}$ finite as $Z\to\infty$. Then $z^{d-1}=\zeta_+^{d-1}+x+o(1)$ and
\beq
\label{E:KSWinnerF}
 F(z)=\frac{x+\zeta_+^{d-3}}{(x+\zeta_+^{d-1})^{(d-3)/(d-1)}}-1+o(1)\,.
\eeq
The imaginary part of the first term is positive for every $x>0$. To see this, define
\beq
 g(x)=\arg(x+\zeta_+^{d-3})-\frac{d-3}{d-1}\arg(x+\zeta_+^{d-1})\,.
\eeq
It obeys $g(0)=g(\infty)=0$ and
\beq
\label{E:KSWinnerDerivative}
 g'(x)=-\frac{r_{\rm N}^{d-3}\sin((d-3)\vartheta)}{|x+\zeta_+^{d-3}|^2}
 +\frac{d-3}{d-1}\frac{r_{\rm N}^{d-1}\sin((d-1)\vartheta)}{|x+\zeta_+^{d-1}|^2}\,.
\eeq
Using $r_{\rm N}^2=(d-3)/(d-1)$, we find $g'(0)=2\sin\vartheta\cos(q\vartheta)/r_{\rm N}^{d-3}>0$. At large $x$, the coefficient of $x^{-2}$ is proportional to
\beq
 c^2\sin((d-1)\vartheta)-\sin((d-3)\vartheta)<0\,.
\eeq
This follows because $\sin y/y$ decreases for $0<y<\pi$. After multiplying~\eqref{E:KSWinnerDerivative} by its positive denominators, its numerator is a quadratic polynomial with a positive constant term and a negative leading coefficient. It therefore has one positive root. Thus $g$ first increases and then decreases to zero, remaining positive for $x>0$. Also $g<(d-3)\vartheta<\pi/2$, so there is no ambiguity from the sine of its argument.

In the outer region $sZ^{d-1}\gg1$, expansion of the exact ratio $F(z(s))/F(Z)$ gives
\beq
\label{E:KSWouterPhase}
 \eta(s)=\frac{1-s}{sZ^{d-1}}\mathcal C_\vartheta
 \left[1+O\bigl((sZ^{d-1})^{-2/(d-1)}\bigr)+O(Z^{-2})\right]\,,
 \qquad
 \mathcal C_\vartheta=\text{Im}\left(\zeta_+^{d-3}-c\zeta_+^{d-1}\right)>0\,.
\eeq
The positivity is the same inequality used for the large-$x$ sign above. The factor $1-s$ is fixed by the exactly real boundary condition and can be factored also in the error terms. This estimate therefore remains valid up to $s=1$, where $\eta=0$.

The two expansions establish positivity at a sufficiently large but finite cutoff. For example, divide the contour at $x=Z$. In the inner region, the limiting phase is positive, has a positive cap limit, and decays as $\mathcal C_\vartheta/x$ at large $x$. The corrections from~\eqref{E:KSWfiniteData} are uniformly $O(Z^{-2})$ for $0<x\leq Z$, so they are smaller than the limiting phase. In the outer region $x\geq Z$, the relative errors in~\eqref{E:KSWouterPhase} tend uniformly to zero. Combining this with~\eqref{E:KSWradialBound} proves the strict KSW inequality in the interior. At the cap, the polar coordinate degeneracy is removed using $\sqrt{s}$ as the disk radius. The phase sum is $\pi-2\epsilon_Z$, and $\epsilon_Z\to0$ as $Z\to\infty$. Thus, this particular limiting representative saturates KSW throughout, rather than only in its Lorentzian future. The result is a sequence of strictly allowable finite-cutoff contours, not an unregulated contour with a strict margin. It is a sufficient construction for the upper branch and does not classify all possible contours.

The finite-cutoff equations can also have roots which do not tend to either of the finite caps~\eqref{E:S1roots}. Squaring~\eqref{E:KSWfiniteData} gives
\beq
\label{E:KSWcutoffPolynomial}
 \bigl((d-1)\zeta_Z^2-(d-3)\bigr)^2
 =-4\varphi^2\zeta_Z^2\left[1-Z^{-2}
 -\frac{\zeta_Z^{d-3}(\zeta_Z^2-1)}{Z^{d-1}}\right]\,.
\eeq
It has degree $d+1$; its four finite roots include the two signs of the smoothness equation, and any root must be checked in the unsquared equations with the chosen sheet. The remaining roots scale as
\beq
 \zeta_Z\sim\left(\frac{d-1}{2\varphi}\right)^{2/(d-3)}
 Z^{(d-1)/(d-3)}e^{2\pi i k/(d-3)}\,.
\eeq
In particular, a positive real cap larger than $Z$ describes a purely Euclidean finite-cutoff filling, with a different circle normalization. Its action includes both bulk and boundary terms and grows positively as $Z^{(d-1)(d-2)/(d-3)}/G$. It is exponentially suppressed in the late-time limit and has no Lorentzian asymptotic region of the form used in the manuscript. These cutoff-dependent roots should therefore be distinguished from the finite late-time saddles studied here; squaring the cap equation alone does not determine their KSW admissibility or their coefficients in the GPI.

%--------------------------------------
\section{KSW contours for Lens-space Taub-Bolt saddles}
\phantomsection\label{VI:A:lens}
%--------------------------------------

We give the finite-cutoff contour construction for the Taub-Bolt saddles of \eqref{E:lensmetric}. The construction is asymptotic in a large boundary radius, at fixed $p$ and fixed squashing away from a degenerate bolt. It supplies a sufficient family of KSW contours, rather than a classification of all possible contours or a determination of the gravitational integration cycle.

%--------------------------------------
\subsection{Bolt roots and action}
%--------------------------------------

Impose real boundary data at a cutoff $\epsilon^{-1}$ by
\beq
\label{E:lenscutoff}
2\nu\sqrt{f(t_c)} = \frac{2\nu_0}{\epsilon}\,,\qquad
\sqrt{t_c^2+\nu^2} = \frac{1}{\epsilon}\,.
\eeq
Both $t_c$ and $\nu$ acquire small imaginary parts at finite $\epsilon$. In the limit $\epsilon\to0$, write $t_h|_{\epsilon=0}=ir_0$ and $m|_{\epsilon=0}=i\mu_0$. The smoothness conditions give
\begin{align}
\label{E:lensroots}
3r_0^2-\frac{pr_0}{2\nu_0}-3\nu_0^2+1 &= 0\,,&
r_{\pm} &= \frac{p\pm\sqrt{p^2-48\nu_0^2+144\nu_0^4}}{12\nu_0}\,,\\
\label{E:lensmass}
\mu_0 &= \frac{r_0^4+(1-6\nu_0^2)r_0^2+\nu_0^2(1-3\nu_0^2)}{2r_0}\,.
\end{align}
For $p\geq3$, the polynomial in \eqref{E:lensroots} has value $1-p/2<0$ at $r_0=\nu_0$, so $r_-<\nu_0<r_+$. The larger root has a negative two-sphere metric at the cap and violates KSW. For $0<\nu_0^2<1/3$, the smaller root has $r_0>0$ and $\mu_0>0$ and is the branch used below.

The exceptional low-$p$ cases need separate treatment. For $p=2$ the factorized roots are $\nu_0$ and $1/(3\nu_0)-\nu_0$. The root $r_0=\nu_0$ has a collapsed two-sphere and is not a regular bolt; it arises when the cap equation is multiplied by $\nu_0^2-r_0^2$. The other root satisfies $0<r_0<\nu_0$ precisely for $1/6<\nu_0^2<1/3$. For $p=1$, the sign argument excluding $r_+$ does not apply. In particular, both algebraic roots have $0<r_{\pm}<\nu_0$ when $(2+\sqrt3)/12<\nu_0^2<1/3$. We do not use a classification of the $p=1$ branches in the lens-space comparison.

With the action conventions of the main text, the bulk and Gibbons--Hawking terms give
\beq
\label{E:lensunrenaction}
S = \frac{4\pi\nu}{Gp}\left[-2t_c^3+2(1-4\nu^2)t_c-t_h^3-3\nu^2t_h-3m+O(t_c^{-1})\right]\,.
\eeq
It is essential to keep fixed the boundary metric, rather than $t_c$ and $\nu$. In particular,
\begin{align}
\label{E:lensfiniteboundary}
\nu &= \nu_0\left[1+\frac{1-4\nu_0^2}{2}\epsilon^2-i\mu_0\epsilon^3+O(\epsilon^4)\right]\,,\\
t_c &= \epsilon^{-1}-\frac{\nu_0^2}{2}\epsilon+O(\epsilon^3)
+i\mu_0\nu_0^2\epsilon^4+O(i\epsilon^6)\,.
\end{align}
These corrections multiply the divergent terms in \eqref{E:lensunrenaction} and change its finite imaginary part. The result is
\beq
\label{E:lensfiniteaction}
W_{\rm E} = 2\,\text{Im}\,S = \frac{8\nu_0S_0}{p}\left(r_0^3-3\nu_0^2r_0-\mu_0\right)\,,\qquad S_0=\frac{\pi}{G}\quad(d=4)\,.
\eeq
Substituting the smaller root gives \eqref{E:lensaction}. Its bounds can also be checked directly. Set $y=\nu_0^2$ and $D=p^2-48y+144y^2$. In the range $0<y<1/3$, $0<D<p^2$, and
\beq
\label{E:lensactionbound}
\frac{W_{\rm E}}{S_0}+\frac{2}{3}=\frac{p^3-D^{3/2}}{108py}>0\,.
\eeq
Moreover, convexity gives $D^{3/2}\geq p^3-72py(1-3y)$ and hence $W_{\rm E}<0$. For $p\geq3$ the endpoints are $W_{\rm E}\to0$ as $y\to0$ and $W_{\rm E}\to-2S_0/3$ as $y\to1/3$. The latter limit also holds on the genuine $p=2$ bolt branch.

%--------------------------------------
\subsection{Contour construction}
%--------------------------------------

Write the exact finite-cutoff endpoints as $t_h=x_h+iy_h$ and $t_c=x_c+iy_c$, and let $\Delta x=x_c-x_h$. Define a contour using a real parameter $s\geq0$ by
\begin{align}
\label{E:lenscontour}
x(s) &= x_h+\Delta x\frac{s}{1+s}\,,\\
y(s) &= y_c+\frac{\mu_0\nu_0^2\epsilon^3}{2}
\left[\frac{1}{x(s)+\nu_0/4}-\frac{1}{x_c+\nu_0/4}\right]+c\frac{\alpha}{\alpha+s}\,,\qquad t(s)=x(s)+iy(s)\,.
\end{align}
with constants $\alpha$ and $c$ that we will fix presently. The constant $c$ fixes the cap endpoint,
\beq
\label{E:lensc}
c = y_h-y_c-\frac{\mu_0\nu_0^2\epsilon^3}{2}
\left[\frac{1}{x_h+\nu_0/4}-\frac{1}{x_c+\nu_0/4}\right]
=r_0+O(\epsilon^2)\,.
\eeq
To fix the initial tangent, let $d_h=-f'(t_h)/4=d_x+id_y$. A smooth positive disk is obtained by taking $t-t_h=d_h\rho^2$ with $\rho$ real near the cap. Choosing
\beq
\label{E:lensalpha}
\alpha = -\frac{c}{\Delta x\left[\mu_0\nu_0^2\epsilon^3/(2(x_h+\nu_0/4)^2)+d_y/d_x\right]}
=r_0\mu_0\epsilon^4\left[1+O(\epsilon^2)\right]
\eeq
aligns the initial tangent with $d_h$. Thus $\alpha>0$ for sufficiently small $\epsilon$. Along the contour,
\beq
\label{E:lensG}
t' = x'(1-iG)\,,\qquad
G = \frac{\mu_0\nu_0^2\epsilon^3}{2(x+\nu_0/4)^2}
+\frac{c\alpha(1+s)^2}{\Delta x(\alpha+s)^2}>0\,,\qquad
\arg(-t'^2)=\pi-2\tan^{-1}G\,.
\eeq
Along the contour $t = t(s)$, the metric is diagonal in a real basis, with coefficients $-t'^2/f(t)$, $4\nu^2 f(t)$, and $t^2 + \nu^2$. The KSW sum is
\beq
\label{E:lenskswsum}
\Theta = \left|\arg\!\left(-\frac{t'^2}{f}\right)\right|
+|\arg(4\nu^2 f)|+2|\arg(t^2+\nu^2)|\,.
\eeq
We check $\Theta<\pi$ in the successive asymptotic regions of the contour.

Near the cap, $s\ll\alpha$, the disk is positive by construction and the remaining two-sphere has squared radius $\nu_0^2-r_0^2>0$. At $s=O(\alpha)$, the contour descends approximately along the imaginary $t$ axis from $ir_0$ toward zero. In terms of $t=ir$, the radial and fiber terms are positive for $0<r<r_0$, since $f(ir)>0$ on that interval. Both regions have a finite margin in the KSW inequality. The tangent turns when\footnote{For this we used that when $1\gg s \gg \alpha$ we have that $G \approx 8\mu_{0}\epsilon^{3}+\frac{\mu_{0}r_{0}^{2}\epsilon^{5}}{s^{2}}$ since $\Delta x \sim \epsilon^{-1}$ and $x$ is near zero in that region for small $\epsilon$.}
\beq
\label{E:lensturn}
s\sim r_0\sqrt{\mu_0}\,\epsilon^{5/2}\,,\qquad x-x_h\sim r_0\sqrt{\mu_0}\,\epsilon^{3/2}\,.
\eeq
At fixed ratio to this transition scale, $t$ is still small and $G$ is of order one, leaving a finite phase margin. Farther along the contour, while $x\to0$ and the tangent becomes almost horizontal, the deficit $2\tan^{-1}G$ is larger than the $O(\epsilon^3)$ corrections to the KSW sum. The region with $x$ fixed is checked explicitly next.

For $x$ fixed and positive as $\epsilon\to0$, the leading terms are
\begin{align}
\label{E:lensintermediate}
y &= \mu_0\epsilon^3\left[\frac{\nu_0^2}{2(x+\nu_0/4)}+\frac{r_0^2}{x}\right]+\cdots\,,\\
G &= \mu_0\epsilon^3\left[\frac{\nu_0^2}{2(x+\nu_0/4)^2}+\frac{r_0^2}{x^2}\right]+\cdots\,,\qquad
\arg\nu^2=-2\mu_0\epsilon^3+\cdots\,.
\end{align}
Here $0<\arg f<\pi$ because the leading imaginary part of $f(x)$ is $2\mu_0x/(x^2+\nu_0^2)>0$. Its phase cancels between the radial and fiber terms in \eqref{E:lenskswsum}. Set $X=x/\nu_0$ and $\eta=r_0/\nu_0<1$. Then
\beq
\label{E:lensF}
\Theta = \pi+\mu_0\epsilon^3 F(X,\eta)+\cdots\,,\qquad
F = -2-\frac{1}{(X+1/4)^2}-\frac{2\eta^2}{X^2}
+\frac{2}{1+X^2}\left|2\eta^2-2+\frac{X}{X+1/4}\right|\,.
\eeq
To verify the sign, put $b=2-X/(X+1/4)$, so $1<b<2$ and $|2\eta^2-b|<b$. Dropping the negative term $-2\eta^2/X^2$ gives
\beq
\label{E:lensFbound}
F < -\frac{2\left(16X^4+8X^3+9X^2-4X+7\right)}{(4X+1)^2(1+X^2)}<0\,,
\eeq
where $9X^2-4X+7=9(X-2/9)^2+59/9$. This checks the phase inequality throughout the intermediate region; its small- and large-$X$ limits overlap the adjoining expansions.

Finally, at $x=O(\epsilon^{-1})$, set $u=s/(1+s)$. To leading order $x=u\epsilon^{-1}$, and
\begin{align}
\label{E:lenslate}
y &= \mu_0\epsilon^4\left[\nu_0^2+\frac{1-u}{u}\left(\frac{\nu_0^2}{2}+r_0^2\right)\right]+\cdots\,,\\
G &= \frac{\mu_0\epsilon^5}{u^2}\left(\frac{\nu_0^2}{2}+r_0^2\right)+\cdots\,,&
\arg f &= \frac{2\mu_0\epsilon^3}{u^3}+O(\epsilon^5)\,,\\
\arg(\nu^2 f) &= 2\mu_0\epsilon^3(u^{-3}-1)+O(\epsilon^5)\,,&
\arg(t^2+\nu^2) &= O(\epsilon^5)\,.
\end{align}
Consequently,
\beq
\label{E:lenslateksw}
\Theta = \pi-2\mu_0\epsilon^3+O(\epsilon^5)<\pi\,.
\eeq
At the endpoint the spatial metric is exactly real by \eqref{E:lenscutoff}. Keeping the finite-cutoff imaginary parts is essential: setting them to zero prematurely would leave a saturated Lorentzian KSW sum in the asymptotic region.

This construction applies at fixed $0<\nu_0^2<1/3$ on the smaller bolt branch for $p\geq3$, and at fixed $1/6<\nu_0^2<1/3$ for the genuine $p=2$ bolt. The expansions are not uniform as the bolt degenerates. For $\nu_{0}^{2} >\frac{1}{3}$, the smaller root becomes negative. A positive disk at the bolt initially decreases $\text{Re}(\nu r)$, whereas reaching the late-time boundary requires an overall increase, contradicting the monotonicity established in the discussion around \eqref{eqargnu1}

\end{appendix}

\bibliography{anisotropy}
\bibliographystyle{JHEP}

\end{document}